\documentclass[fleqn,usenatbib]{mnras}
\usepackage{newtxtext,newtxmath}
\usepackage{float}
\usepackage{movie15}
\usepackage{animate}

\usepackage[T1]{fontenc}
\usepackage{amsmath}
\usepackage{subcaption}
\DeclareRobustCommand{\VAN}[3]{#2}
\let\VANthebibliography\thebibliography
\def\thebibliography{\DeclareRobustCommand{\VAN}[3]{##3}\VANthebibliography}

\usepackage{graphicx}	

\title[SIDM subhalo]{Core-collapse, evaporation and tidal effects: the life story of a self-interacting dark matter subhalo}

\author[Z. Zeng et al.]{
Zhichao Carton Zeng,$^{1, 2}$\thanks{E-mail: zeng.408@osu.edu}
Annika H. G. Peter,$^{1, 2, 3}$
Xiaolong Du,$^{4}$
Andrew Benson,$^{4}$ 
\newauthor
Stacy Kim,$^{5}$
Fangzhou Jiang,$^{4,6}$
Francis-Yan Cyr-Racine$^{7}$ and
Mark Vogelsberger$^{8}$
\\
$^{1}$Department of Physics, The Ohio State University, 191 W. Woodruff Ave., Columbus OH 43210, USA\\
$^{2}$Center for Cosmology and Astroparticle Physics, The Ohio State University, 191 W. Woodruff Ave., Columbus OH 43210, USA\\
$^{3}$Department of Astronomy, The Ohio State University, 140 W. 18th Ave., Columbus OH 43210, USA\\
$^{4}$ Carnegie Observatories, 813 Santa Barbara Street, Pasadena CA 91101, USA\\
$^{5}$Department of Physics, University of Surrey, Guildford, GU2 7XH, United Kingdom\\
$^{6}$ TAPIR, California Institute of Technology, Pasadena, CA 91125, USA \\
$^{7}$ Department of Physics and Astronomy, University of New Mexico, 210 Yale Blvd NE, Albuquerque NM 87106, USA
\\
$^{8}$ Department of Physics, Massachusetts Institute of Technology, 77 Massachusetts Avenue, Cambridge, MA 02139, USA
}

\date{Accepted XXX. Received YYY; in original form ZZZ}

\pubyear{2021}

\begin{document}
\label{firstpage}
\pagerange{\pageref{firstpage}--\pageref{lastpage}}
\maketitle

\begin{abstract}
Self-interacting dark matter (SIDM) cosmologies admit an enormous diversity of dark matter (DM) halo density profiles, from low-density cores to high-density core-collapsed cusps. The possibility of the growth of high central density in low-mass halos, accelerated if halos are subhalos of larger systems, has intriguing consequences for small-halo searches with substructure lensing. However, following the evolution of $\lesssim 10^8 M_\odot$ subhalos in lens-mass systems ($\sim 10^{13}M_\odot$) is computationally expensive with traditional N-body simulations. In this work, we develop a new hybrid semi-analytical + N-body method to study the evolution of SIDM subhalos with high fidelity, from core formation to core-collapse, in staged simulations. Our method works best for small subhalos ($\lesssim 1/1000$ host mass), for which the error caused by dynamical friction is minimal. We are able to capture the evaporation of subhalo particles by interactions with host halo particles, an effect that has not yet been fully explored in the context of subhalo core-collapse. 
We find three main processes drive subhalo evolution: subhalo internal heat outflow, host-subhalo evaporation, and tidal effects. 
The subhalo central density grows only when the heat outflow outweighs the energy gain from evaporation and tidal heating. Thus, evaporation delays or even disrupts subhalo core-collapse. We map out the parameter space for subhalos to core-collapse, finding that it is nearly impossible to drive core-collapse in subhalos in SIDM models with constant cross sections. Any discovery of ultra-compact dark substructures with future substructure lensing observations favors additional degrees of freedom, such as velocity-dependence, in the cross section.

\end{abstract}




\section{Introduction} \label{sec:intro}

Even after decades of being widely accepted as the most plausible answer to astrophysical anomalies such as the mass deficit in galaxy rotation curves \citep{Rubin_1978} and the spatial offset of the total mass center from the luminous matter center during cluster collisions \citep[e.g., the Bullet Cluster;][]{Clowe_2006}, dark matter's physical nature remains essentially unknown to us. 
The standard cold dark matter (CDM) paradigm has been well-tested on large, cosmological scales \citep{Planck2018, SDSS}, but is challenged on galactic scales \citep{Bullock_2017,Buckley:2017ijx}. The tensions between CDM predictions and galactic observations can be sorted into three categories. First, there is the 2000's-era mismatch between the small number of observed satellite galaxies and the large number of galaxy-scale subhalos in simulation --- the missing satellites problem \citep{Klypin_1999, Moore_1999}.  Recently, the discovery of many new satellites, a careful accounting to survey selection functions, and insights from simulation work instead now hint that, CDM may underpredict the abundance of luminous satellites \citep[][]{willman2005a, zucker2006, Tollerud_2008, Koposov_2009, Walsh_2009, 2009MNRAS.397.1748B, drlica-wagner2015, Laevens15, Torrealba16b, GK_2017, Newton:2017xqg, Kim_2018, 2018MNRAS.473.2060J, 2018MNRAS.475.5085T, Kelley_2019, 2019PASJ...71...94H, Nadler_2020b, Kim_2021}. Second, the dynamical mass in the inner region of dwarf galaxies (in both the dwarf spheroidal satellites of the Milky Way as well as rotation-supported field dwarfs) is low compared to CDM simulations without baryons, which is classically understood as an issue of core vs. cusp in the shape (slope) of density profiles \citep{Moore_1994,Adams_2014, Oh_2015}.  Recently,  
this problem has expanded, with discoveries that observed dwarf galaxies display a large diversity in rotation curves as well as stellar dynamics \citep[diversity problem;][]{KdN_2014, Oman_2015,2018MNRAS.481.5073E,2019MNRAS.484.1401R, Relatores_2019, Zavala_2019, Santos-Santos_2020, Li_2020, Hayashi_2020}. The two former problems converge in the third problem, in the dynamical properties of the most luminous satellites of the Milky Way, showing that CDM halos whose central densities match the bright satellites are much lower in mass than expected from abundance matching \citep[too-big-to-fail problem;][]{BK_2011, BK_2012, Tollerud_2014, Jiang_2015, Kaplinghat_2019}.

Introducing baryons into simulations has been shown to have the potential of addressing the small scale problems within the CDM framework. Baryon physics determines which halos contain luminous baryons at all \citep{White_1978, 2000ApJ...542..535G,2002MNRAS.333..177B,2008MNRAS.390..920O}. Various baryonic feedback processes, such as stellar feedback and supernova explosion, help with the redistribution of dark matter and core formation in the halo center \citep{Governato_2012, 2014ApJ...786...87B, Chan_2015, 2016MNRAS.459.2573R, Wetzel_2016, Fitts_2017}. The shallower potential, as a consequence, also leads to more satellite galaxies being tidally disrupted, decreasing the abundance of satellites predicted by CDM \citep{2010ApJ...709.1138D, 2013ApJ...765...22B, Zhu_2016, GK_2017, Despali_2017}. On the other hand, if the mass fraction of stellar objects is high ($\sim10\%$), this extra baryonic potential could possibly compensate for the feedback and make a cuspy central density, potentially fitting in the cuspy side of the diversity problem \citep{DC_2014, Tollet_2016, Hopkins_2018, Lazar_2020}. 

Alternatively, modifications to CDM have been proposed to alleviate the tension on small scales while preserving the large scale success. A large category of such modifications is  hidden-sector models, in which non-gravitational beyond-standard-model interactions among dark matter particles are generic features \citep{Spergel_2000, DM_particle_2017}.
In these self-interacting dark matter models (SIDM) the scattering, often (but not always) elastic, leads to momentum exchange between dark matter particles, thermalization of particles in halos, and thus the formation of cored density profiles in the center of the dark halo.  This halo phenomenology has been demonstrated with numerical N-body simulations with only dark matter \citep{Dave_2001, Colin_2002, Vogelsberger_2012, Rocha_2013, Zavala_2013, Elbert_2015, Vogelsberger_2019}, and  analytically via isothermal Jeans equations \citep[][Jiang et al in prep.]{Kaplinghat_2014, Valli_2018, Robertson_2021}. The shallower gravitational potential wells of subhalos result in them being more vulnerable to tidal stripping, potentially alleviating the tension in satellite abundance \citep{Colin_2002, Vogelsberger_2012, Robles_2019, Dooley_2016}, or otherwise causing this tension to grow with the satellite completeness correction \citep["too-many-satellites" problem;][]{Kim_2021, Souradip_2021, Nadler_2020}.

The cross section per unit mass, $\sigma/m$, is the key parameter to characterize a specific SIDM model. To resolve the density problems on galaxy scales, $\sigma/m \gtrsim 1\ \rm cm^2/g$ is needed \citep[][]{Zavala_2013, Kaplinghat_2016, Valli_2018, Ren_2019}, which is in conflict with observations on cluster scales: cluster central density \citep[$\sim 0.1\ \rm cm^2/g$;][]{ Rocha_2013, Elbert_2018}, cluster ellipticity \citep[$\lesssim 1\ \rm cm^2/g$;][]{Peter_2013, Kaplinghat_2014, Robertson_2019, McDaniel_2021}, satellite galaxy structure \citep[$\lesssim 0.3~\rm cm^2/g$;][]{gnedin2001,Natarajan_2002}, and the mass-to-light-ratio and miscentering between the stellar and dark matter of cluster mergers \citep[$\lesssim \mathcal{O}(1)\ \rm cm^2/g$;][]{Randall_2008, Robertson_2017, Kim_2017}. Therefore, SIDM models with velocity-dependent cross sections have been suggested, with both theoretical motivation from particle physics in the context of hidden-sector models \citep[][]{Feng_2009,Tulin:2012wi, Tulin_2013, Boddy_2014, Cline_2014, Cyr-Racine:2015ihg, Tulin_2018}, and simulation efforts on cluster and galaxy scales \citep[][]{Vogelsberger_2012, Zavala_2013, Nadler_2020, Banerjee_2020, Turner_2020, Correa_2021}. Other variants of SIDM that have more degrees of freedom are being studied, such as energy dissipation during the dark matter two-body scattering \citep[][]{Schutz:2014nka,Essig_2019, Vogelsberger_2019, Huo_2020, Shen_2021, Chua_2021}, anisotropic scattering \citep[][]{Kahlhoefer_2014, Robertson_2017b}, and multi-state scattering \citep[][]{Vogelsberger_2019}.

Although low-density cores were long considered the trademark signature of SIDM, a dramatically different phenomenology may take over the central density of the SIDM halo, long after the core-formation phase. 
SIDM scatterings lead to heat outflow from the inner, hotter region to the outer, cooler layers of the halo. Such heat loss in the inner region results in the infall of dark matter particles to more bound orbits, where they become even hotter than before. Due to this negative heat capacity of the self-gravitating system, the trend of dark matter in-fall accelerates itself as the negative heat gradient steepens, leading to a denser and more cuspy central density, known as the core-collapse process or gravothermal catastrophe \citep{LB1968, Kochanek_2000, Colin_2002, Balberg_2002, Koda_2011, Pollack_2015, Zavala_2019, Essig_2019, Nishikawa_2020, Sameie_2020, Turner_2020, Correa_2021}. Indeed, while low-density cores may be formed in other non-CDM cosmologies, core-collapse is a unique signature of SIDM, distinguishing it from other alternative DM models. The two phases of core-formation and core-collapse suggest that dark matter halos can have a large diversity of central densities and profiles in an SIDM Universe, thus potentially capable of solving the diversity problem of galaxy rotation curves \citep{Kamada_2017, Zavala_2019, Sameie_2020, Kaplinghat_2020}.  The high density also makes subhalos robust to destruction, avoiding a ``too many satellites" problem \citep[][]{Kim_2018, Kelley_2019, Kim_2021}. 

However, the time scale for SIDM core-collapse to happen is much longer than the age of the Universe unless the cross section is large \citep[][see also Sec. \ref{sec:result1} in this work]{Balberg_2002, Koda_2011, Elbert_2015, Essig_2019}. Excitingly, it has been recently reported that tidal stripping by the host halo can noticeably accelerate core-collapse of the subhalo, since the removal of dark matter from the outer layers helps with the formation of a negative temperature gradient and makes heat outflow more efficient \citep[][also Sec. \ref{sec:result2} in this work]{Nishikawa_2020, Correa_2021}. This tidal acceleration of core-collapse has been used to explain the diversity of the Milky Way's dwarf spheroidal galaxies \citep{Kahlhoefer_2019, Sameie_2020}. 

Apart from observations of the Milky Way's satellite galaxies, another way to detect core-collapse is through substructure lensing. Substructure lensing is the perturbation to a strongly lensed system, due to the existence of dark matter substructures/subhalos in the foreground lens  \citep[with in-field halos along line-of-sight as a systematic; ][]{Mao_1998, Dalal_2002, Chiba_2002}. Properties of subhalos can be inferred from the perturbed lensing image, either by flux anomalies of lensed quasars \citep{Dalal_2002, Xu_2009, Gilman_2018, Gilman_2020a, Gilman_2020b}, or distorted images of lensed galaxies \citep{Koopmans_2005, Vegetti_2009, Vegetti_2010, Vegetti_2014, Minor_2020}, thereby setting constraints on dark matter physics. Observed substructure lensing systems so far have shown subhalo mass function in consistency with CDM \citep{Dalal_2002, Vegetti_2014, Hsueh_2020, Gilman_2020a}. Core-collapsed subhalos, if they do exist, are promising targets because their highly concentrated mass distributions leave greater distortions on strongly lensed images than cored or CDM substructures \citep{Gilman_2021, Yang_2021}.

A problem for theoretical predictions, though, is that substructure lensing can be used to probe subhalos down to $10^6 M_\odot$---a factor of $10^7$ smaller than typical host halos as main lenses---which is too computationally expensive to resolve in cosmological simulations, especially since ensembles of simulated systems are required. We are thus motivated to build a hybrid semi-analytical + N-body method capable of simulating the evolution of such small SIDM subhalos in this work, where we implement the host halo in the form of an analytic density profile, while tracking the subhalo with N-body particles. Compared to previous works with a similar approach \citep{Kahlhoefer_2019, Sameie_2020}, our method includes not only gravitational interaction between the host and subhalo, but also the dark matter self-interaction. The latter, which we denote as ``evaporation" for the rest of this paper (also known as ``SIDM ram pressure" in \citealt{Kummer_2018}, \citealt{Nadler_2020} and \citealt{Jiang_2021}), leads to energy gain and mass loss in the subhalo, and we show to be significant for the onset of the subhalo core-collapse  (Sec. \ref{sec:val2}  and Sec. \ref{sec:result3}). 

Unlike cosmological simulations (or zoom-ins of them) with SIDM which mostly concentrate on the host halo, or statistics of subhalos such as the mass function and radial distribution \citep{Robles_2019, Banerjee_2020}, in this study we mainly focus on the evolution and possible core-collapse of individual subhalos, using idealized ensembles of one-host-one-subhalo systems. 
Our work is structured as follows: in Sec.~\ref{sec:method} we introduce our semi-analytic method for subhalo evaporation.  In Sec.~\ref{sec:val} we present our code validation procedure, which includes comparing to live host simulations for SIDM scenarios of  constant and velocity-dependent cross sections, and for subhalos that remain cored or are core-collapsing. 
Sec.~\ref{sec:results} is the heart of the paper, in which we present results from a series of production runs with our method.  The major theme of this section is tracking the detailed evolution of SIDM subhalo central density, as a sign of whether and when the subhalo core-collapses, and describing how the relevant physical processes imprint themselves on the evolution of the central density.  We also map out the parameter space needed for subhalo core-collapse, which generally requires an initial subhalo concentration much higher than cosmological values, when the SIDM model is velocity-independent.
 In Sec.~\ref{sec:s&d} we summarize and discuss our work.

\section{Method} \label{sec:method}

There are two major effects from the host halo on the subhalo:  the gravitational/tidal effects (including tidal heating and stripping), and the host-subhalo evaporation, rooted from the self-interaction between the host and subhalo dark matter. The latter, which matters because of the large gap between the velocities of host and subhalo dark matter particles in a paired two-body scattering, has not been included in previous semi-analytical approaches \citep{Kahlhoefer_2019, Sameie_2020}, nor systematically studied in the context of subhalo core-collapse. The evaporation effect blows away particles throughout the whole subhalo (note: we use the term 'evaporation' to denote the host-subhalo dark matter interaction throughout this paper, because typically most of the subhalo particles will become unbound --- but not necessarily all of them; see Sec. \ref{sec:result3} and Appendix \ref{appendix:heating} for more detailed discussion), leading to its additional mass loss and halting the process of core-collapse.

In this work, we use a quasi-analytic approach to model this evaporation effect. For each subhalo N-body particle, we evaluate the evaporation probability based on knowledge of the particle's position and velocity with respect to the host center.  
This approach, with any user-specified analytic host profile, can also be easily extended to embed additional analytic potentials, such as those due to baryons. The probability that a certain subhalo particle scatters off a virtual host particle (`virtual' because the simulation does not really include any host particles, we only sample them when evaluating the host-sub scatterings) is evaluated as

\begin{equation}
\label{eqn:ph}
P_h = \delta_t \frac{\sigma_{\rm T}}{m} \rho_h \overline{|\textbf{v}_h - \textbf{v}_s|},
\end{equation}
where $\delta_t$ is the time between the previous and the current timestep, $\sigma_{T}/m$ is the self-interaction transfer cross section per unit mass (since SIDM particles are identical in the simulation, both forward and backward scatterings are suppressed in momentum transfer, thus we consider the definition of $\sigma_T = \int\frac{d\sigma}{d\Omega} (1-{\rm cos}^2 \theta) d\Omega$; see \citealt{Tulin_2013, Cline_2014, Boddy_2016, Tulin_2018}), $\rho_h$ is the local SIDM mass density of the host at the position of the target subhalo particle, and $\overline{|\textbf{v}_h - \textbf{v}_s|}$ is the mean relative velocity between this subhalo particle and virtual host particles in its vicinity. Since the scattering probability between any two DM particles depends linearly on their relative velocity, the expected evaporation probability $P_h$ also scales linearly with the mean relative velocity $\overline{|\textbf{v}_h - \textbf{v}_s|}$. Here $\textbf{v}_s$ is the velocity of the well-tracked subhalo particle.  Since we do not have real host particles in the simulation, we evaluate this mean relative velocity with a statistical approach. We have explicitly checked the isotropy of host particle velocities with host-only runs, and found that $\sigma_{vx} \approx \sigma_{vy} \approx \sigma_{vz} \equiv \sigma_{vh}$ and that $\bar{v}_x \approx \bar{v}_y \approx \bar{v}_z \approx0$. Thus we assume $\textbf{v}_h$ to be a Gaussian distribution centered at 0 with standard deviation value $\sigma_{vh}$, directly measure the radial distribution $\sigma_{vh}(r)$ in these host-only runs, and load it in a tabulated form in the simulation to parameterize the velocity of the host halo particles. For convenience of calculation, we build the coordinate system such that $z$-axis is aligned with $\textbf{v}_s$, finding

\begin{equation} \label{eqn:vrel}
\begin{split}
    \overline{|\textbf{v}_h - \textbf{v}_s|} &= \int |\textbf{v}_h - \textbf{v}_s| P(\textbf{v}_h) d^3 \textbf{v}_h \\
    &= \iiint \sqrt{v^2_h \sin^2{\theta} + (v_h \cos\theta - v_s)^2}\frac{1}{(2\pi)^{3/2} \sigma_{vh}^3}  \\
    &\ \ \times \exp\left({-\frac{v^2_h}{2\sigma_{vh}^2}}\right)  v^2_h \sin{\theta} d\theta d\phi dv_h \\
    &= \sqrt{\frac{2}{\pi}} \sigma_{vh} \exp{\left(\frac{-v^2_s}{2\sigma_{vh}^2}\right)} +\left(v_s + \frac{\sigma_{vh}^2}{v_s}\right) \textrm{Erf} \left(\frac{v_s}{\sqrt{2}\sigma_{vh}}\right),
\end{split}
\end{equation}
where $v_s$ and $v_h$ are the norm of $\textbf{v}_s$ and $\textbf{v}_h$, and $\rm Erf$ refers to the Gauss error function.

We incorporate our semi-analytical treatment of the evaporation process as a patch to the \texttt{Arepo} code \citep{Springel_2010}, which already has a well-tested SIDM module built in \citep{Vogelsberger_2012, Vogelsberger_2013, Vogelsberger_2014}. At each timestep in our simulation, for each gravitationally active subhalo particle, it is probabilistically determined whether it scatters with another subhalo particle (probability $P=P_s$), is `evaporated' via scattering with a virtual host particle ($P=P_h$), or does not scatter at all ($P=1-P_s-P_h$).  To determine this, we compare the corresponding probabilities with a random number $X\in (0, 1)$.  We evaluate $P_s$ of a subhalo particle labelled $i$ with the default SIDM module of \texttt{Arepo}, by summing up the two-body scattering probability over its neighbors. $P_h$ is evaluated according to Eqs.~\eqref{eqn:ph} and \eqref{eqn:vrel}. In terms of the choice of the scattering partner of particle $i$, if $ X < P_s$, the first neighboring subhalo particle $j$ that satisfies $\sum_{j} P_{ij} \geq X$ is assigned to be paired with particle $i$ \citep{Vogelsberger_2012}.  If $P_s< X < P_s + P_h$, we sample the 3D velocity of a virtual host particle with a Gaussian distribution centered at 0 and has the standard deviation $\sigma_{vh}$ measured from our host-only SIDM simulation, to be the partner of subhalo particle $i$. During the two-body scattering process, both particles have their velocities redistributed spherically symmetrically relative to the center of mass of this two-body system. After the scattering, the sampled virtual host particle velocity is discarded, therefore the analytic host is always static in our simulation and there is no feedback from the subhalo on the host, which is a reasonable approach when we choose a subhalo that is small compared to the host. In this work, we have only considered elastic, isotropic, and single-state scattering processes.  

Because our goal is to determine when (sub)halos start core-collapsing, we must determine a time-stepping algorithm to resolve the core-collapse, and define a criterion for stopping the simulation before the core goes too far into the fluid regime.  The central density of a halo going into the core-collapse phase can be high, and so is the scattering probability per unit time. Thus we dynamically confine the simulation timestep $\delta_t$ of each particle to make sure that its total scattering probability is always smaller than 1. The default SIDM module of \texttt{Arepo} has such a setup built in, evaluating the SIDM timestep $\delta_t ({\rm sidm})$ and gravity timestep $\delta_t({\rm grav})$, and selecting the smaller of them to be $\delta_t$. But it also limits how small $\delta_t(\rm sidm)$ can be, relative to $\delta_t(\rm grav)$, preventing the timestep from going arbitrarily small. In this work we effectively disable this lower limit on $\delta_t(\rm sidm)$ and terminate the simulation if the halo central density grows to 100 times the initial central density. 
This termination is necessary for these particle-type simulations for two reasons.  First, although we can currently afford the reduction of the timesteps by 3 to 4 orders of magnitude compared to a similar-scale CDM simulation, the computational cost is formidable when the central density gets much higher.  Second, when the central density is too high, a larger number of particles are located near or under the force resolution limit, characterized by $\sim 3$ times the softening length, as we will show in the code validation part, and also discussed in \citealt{Turner_2020}.
We argue that additional analytical methods or hydrodynamic simulation techniques (such as in \citealt{Kummer_2019}) should be used to better understand the physics in this ultra-dense region, which is beyond the scope of this work. 

To track the process of SIDM (sub)halo core-collapse, we measure the central density of the (sub)halo by marking the 50 simulation particles that have the greatest local densities, and defining the average as $\rho_{\rm cen50}$. This set of the top 50 particles ranked by local density is updated at each simulation timestep, not each snapshot output time. For our current core-collapse simulations, we define a tentative collapse criterion of
\begin{equation}\label{eq:cc}
    \rho_{\rm cen50} = 100\rho_{\rm cen50}(t=0),
\end{equation}
where we terminate the simulation.

Compared to fully live host runs, our semi-analytic method reduces the computational time by over two orders of magnitude for cored subhalos with $1/1000$ the mass of the host, and by a factor of a few tens for core-collapsing subhalos.

\section{Code Validation} \label{sec:val}

\begin{figure}
    \begin{subfigure}{\columnwidth}
        \centering
    	\includegraphics[width=\textwidth]{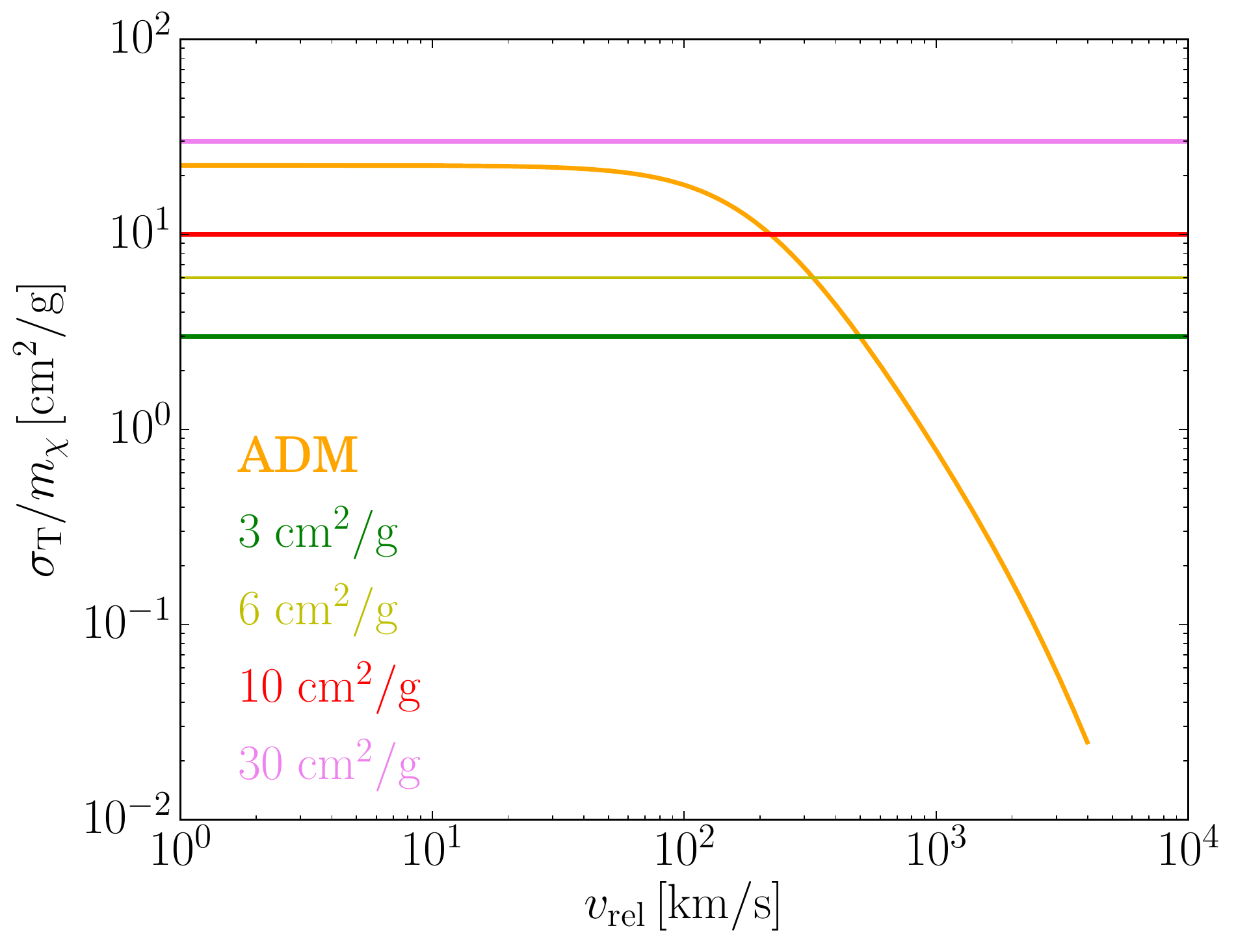}
	    \caption{}
	    \label{fig:xsectiona}
	\end{subfigure}
	\hfill
	\begin{subfigure}{\columnwidth}
	    \centering
	    \includegraphics[width=\textwidth]{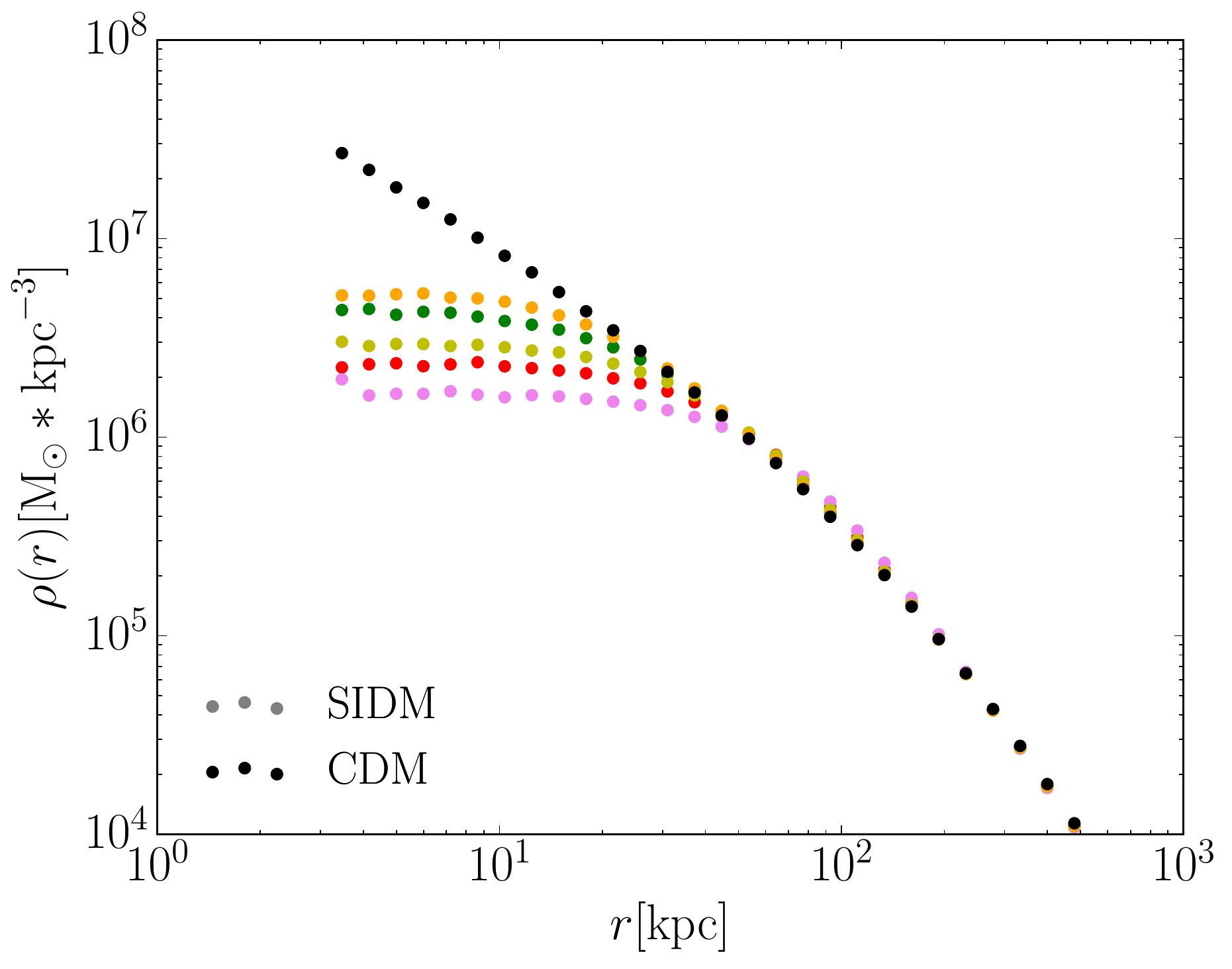}
    	\caption{}
    	\label{fig:xsectionb}
	\end{subfigure}
	\caption{SIDM cross sections and corresponding host density profiles. \textbf{Top:} SIDM cross sections $\sigma_{\rm T}/m=3~{\rm cm^2/g}, 6~{\rm cm^2/g}, 10~{\rm cm^2/g}, 30~{\rm cm^2/g}$ are shown in green, yellow, red and pink colors. One specific choice of velocity-dependent SIDM model, the atomic dark matter model (ADM) from \protect\citealt{Cline_2014}, is shown in orange. The fine-tuned parameters of the ADM model are described in Sec. \ref{sec:valv}. \textbf{Bottom:} The corresponding host density profiles after evolution for 5 Gyr. Colors are the same as above. The CDM counterpart is shown by black dots. } 
    \label{fig:xsection}
\end{figure}

In this section, we validate our evaporation model against fully live host simulations, testing both constant cross sections and a velocity-dependent cross section. A subhalo is injected into either an analytical or live host at its orbit apocenter. 
The evolution of its bound mass and its detailed density/mass profile are tracked and compared. Throughout this work, we use $R_{200m}$ to define the boundary of a halo, within which the mean density is 200 times the mean matter density of the Universe. We assume a flat $\Lambda$CDM cosmology with Hubble parameter $H = 70$ km/s/Mpc and $\Omega_m = 0.3$. We as well fix the host halo mass $M_{\rm host} = 10^{13.5}M_\odot$ ($R_{200m}\sim 1000\ \rm kpc$), which is typical of a strong gravitational lens system \citep{Birrer_2017}.

To implement the analytic host consistently, for each SIDM cross section, we measure the density profile of the live host halo after it reaches equilibrium with DM self-interaction and has a cored host density profile (see Sec. \ref{sec:val2} for details). Some of the SIDM cross sections we use in this paper and corresponding host density profiles are presented in Fig. \ref{fig:xsection}. We then load these host density profiles in tabulated form to \texttt{Arepo}, and log-linearly interpolate (i.e. log$[\rho_h$] scales linearly with log$[r]$) between these measured data points, as the input $\rho_h(r)$ in Eq. \ref{eqn:ph}. This external input of $\rho_h(r)$ also accounts for the host gravitational field.  To test SIDM models with a velocity-dependent cross section (which we plan to explore in detail in future work), we select an atomic dark matter model (hereafter `ADM') from \citealt{Cline_2014} for the purpose of validation. We present its $\sigma_{\rm T}/m - v_{\rm rel}$ relation in Fig. \ref{fig:xsectiona} and the host density profile as the orange scattered dots in Fig. \ref{fig:xsectionb}. 

We note that in this work, the analytic host density profile remains static in time, so effectively we are simulating the evolution of late-accreted subhalos, after the thermal equilibrium within the host center has been reached. A time-dependent analytic host density will be used for a more realistic modelling of SIDM subhalo population in the future.

As we show in the next two sections, our semi-analytic evaporation method is in good agreement with the live host simulations, in both the low-density-cored and core-collapsing subhalos. 
However, we find two subtle systematics that, while not affecting our overall results, are important to highlight and describe here before we show our validation results.  First, in the live host simulations, the subhalo accretes dark matter from the host as it travels along its orbit. 
However, we are able to distinguish the captured host dark matter from the original subhalo dark matter when measuring the mass profile of the subhalo. This systematic of host-particle-capturing accounts for only $\sim 1\%$ for subhalos with mass $1/1000$ of the host. 

Second, dynamical friction in the live host runs results in the shrinking of subhalo orbits, thus affecting both evaporation and tides. 
When the whole host is modelled as an analytical field, dynamical friction cannot be easily represented without introducing more complex semi-analytical models \citep{Penarrubia_2005, BK_2008, Petts_2015} that would add to the complexity of this work. Dynamical friction slows down the bulk motion of the subhalo as it orbits around the host, leading to a stronger orbit decay and earlier arrival at pericenter, where both evaporation and tidal effects are strongest. However, its importance is significantly reduced when the subhalo mass is small relative to the host. We present a comparison between sub-host mass ratios of 1:1000 and 1:100 in the next section and show that the discrepancy is much smaller for the former case, and that the accuracy of the $1:100$ case is noticeably improved with inclusion of the dynamical friction model from \citealt{Petts_2015} (see Appendix \ref{appendix:dyn} for details). 

Thus we find that our approach of including an analytic host works best for the evolution of relatively small subhalos. This is the regime that is computationally challenging for simulations with live hosts, and modelling these small halos is a main reason behind our analytic approach.  In short, we show that the systematic errors in our approach become vanishingly small for our target use cases.  

\subsection{SIDM with constant cross section}\label{sec:val2}
In this section, we validate our method of subhalo evolution for SIDM with a constant cross section $\sigma_{\rm T}/m=6\ \rm cm^2/g$.
The simulated system is composed of a host halo with mass of $M_{200m}=10^{13.5} M_\odot$, and a subhalo with mass ratio 1:1000 (or 1:100 for a comparison case) with respect to the host. This choice of host-subhalo mass ratio is aimed to reduce dynamical friction and accretion of host dark matter as much as possible, while keeping the live host simulations computationally inexpensive.

Both the host and the subhalos are generated with initial NFW profiles using the \texttt{SpherIC} code \citep{GK_2013}. The concentration $c_{200\rm m}$ of the host is 6.5, while that of the subhalo is set to be 14.8, 40 and 80 for three control groups. As a reference, the cosmological value of $c_{200 \rm m}$ for the subhalo at redshift $z=0$ is around 24 \citep{Duffy_2008, WMAP}. Thus our three initial concentrations of subhalos include one less concentrated than the cosmological mean, one more concentrated, and the extremely high concentrations of 80 for the subhalo is artificially set, meant to test our code in the context where core-collapse of the SIDM subhalo may occur for relatively low values of the self-interaction cross section.
We evolve the host halo in isolation with dark matter self-interactions for 5~Gyr before we insert the subhalo, so that the host comes to equilibrium, and its density profile is evolved from the initial cuspy NFW to a cored state. The density profile and velocity dispersion of the host are then measured and used as host model for the analytic evaporation modelling. The SIDM subhalos are not pre-evolved prior to infall in the same way, because the timescales for core-formation and core-collapse are different for subhalos with different concentrations \citep{Essig_2019}. Such pre-evolution could introduce bias between control groups. In future work, we will treat this point more holistically in a semi-cosmological context, introducing different infall times and thus various pre-evolution times of subhalos.  

The subhalo is initially placed at its apocenter at $R_{\rm apo} = 0.7R_{200m}$ from the center of the host, following the subhalo population model of \citealt{Pennarrubia2010}.
The magnitude of its velocity is determined such that the distance between the pericenter of its orbit and the host center, $R_{\rm peri}$, is $1/10$ of its initial displacement $R_{\rm apo}$. In later sections of our production runs, we will fix this apocenter distance $R_{\rm apo}$ and vary the pericenter-apocenter ratio, $R_{\rm peri} / R_{\rm apo}$, to characterize different subhalo orbits. 

The mass of each simulation particle is set to be $m_p=10^6 M_\odot$ for our validation runs. For the force resolution, we follow the criteria of Equation (20) and (21) of \citealt{van_den_Bosch_2018} to set the softening length $\epsilon$ such that these two criteria are comparable:
\begin{equation}\label{eqn:soft}
    \epsilon = r_s  [\ln{(1+c)} - \frac{c}{1+c}]  \sqrt{\frac{0.32 (N/1000)^{-0.8}}{1.12c^{1.26}}}.
\end{equation}
where $N$ is the initial number of particles in a halo, $c$ is the concentration and $r_s = R_{200m} / c$ is the scale radius. Because our main goal is to make sure that the subhalo is well resolved, these $N$, $c$ and $r_s$ are all drawn from the initial condition of the subhalo instead of the host. For example, the simulations with a subhalo of $10^{10.5}M_\odot$ with initial $c=14.8$ have $\epsilon = 0.294$ kpc while for the subhalos with $c=80$ we have $\epsilon = 0.035$ kpc. To test the robustness of our evaporation mechanism in terms of the particle mass resolution, we have also prepared corresponding runs with 10 times more particles, denoted as `$10N_p$' hereafter, where the particle mass is set to $M_p = 10^{5}M_\odot$ instead of the fiducial choice of $10^{6}M_\odot$.  Note that we only run the high-resolution tests for the analytical potential cases, and not the live host simulation runs. 

As a (sub)halo core-collapses, its central region contracts to become an ultra-dense structure, of which the length scale might be comparable to, or even smaller than, the gravitational softening length. A typical/average particle distance within the dense core of a core-collapsing (sub)halo, for example for the subhalo with $[M_{\rm sub}=10^{10.5}M_\odot, c=80]$ and simulation particle mass $M_p=10^5M_\odot$, is about 7 pc, while the softening length of gravity is $\epsilon=14$ pc as set by Eq.~\eqref{eqn:soft}.
Generally any structure at the subhalo center is not numerically robust given that the average particle separation is below the gravitational softening length. We prepared a companion run with a smaller softening length by an order of magnitude, $\epsilon=1.4$ pc, to test the convergence of our force resolution. We find that the evolution of the central densities of these two runs are nearly indistinguishable, with only $<3\%$ difference in their core-collapse times, suggesting that the default force resolution is reasonable up to the termination point. This is probably because the (sub)halo only enters such a dense regime after the onset of core-collapse, which is a quick, runaway process and discrepancy of force resolution does not have the time to accumulate yet. Therefore, we suggest that our default force resolution set by Eq.~\eqref{eqn:soft} should be generally good enough for SIDM subhalo evolution (see more discussion in Sec. \ref{sec:s&d}).

In this section, we fix the cross section of the dark matter self-interaction to be $\sigma_{\rm T}/m = 6\ {\rm cm}^2/{\rm g}$, which is much higher than the constraint of $\sigma_{\rm T}/m \lesssim 1\ {\rm cm}^2/{\rm g}$ for group-scale halos such as our hosts \citep{Yoshida_2000, Natarajan_2002, Rocha_2013, Peter_2013, Tulin_2018, Elbert_2018}, but plausible to produce core-collapsed subhalos \citep{Nishikawa_2020, Sameie_2020, Correa_2021}. We will explore a wider range of cross sections in Sec. \ref{sec:results}. 

\begin{figure}
    \begin{subfigure}{\columnwidth}
    	\includegraphics[width=\columnwidth]{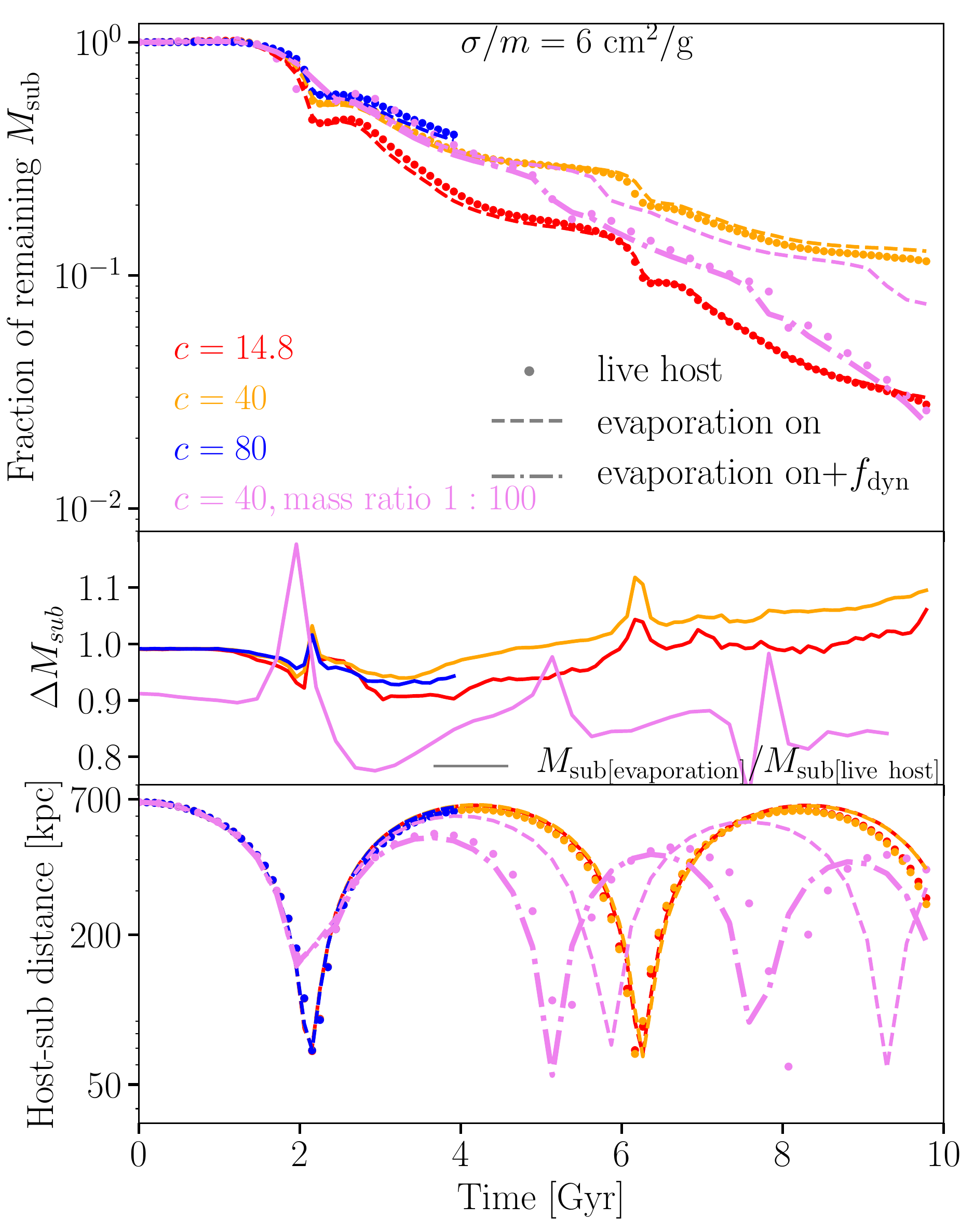}
    	\caption{}
        \label{fig:val-m1}
    \end{subfigure}
    \begin{subfigure}{\columnwidth}
        \includegraphics[width=\columnwidth]{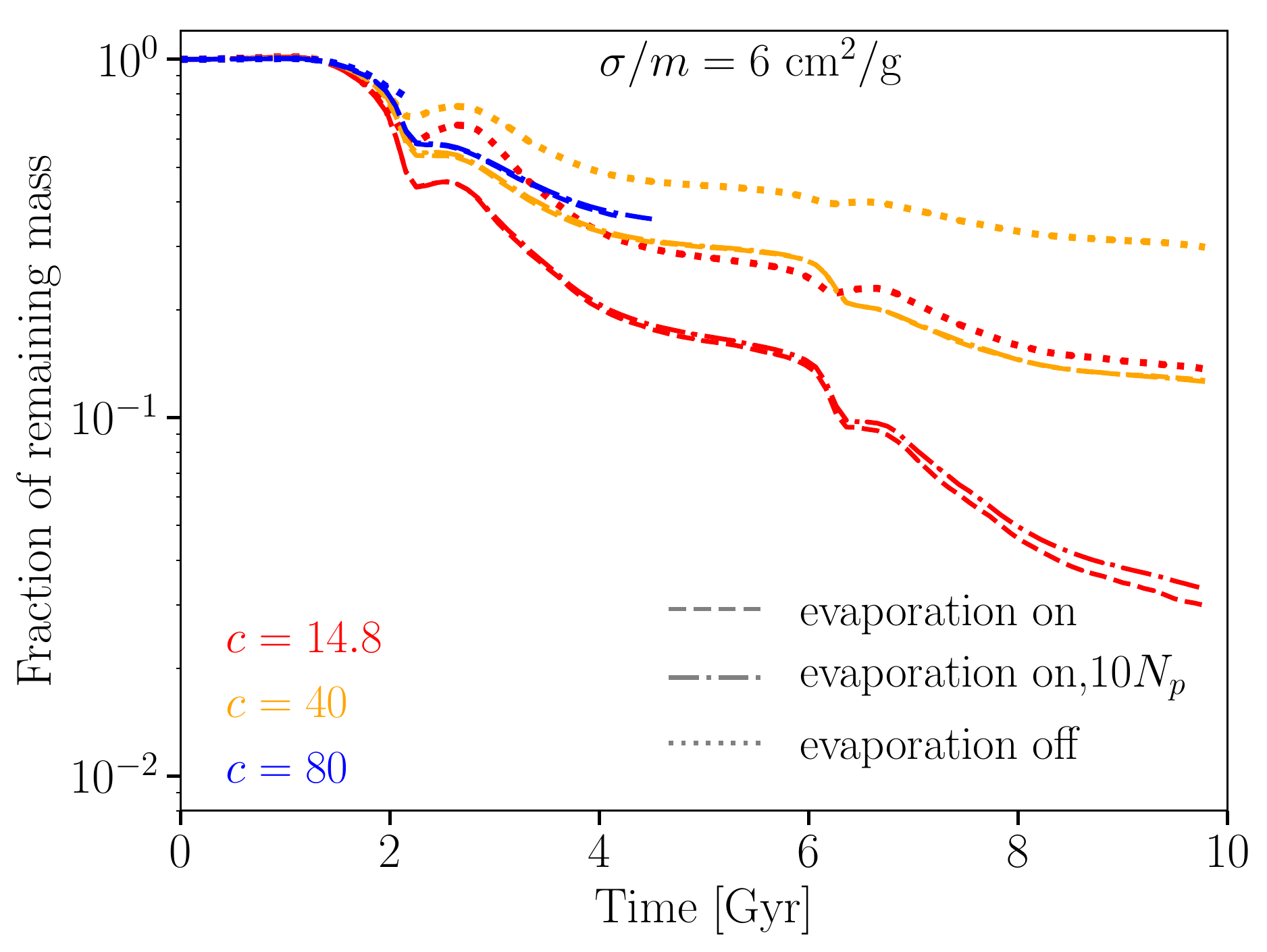}
        \caption{}
        \label{fig:val-m2}
    \end{subfigure}

    \caption{The mass-loss history and orbital distance between subhalo and host for different control groups. a) \emph{Top:} Remaining mass fraction of the subhalos vs. time. Subhalos with initial concentrations of $c=14.8, 40, 80$ are in red, orange and blue colors separately, and an extra group for $c=40$ but with a 10 times more massive subhalo is shown in pink, to highlight the importance of dynamical friction $f_{\rm dyn}$. The blue data set ends much earlier, because the $c=80$ subhalo core-collapses and the simulation is terminated. The live host simulations are shown in scattered dots, while our evaporation modelling in dashed lines.
    \emph{Middle:} The ratio between the remaining mass fraction of subhalos within our evaporation model and the ones from live host runs. 
    \emph{Bottom:} The host-subhalo separation as a function of time. b) Comparison between cases with evaporation model turned on (dashed lines) and off (dotted lines), and also the convergence test over the particle mass resolution (dashed lines for default resolution vs. dot-dashed lines for 10 times more particles.)}
    \label{fig:val1}
\end{figure}

\begin{figure*}
    \centering
    \begin{subfigure}[t]{0.32\textwidth}
        \centering
        \includegraphics[width=\textwidth, clip,trim=0.3cm 0cm 0.3cm 0cm]{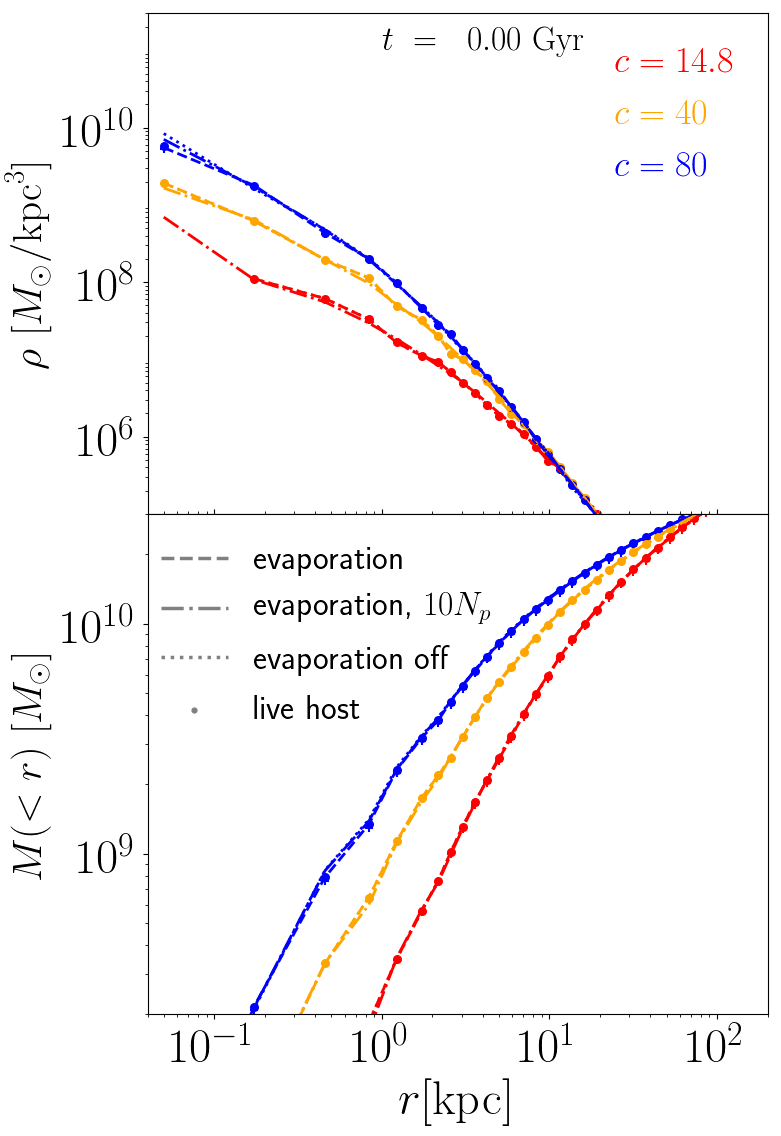}
        \caption{}
        \label{fig:val-proa}
    \end{subfigure}
    ~
    \begin{subfigure}[t]{0.32\textwidth}
        \centering
        \includegraphics[width=\textwidth, clip,trim=0.3cm 0cm 0.3cm 0cm]{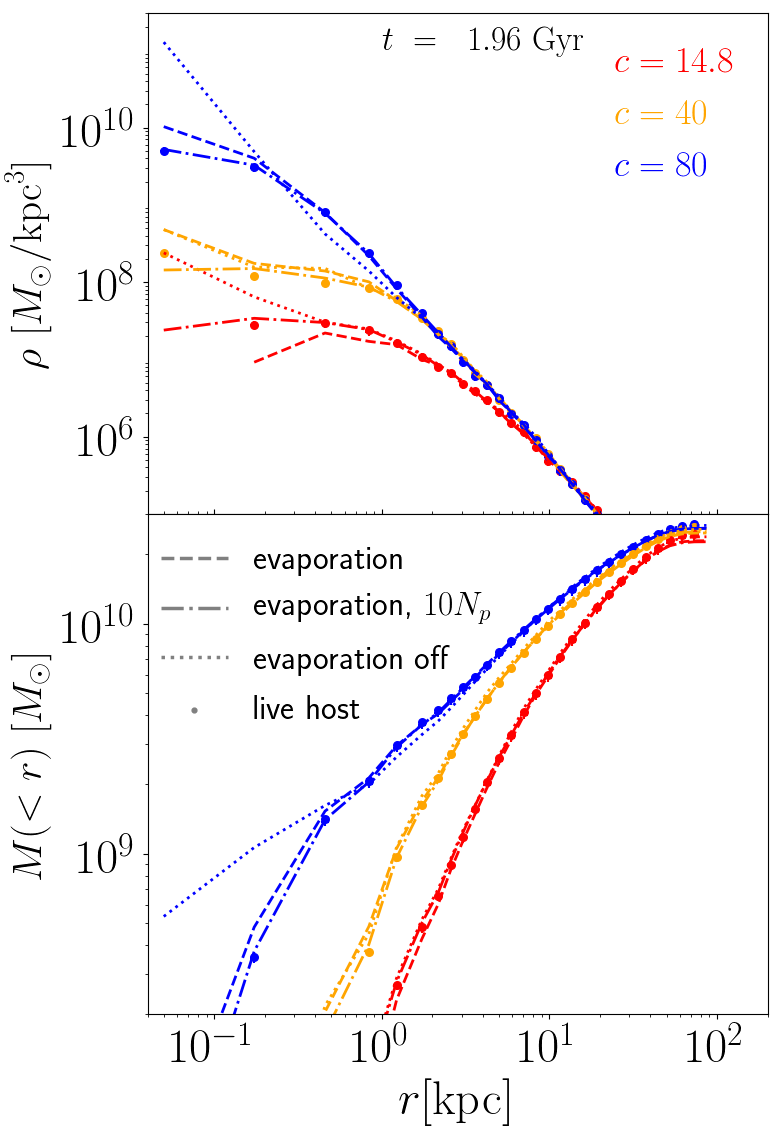}
        \caption{}
        \label{fig:val-prob}
    \end{subfigure}
    ~
    \begin{subfigure}[t]{0.32\textwidth}
        \centering
        \includegraphics[width=\textwidth, clip,trim=0.3cm 0cm 0.3cm 0cm]{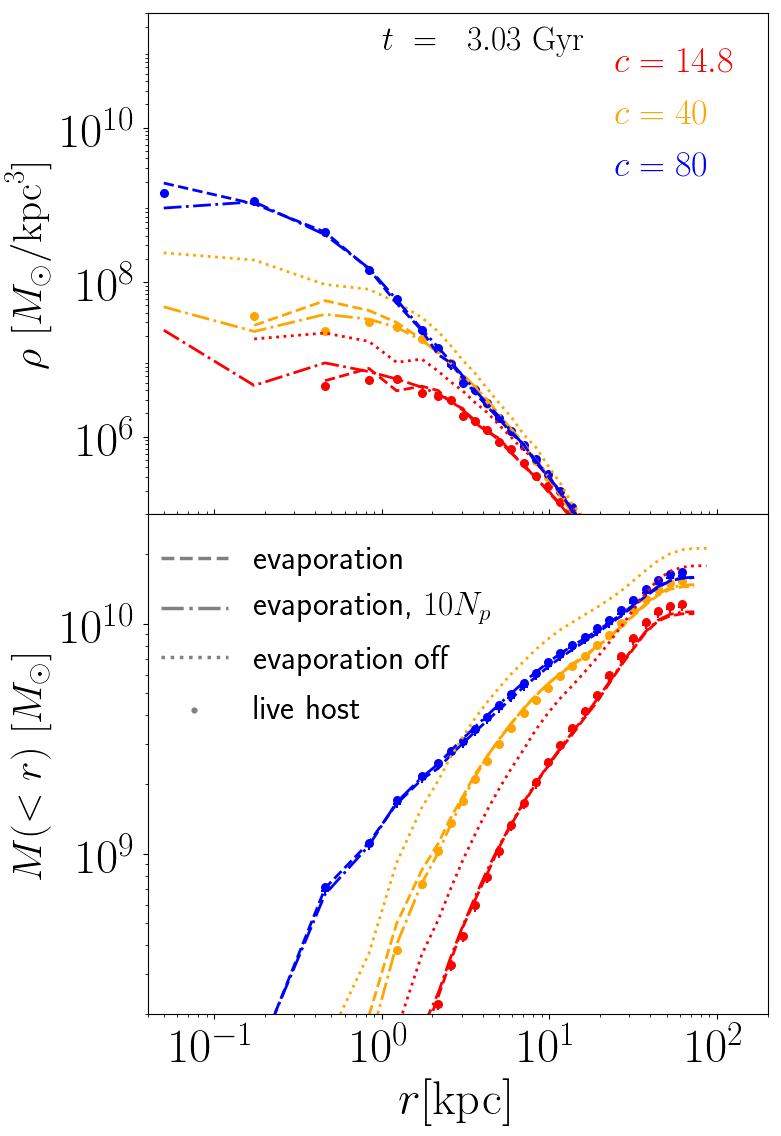}
        \caption{}
        \label{fig:val-proc}
    \end{subfigure}
     ~
    \begin{subfigure}[t]{0.32\textwidth}
        \centering
        \includegraphics[width=\textwidth, clip,trim=0.3cm 0cm 0.3cm 0cm]{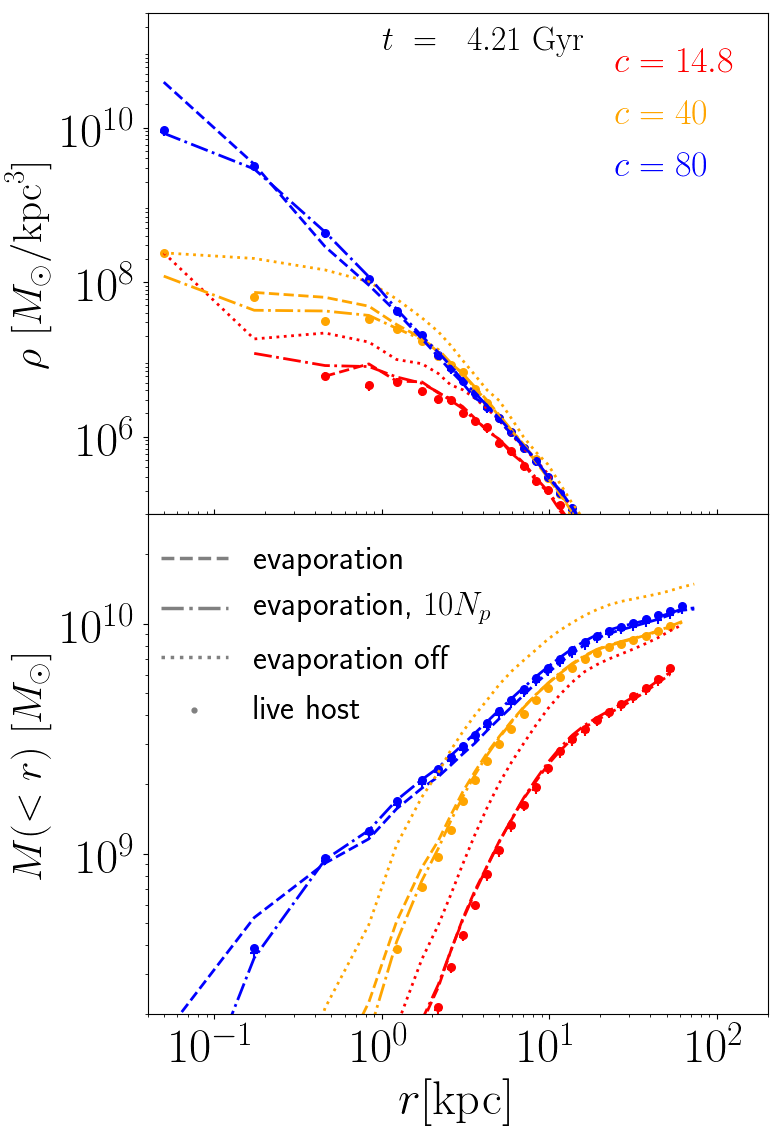}
        \caption{}
        \label{fig:val-prod}
    \end{subfigure}
    ~
    \begin{subfigure}[t]{0.32\textwidth}
        \centering
        \includegraphics[width=\textwidth, clip,trim=0.3cm 0cm 0.3cm 0cm]{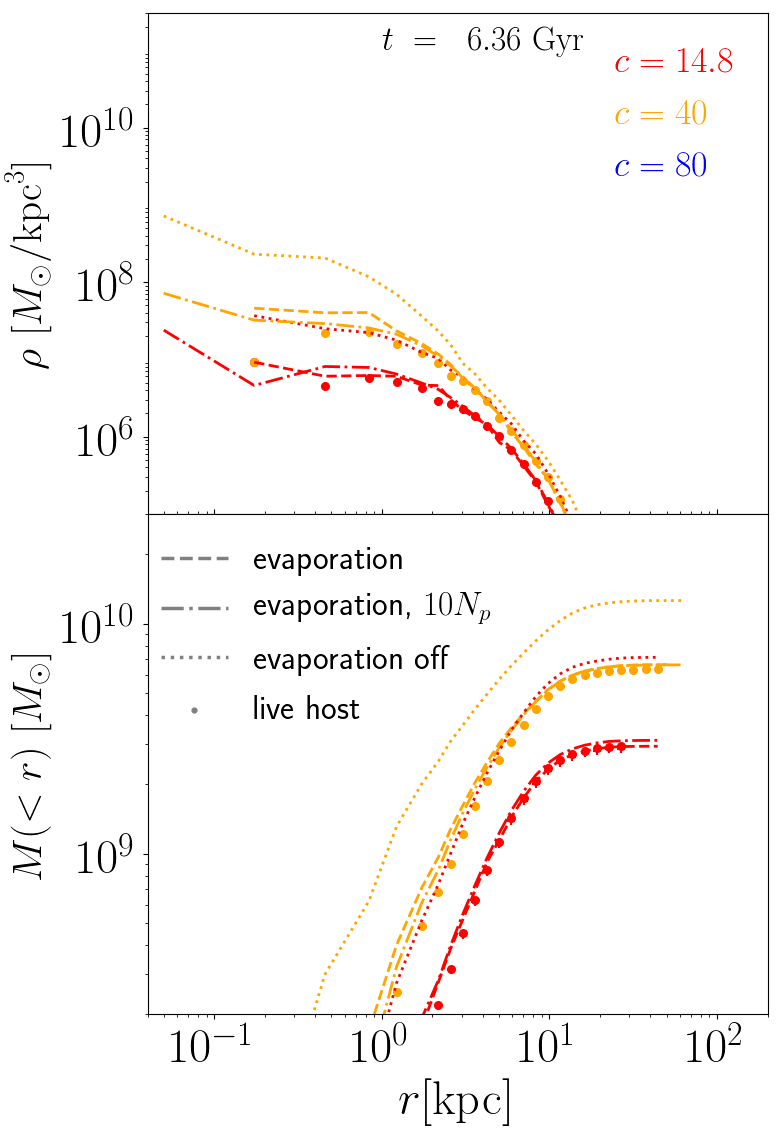}
        \caption{}
        \label{fig:val-proe}
    \end{subfigure}
    ~
    \begin{subfigure}[t]{0.32\textwidth}
        \centering
        \includegraphics[width=\textwidth, clip,trim=0.3cm 0cm 0.3cm 0cm]{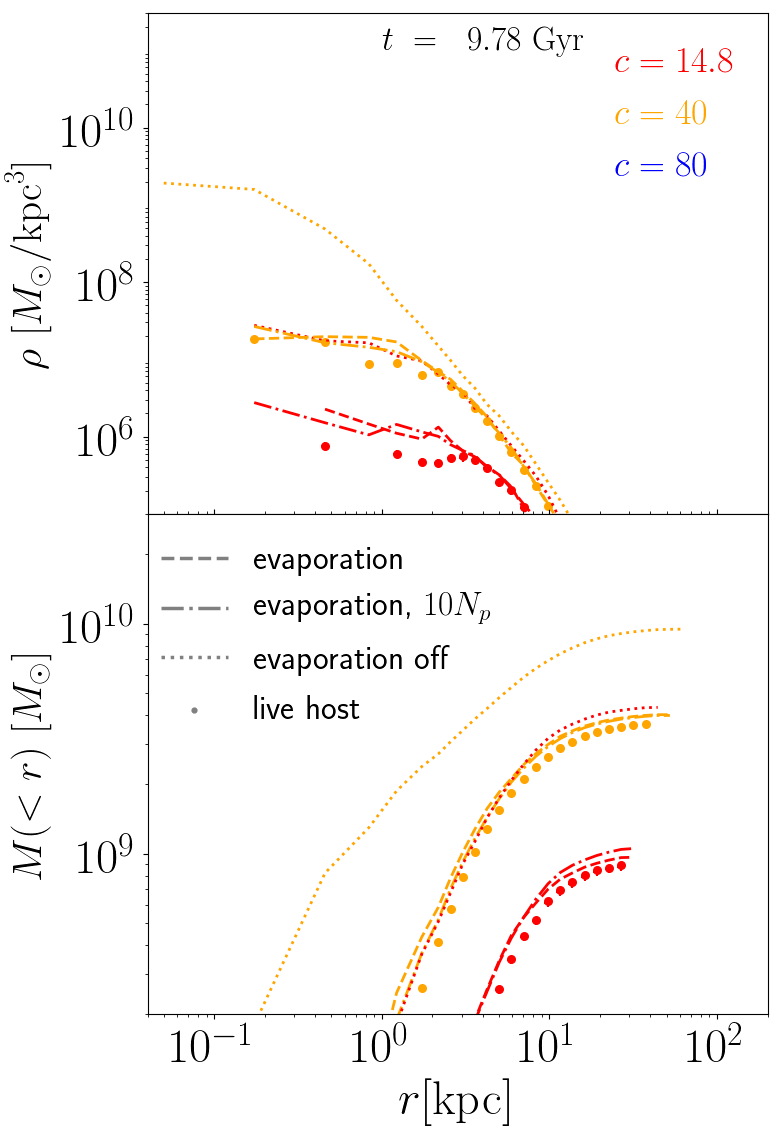}
        \caption{}
        \label{fig:val-prof}
    \end{subfigure}    
    \caption{The density profiles of subhalos with $\sigma_{\rm T}/m=6\ {\rm cm^2/g}$, sub-host mass ratio 1:1000 and $r_{\rm peri}:r_{\rm apo}=1:10$, for six selected time stamps. Subhalos with initial $c=14.8, 40, 80$ are in red, orange and blue colors. Different control groups with a live host; with an analytical host plus our evaporation model; with an analytical host but without evaporation; and with our evaporation model and also with particle mass resolution increased by a factor of 10, are shown in scattered dots, dashed lines, dotted lines and dash-dotted lines respectively. The '+' data sets for $c=14.8$ and $c=80$ subhalos in the $M(<r)$ vs $r$ panels (may need to zoom in to see) represent the mass profile of subhalos after removing particles that were bound to the host at the beginning but are captured by the subhalo in later times. The six times are: a) the beginning of the simulation; b) near the core-collapse time for the $c=80$, evaporation-off case (blue-dotted line); c) after the first pericenter, halted core-collapse process for other $c=80$ cases with evaporation (note that central $\rho$ decreases by nearly an order of magnitude compared to panel b); d) the core-collapse time for remaining $c=80$ cases; e) near the second pericenter; f) the end of the simulation. Readers are referred to Appendix \ref{appendix:anime-val} for a full evolution movie.}
	\label{fig:val-pro}
\end{figure*}

 The mass-loss histories of the control groups of subhalos, the discrepancy between the mass loss of subhalos with live hosts and with our evaporation model, and the corresponding host-subhalo separations are plotted in Fig. \ref{fig:val-m1}. We use Amiga Halo Finder \citep[\texttt{AHF}][]{AHF} to track the subhalo evolution in each snapshot. We compare our evaporation modelling against live host simulations (dashed lines vs scattered dots) with cases of sub-host mass ratio $1:1000$ and initial concentration $c=14.8, 40, 80$ (red, orange, blue), and mass ratio $1:100$ with $c=40$ (pink). The $c=80$ group simulations (blue) all end well before the time limit of 10 Gyr because we terminate them when the halo central density reaches the termination criterion described in Eq. (\ref{eq:cc}), just as the subhalo approaches the fluid phase of core-collapse. We find that for the mass ratio $1:1000$ cases, the discrepancy is generally smaller than 10\%.  The spike-like features in the discrepancy panel at each pericenter are clear signatures of missing dynamical friction in our model, that the insufficient orbit decay results in an overall time delay in the subhalo mass loss history in our modelling.

 We validate this hypothesis by increasing the subhalo mass by a factor of ten, so that the mass ratio is $1:100$ (pink, dots vs dashed line).
 Compared to $1:1000$ subhalos, this $1:100$ subhalo suffers from a much larger, yet also diverging, discrepancy in the mass-loss history, due to the stronger orbital decay caused by enhanced dynamical friction. We then add a tentative correction for dynamical friction $f_{\rm dyn}$ on top of our model, based on \citealt{Petts_2015} (see Appendix \ref{appendix:dyn} for details), as shown in the pink dash-dotted line. With the inclusion of this dynamical friction model, our evolution model for the $1:100$ subhalo is noticeably improved in its accuracy, in terms of both the orbit and mass loss history. However, we do not implement this model of dynamical friction as a regular part of our semi-analytical model, because dynamical friction remains small for our main targets, the small (mass ratio 1:1000 and smaller) subhalos, where our evolution model has achieved satisfactory accuracy. Furthermore, our implementation of this dynamical friction model itself is not perfect (see Appendix \ref{appendix:dyn} for more discussion), and potentially introduces additional uncertainties, such as the orbit over-decay we already see in the $1:100$ subhalo (see the orbit panel of Fig. \ref{fig:val-m1}). Thus we only use the dynamical friction model as a demonstration here, and note that it is not included in the rest part of this work.

 Another source of discrepancy between our subhalo evolution model and a live host simulation is the capture of host dark matter by the subhalo. Its initial offset is shown in the middle, $\Delta M_{\rm sub}$, panel of Fig. \ref{fig:val-m1}, at time 0, denoting the host particles that immediately become bound to the subhalo when it is placed in the host. We can see that for subhalos with sub-host mass ratio $1:1000$, captured host dark matter only counts for around $1\%$ of its total mass, while it is $\sim 10\%$ for the $1:100$ subhalo. Thus we expect that for higher mass ratios, greater than 1:1000 (e.g. we consider the mass ratio of 1:100 to be 'smaller' than 1:1000), which is the physical case driving the creation of this semi-analytic evaporation model, both dynamical friction and the accretion of host dark matter would be a smaller factor, and the discrepancy between our code and a full live host SIDM simulation would be further reduced. Note that the subhalo continues to capture dark matter from the host while it orbits, but this still only counts for a small portion of the subhalo mass, as we will see later in the subhalo's density profiles in Fig. \ref{fig:val-pro}.

In Fig. \ref{fig:val-m2} we show convergence tests for the particle mass resolution of our evaporation code, comparing the default resolution cases with ones with 10 times more particles (dashed vs dash-dotted).  The level of convergence is remarkable for all the groups of simulations. The mass loss histories of the higher resolution runs are nearly indistinguishable from the ones with lower resolution, for $c=40$ and 80. We observe differences of around 10\% for the subhalo with lowest concentration at late times, when the subhalo has already lost more than 95\% of its initial mass. The convergence regarding particle mass resolution can be also seen later in the detailed evolution of the subhalo's density profile in Fig. \ref{fig:val-pro}. 

In Fig. \ref{fig:val-m2}, we also show how the mass-loss history of the subhalo changes if our evaporation model is turned off (dotted) compared to our fiducial case in which both tidal and evaporative mass loss are included (dashed lines).
The differences in mass-loss history start near the first pericenter, where both tidal mass loss and evaporation are at their highest rates, regardless of the subhalo's initial concentration. 
The mass-loss histories diverge with increasing time.  For our particular choice of orbit and cross section, the simulations that include evaporation indicate that the subhalos are much less massive than in the tidal-only cases at the end of the simulation (10 Gyr).  The largest difference at the end of the simulation is for the lowest concentration halo---a difference of nearly an order of magnitude. This highlights the necessity of including the physics of host-subhalo evaporation properly.

We explore the evolution of the density profile in Fig. \ref{fig:val-pro}.  We present the detailed evolution of the subhalo density profiles and mass profiles for six selected snapshots: a) the beginning of the simulations; b) near the core-collapse time of the `evaporation off' semi-analytical case with $c=80$ (blue dotted line); c) after the first pericenter passage, the disruption of core-collapse of other $c=80$ cases by the host-sub evaporation (their central densities decrease by about an order of magnitude); d) core-collapse of other $c=80$ cases (blue dots, dashed, dash-dotted); e) the second pericenter passage; f) the end of the simulation. Readers may refer to Appendix \ref{appendix:anime-val} for a full evolution movie of density/mass profiles.

As can be seen in Fig. \ref{fig:val-pro}, we find that the density/mass profiles of subhalos in our semi-analytical modelling of the evaporation are in overall good agreement with the live host simulations (dashed lines vs dots). Our model is able to capture the fast growth of central density of the core-collapsing subhalos, as well as the evolution of cored subhalos. By contrast with the tidal-field-only cases (dashed vs dotted lines), we can see the necessity of including evaporation. In this figure we show again the robustness of our method in terms of particle mass resolution, where the control groups with higher and lower resolutions (dash-dotted vs dashed lines) show no systematic disagreement. We notice some discrepancy between our model and the live host runs in Fig. \ref{fig:val-prof}, at late times in the simulation, especially for the lowest concentration subhalos. One reason for this is that, as shown in the orbit panel of Fig. \ref{fig:val-m1}, there is small but non-negligible extra orbit decay induced by dynamical friction in the live host runs, which results in stronger evaporation at this end time. 

Overall we note that our method shows good agreement with live host simulations in terms of both the total mass and detailed density/mass profile, for SIDM with constant cross sections.

\subsection{SIDM with velocity-dependent cross section}\label{sec:valv}

\begin{figure*}
    \centering
    \begin{subfigure}{0.32\textwidth}
        \centering
        \includegraphics[width=\textwidth, clip,trim=0.3cm 0cm 0.3cm 0cm]{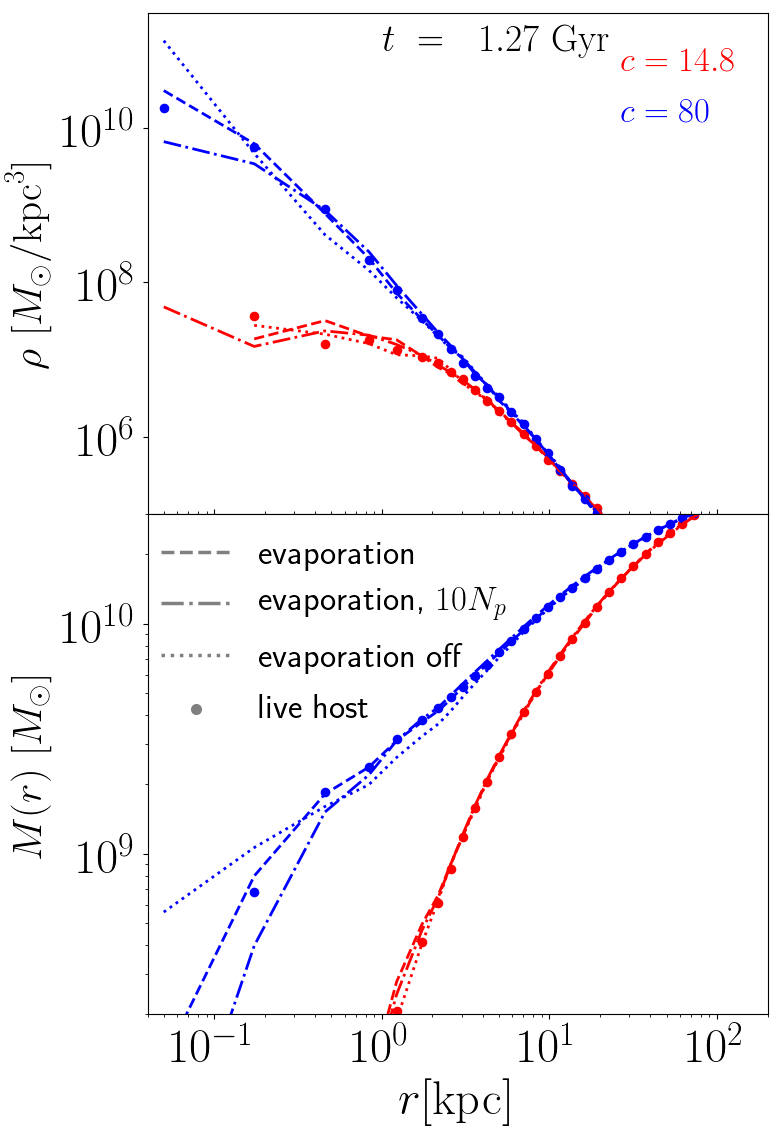} 
        \caption{}
    \end{subfigure}
    ~
    \begin{subfigure}{0.32\textwidth}
        \centering
        \includegraphics[width=\textwidth, clip,trim=0.3cm 0cm 0.3cm 0cm]{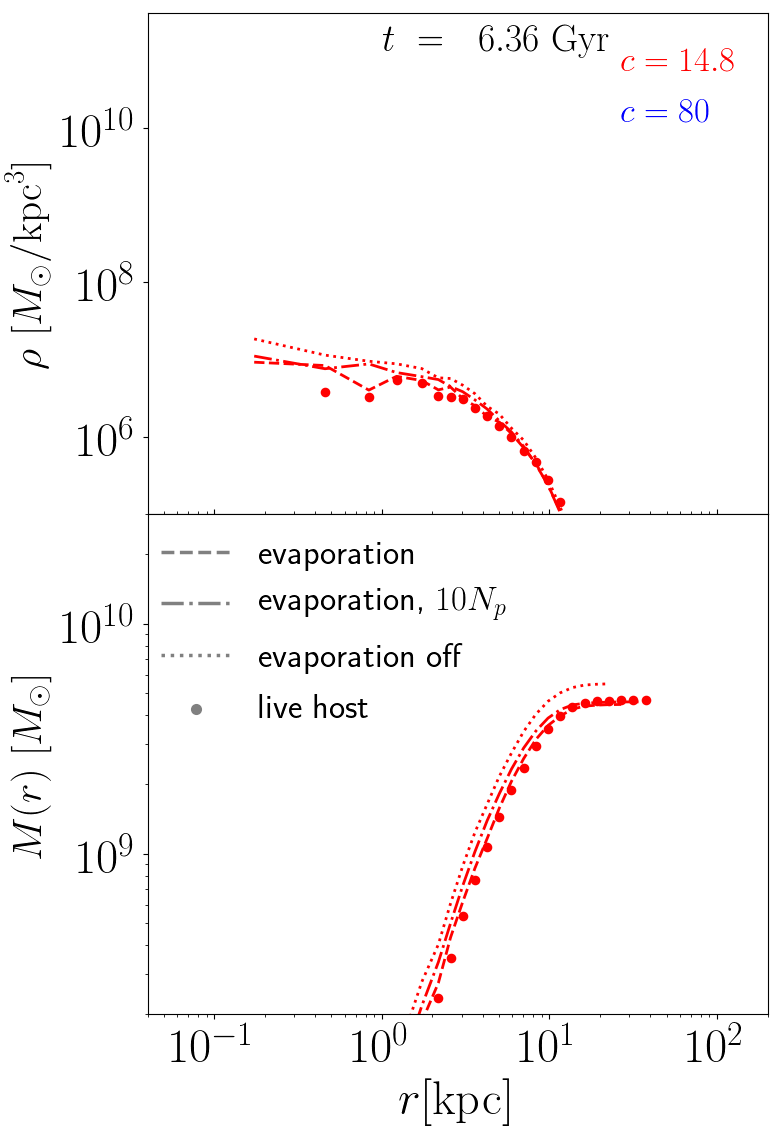} 
        \caption{}
    \end{subfigure}
    ~
    \begin{subfigure}{0.32\textwidth}
        \centering
        \includegraphics[width=\textwidth, clip,trim=0.3cm 0cm 0.3cm 0cm]{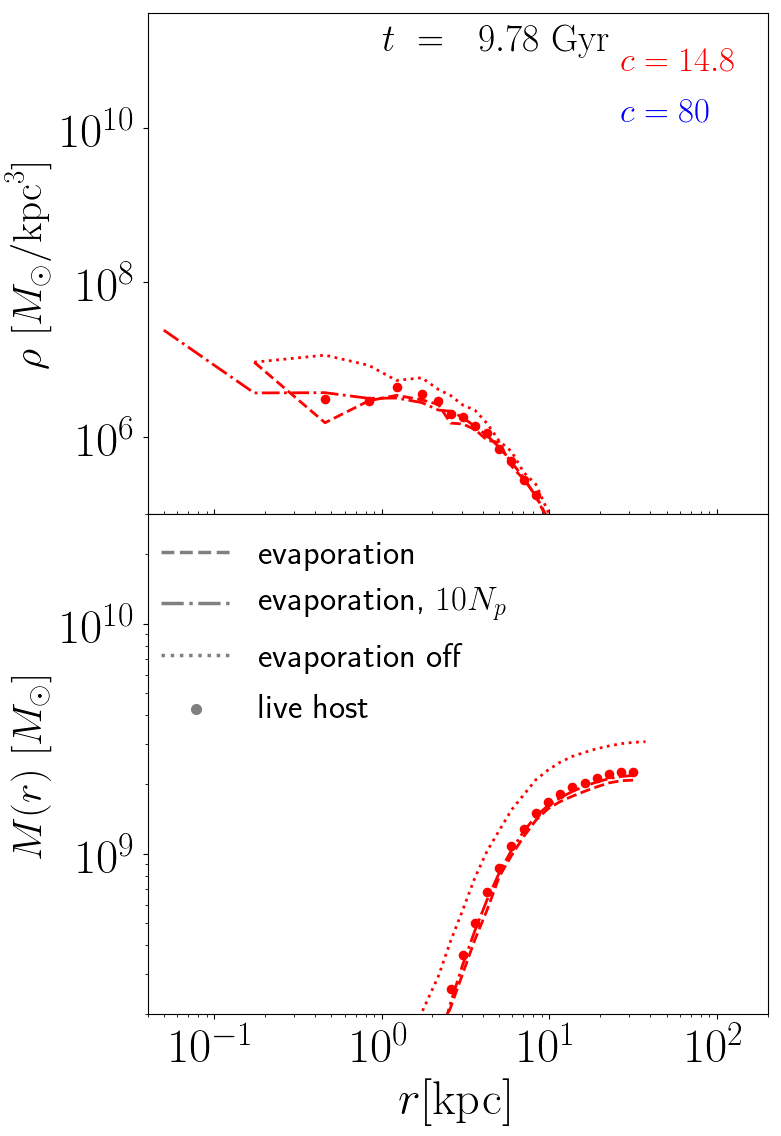} 
        \caption{}
    \end{subfigure} 
    \caption{The density profiles of subhalos with a velocity-dependent cross section from the ADM model (\protect\citealt{Cline_2014}), as shown in the orange line in Fig. \ref{fig:xsection}, with sub-host mass ratio 1:1000 and $r_{\rm peri}:r_{\rm apo}=1:20$. Subhalos with initial $c=14.8$ and 80 are in red and blue colors. Different control groups are as in Fig. \ref{fig:val-pro}. The three time stamps are: a) core-collapse time for the $c=80$ cases; b) near the second pericenter; c) the end of the simulation. }
	\label{fig:val-pro-ADM}
\end{figure*}

Although we use constant cross sections for most parts of this work, we show that our analytic evaporation model also extends to velocity-dependent cross sections. This extension will be important for our future work because, as we show later in Sec. \ref{sec:results},  SIDM with a constant cross section is unlikely to produce core-collapsed subhalos, even in the absence of known cross section constraints.  And, it is challenging to fit observations from clusters to satellite galaxies with a single constant cross section \citep{Tulin_2018}. Moreover, most theoretical models from particle physics predict velocity-dependent cross sections \citep{Feng_2009, Boddy_2014, Tulin_2018}. Therefore, we validate our code with a specific model of velocity-dependent SIDM in this sub-section.

We choose an atomic dark matter (ADM) velocity-dependent cross section model from \citealt{Cline_2014} (see also \citealt{Goldberg:1986nk,Kaplan:2009de,Kaplan:2011yj,Cyr-Racine:2013ab}) where dark matter consists of `dark atoms' composed of a dark proton and a dark electron. We manually fine-tune the input parameters, the dark atom fine structure constant $\alpha = 0.05$, dark atom mass $m_{\rm d} = 50$ GeV and the dark proton-to-electron mass ratio $R_{\rm d} = 400$, in order to generate a velocity-dependent cross section profile that is of order 10 $\,\text{cm}^2/\text{g}$ for velocities close to the subhalo velocity dispersion of $\mathcal{O}(10)$ km/s, and of order 1$\,\text{cm}^2/\text{g}$ near the orbital velocity of $\mathcal{O}(100)$ km/s (see the orange line of Fig. \ref{fig:xsectiona}). The subhalo velocity dispersion is the key velocity scale for subhalo core-collapse, while the subhalo orbital velocity is key to host-subhalo evaporation strength \citep{Nadler_2020, Jiang_2021}. This is meant to test our method in a scenario where the self-interaction among subhalo dark matter is efficient, thus plausible for driving core-collapse, yet also with a relatively weak but non-negligible evaporation from the host.

The setup of ADM subhalos is nearly the same as in Sec. \ref{sec:val2}, only that we set $R_{\rm peri}/R_{\rm apo} = 1/20$ for the orbit because the evaporation effect is stronger at a closer pericenter, and that the ADM host halo has a smaller core than hosts with constant cross sections (see Fig. \ref{fig:xsectionb}). In Fig. \ref{fig:val-pro-ADM}, we present the density evolution of ADM subhalos for selected time stamps: a) the snapshot right before the core-collapse of $c=80$ subhalos; b) the second pericenter encounter, where both evaporation and tidal effects are strongest; c) the last snapshot of the simulation. As expected from the low cross section around the typical bulk velocity of the subhalo (of order $100~$km/s), the evaporation is relatively low in the ADM cases compared to those in Sec. \ref{sec:val2}, thus including it or not (dashed vs dotted lines) only makes a small difference in both the core-collapsing and cored subhalos for our choice of initial conditions. For this particular set of initial conditions with high concentration, the weak evaporation does not disrupt the core-collapse for the $c=80$ subhalo, compared to the constant cross section case $\sigma_{\rm T}=6~{\rm cm^2/g}$ above (Fig. \ref{fig:val-prob} to Fig. \ref{fig:val-prod}). However, for the cored subhalo of $c=14.8$, including evaporation or not leads to a $\sim 50\%$ difference in its final mass in the ADM case. 

This set of validation tests with velocity-dependent cross section further demonstrates the remarkable accuracy of our method, with an even smaller discrepancy from the live host simulation compared to the constant cross sections. 

\section{Results} \label{sec:results}

In this section, we present results from our production simulation runs, with which we study how different physical processes drive the evolution of a (sub)halo central density, including the intrinsic core-collapse of an SIDM halo due to its internal heat outflow, the tidal effects of a host potential, and host-subhalo evaporation. We then scan the relevant parameter space to show how these processes depend on SIDM subhalo properties, and eventually determine whether a subhalo core-collapses or not.  We map out the critical boundary for subhalo core-collapse in the parameter space. All simulations in this section are conducted with the semi-analytical treatment described in the previous two sections, hence we note that analysis on subhalos below are most accurate for small subhalos (< 1:1000 host mass).

\begin{table}
	\centering
	\begin{tabular}{lccc} 
		\hline
		Type of sims & Isolated & Tidal field & Tidal + evaporation\\
		\hline
		Sec. \protect\ref{sec:result1} & \checkmark & - & - \\
		\hline
		Sec. \protect\ref{sec:result2} & \checkmark & \checkmark & -\\
		\hline
		Sec. \protect\ref{sec:result3} & - & \checkmark & \checkmark\\
		\hline
		Sec. \protect\ref{sec:result4} & - & - & \checkmark \\
		\hline
		Sec. \protect\ref{sec:result5} & - & - & \checkmark \\
		\hline
	\end{tabular}
	\caption{Types of simulations we run in each subsection of Sec. \protect\ref{sec:results}. \label{table2}}
\end{table}

We summarize our simulation suite in Table \ref{table2}. In Sec. \ref{sec:result1}, we simulate SIDM halos in isolation and evolve them until core-collapse, in order to provide a numerical scaling relation of halos' intrinsic core-collapse time $t_{\rm c}$ as a function of input parameters. Our time scaling is compared to a similar one from \citealt{Essig_2019}.  This is an additional way to validate our simulations, and to explore a regime that has previously been missed in similar numerical fits for scaling relations for core-collapse times. 

We scrutinize the acceleration of subhalo core-collapse by tidal effects  \citep{Kahlhoefer_2019, Nishikawa_2020, Sameie_2020, Correa_2021} in Sec. \ref{sec:result2}, including tidal stripping and tidal heating. Tidal stripping refers to the process by which outlying dark matter is removed from the subhalo by tidal forces from the host halo. Tidal heating, which is, to  first order, the work done by the tidal force on the subhalo dark matter relative to the subhalo center of mass, heats up the subhalo as it moves along its orbit \citep{Gnedin_1999b, Gnedin_1999a,Gnedin_1999c, Taylor_2001, Pullen_2014, vdB_2018b, Yang_2020}. The strength of tidal heating at different layers of the subhalo increases with distance from the subhalo center. We show that the acceleration of subhalo core-collapse by tidal effects is significant for subhalos with relatively low initial concentrations, but reverses (i.e., tides decelerate core-collapse) at ultra-high concentrations ($c\gtrsim75$ for our choice of $M_{\rm sub}, \sigma_{\rm T}/m$ and orbit). We argue that the net impact of the tidal field arises from the competition between $t_{\rm c}$ and the orbital time.

Sec. \ref{sec:result3} serves two purposes. First, we contrast the subhalos with full evaporation against ones with the same tidal field but no host-subhalo evaporation.  This allows us to highlight the importance of properly modelling evaporation for SIDM subhalo evolution. Second, by putting these subhalos on circular orbits in this subsection, we make the interaction among different physical processes clearer by maintaining nearly time-invariant strength of both evaporation and tides, before we dive into more realistic and complex cases later. We demonstrate three relevant processes as heating/cooling (energy gain/loss) terms: internal heat outflow from the subhalo core as a cooling term, which is driven by the self-interaction of the subhalo dark matter; and evaporation and tidal heating as heating terms. We correlate the evolution of these heating/cooling terms with that of the central density $\rho_{\rm cen50}$, the indicator of whether or when the subhalo core-collapses, and come to our primary conclusion that the subhalo central density grows only when there is a net cooling (energy loss) within the subhalo core. 

We generalize our analysis in Sec. \ref{sec:result4}, generating subhalos with four varying parameters: SIDM cross section $\sigma_{\rm T}/m$, the subhalo mass $M_{\rm sub}$, the subhalo initial concentration $c$, and its orbit characterized by $r_{\rm peri}:r_{\rm apo}$. We confirm our conclusion from the circular orbit runs, that the evolution of the subhalo central density is driven by the net heating/cooling within its core.  We show that whether an SIDM subhalo can eventually core-collapse is extremely sensitive to its initial parameters. Therefore, in Sec. \ref{sec:result5} we scan the multi-dimensional parameter space with a few hundred simulations, and map a boundary of critical parameters for SIDM subhalos with constant cross sections to survive the host-sub evaporation and eventually core-collapse.

\subsection{Core-collapse time scaling for an isolated halo}\label{sec:result1}

In this section, we provide fits for a scaling relation of the core-collapse time $t_{\rm c}$ of an isolated SIDM halo using our particle-based simulations, and compare them with similar ones from \citealt{Essig_2019}.   The results from \citealt{Essig_2019} are based on a completely different method---1D analytical modelling of the heat transport equations---and focus on a slightly different regimes of halos.  The overlap in our regimes makes a comparison between the two methods useful for cross-validation.

\citealt{Essig_2019} studied the time scale of core-collapse $t_{\rm c}$ for dissipative and non-dissipative SIDM.  For non-dissipative SIDM with a constant cross section, such as the case we consider in our simulations, their scaling relation for core-collapse time is (see Eq.~(3) and Fig. 2 of \citealt{Essig_2019})

\begin{equation}\label{eqn:tc1}
    t_{\rm c} \approx \frac{150}{\beta} \frac{1}{r_s \rho_s \sigma/m} \frac{1}{\sqrt{4\pi {\rm G} \rho_s}},
\end{equation}
where $r_{\rm s}$ and $\rho_{\rm s}$ are the scale radius and normalization density of NFW profile, and $\beta$ is a constant factor.  Alternatively, the collapse time can be described by virial quantities, 

\begin{equation}\label{eqn:tc2}
    t_{\rm c} \propto (\sigma/m)^{-1} M_{200c}^{-1/3} c_{200c}^{-7/2}.
\end{equation}

 In this work, we prepare a set of similar isolated SIDM halos by sampling $\sigma_{\rm T}/m$, $M_{\rm 200c}$ and $c_{\rm 200c}$ to check with the power-law exponents of Eq.~\eqref{eqn:tc2}. Note that our overall definition of halo mass $M_{\rm 200m}$ throughout this work is counted in the region within which the average density is 200 times cosmological mean mass density. For this section only, we use a halo mass definition of $M_{200c}$, the `200 times the critical density' definition of \citealt{Essig_2019}, for a fair comparison. 

As specified in Sec. \ref{sec:method}, we measure the central density of the halo $\rho_{\rm cen50}$ at each simulation timestep to track the process of SIDM halo core-collapse. We define the halo core-collapse state when $\rho_{\rm cen50}$ grows by two orders of magnitude, and terminate the simulation.  This criterion is approximately comparable to the $Kn\sim1$ core-collapse state in \citealt{Essig_2019} (see their Fig. 1), where $Kn \equiv \lambda / H = (1/n \sigma) /\sqrt{\sigma_v^2/4\pi {\rm G}\rho}$ is the Knudsen number, with $n$, $\sigma_v$ and $\rho$ being the 1D number density, velocity dispersion and mass density. In \citealt{Essig_2019}, the halo core-collapse state is defined at $Kn\sim0.1$.  However, since the time difference between the two states $Kn\sim1$ and $Kn\sim0.1$ is negligible relative to $t_{\rm c}$, as the core-collapse process exponentially accelerates itself once initiated, our criterion of halo core-collapse is comparable to that of \citealt{Essig_2019}.

\begin{figure*}
    \centering
    \begin{subfigure}{0.33\textwidth}
        \centering
        \includegraphics[width=\textwidth]{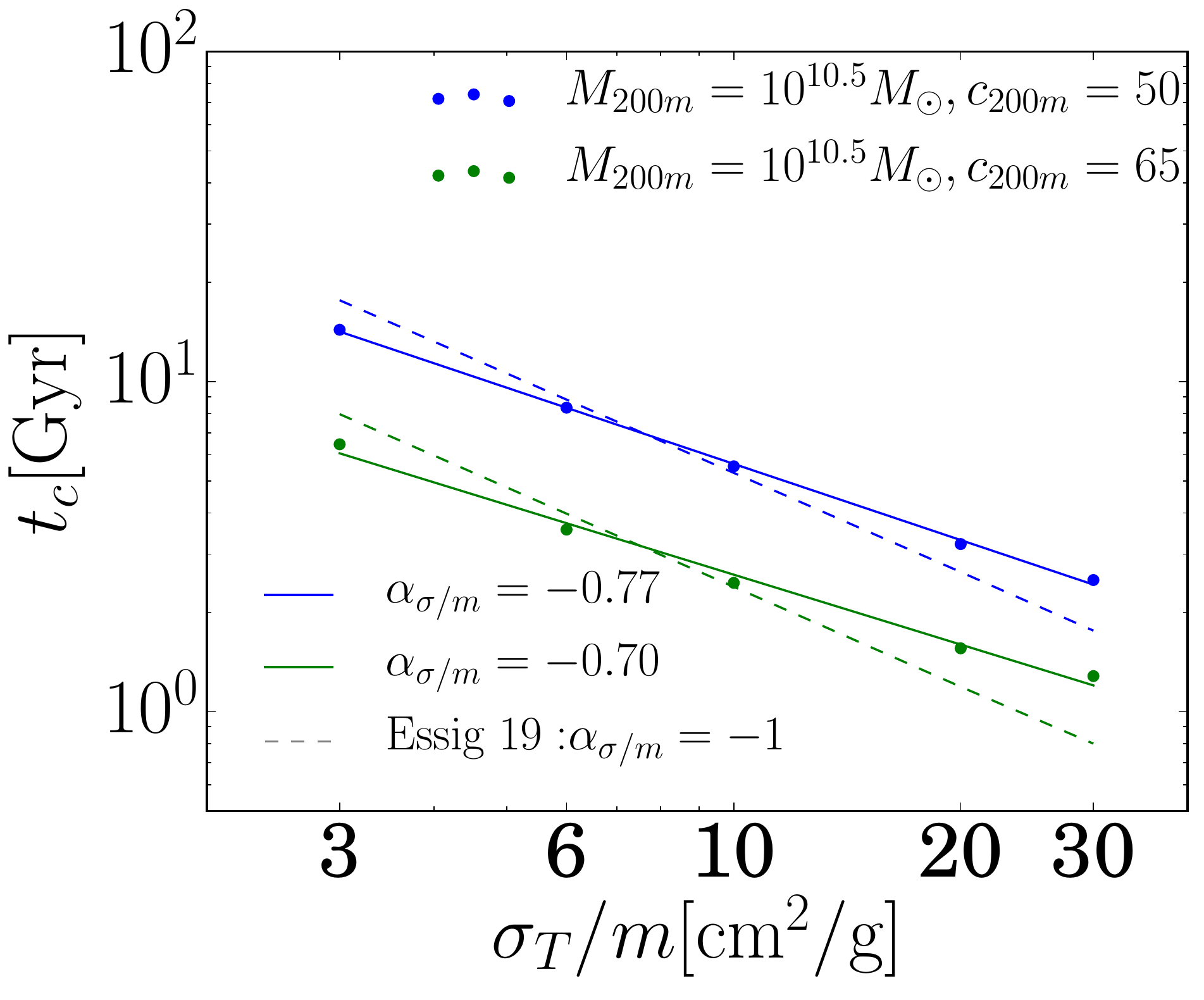} 
        \caption{}
    \end{subfigure}
    ~
    \begin{subfigure}{0.3\textwidth}
        \centering
        \includegraphics[width=\textwidth]{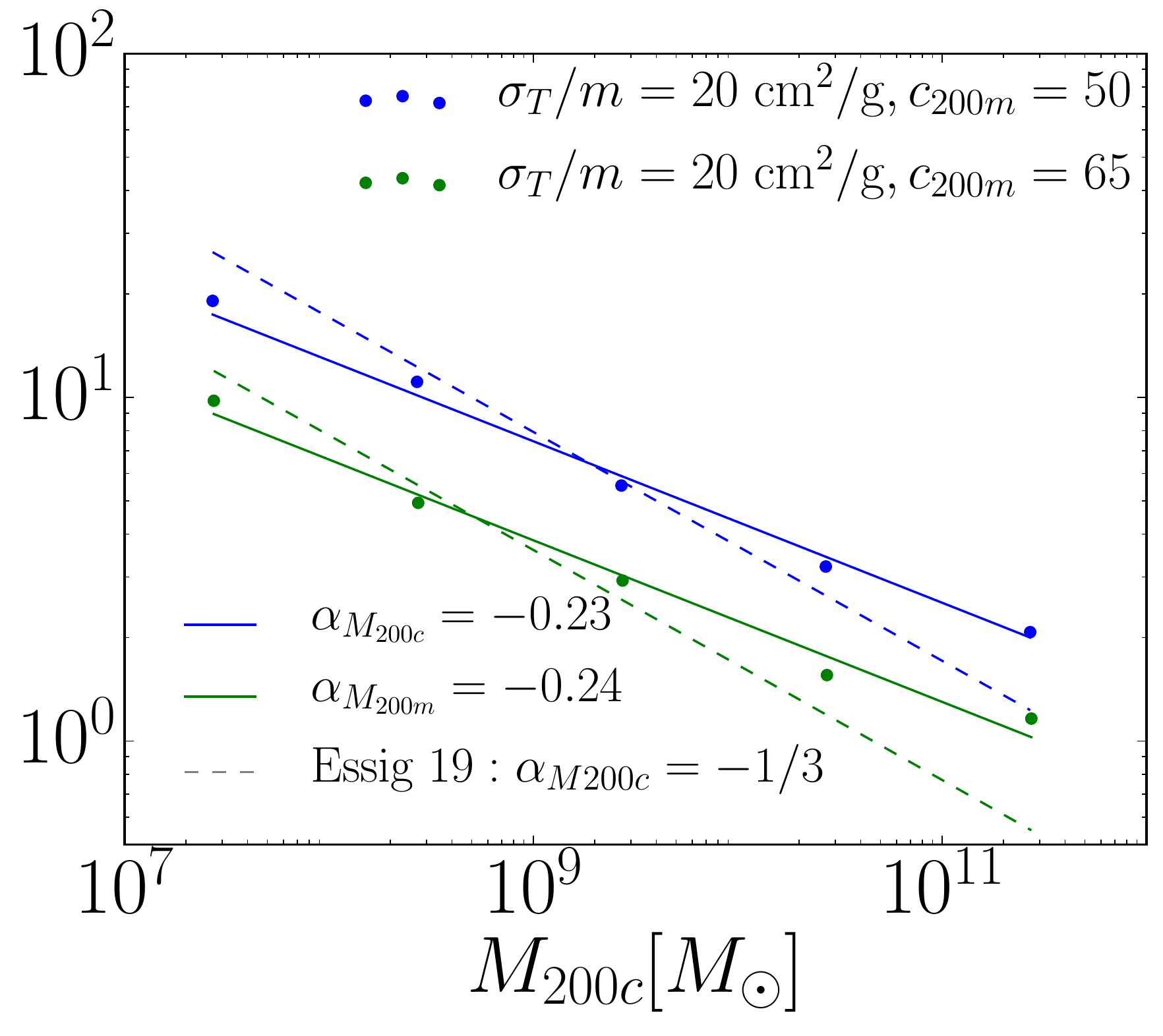} 
        \caption{}
    \end{subfigure}
    ~
    \begin{subfigure}{0.3\textwidth}
        \centering
        \includegraphics[width=\textwidth]{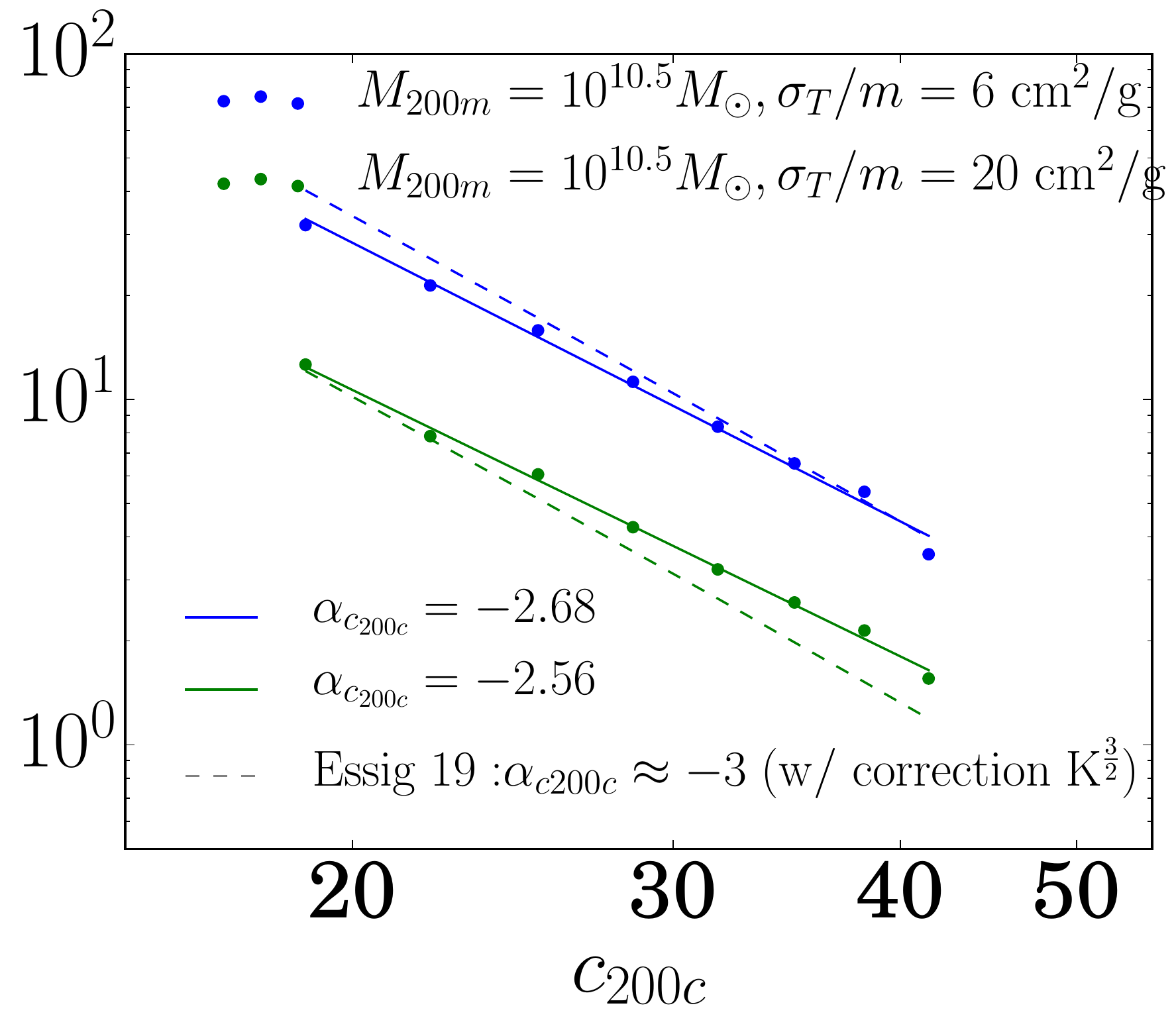} 
        \caption{}
    \end{subfigure} 
    \caption{The scaling relation between the isolated halo core-collapse time $t_{\rm c}$ and three variables: $\sigma_{\rm T}/m$, $M_{\rm 200c}$ and $c_{\rm 200c}$. In each sub-figure two of the three parameters are fixed and the other is varied to fit for the corresponding power law. The simulation results are shown in scattered dots and their fitting in lines, and the similar scaling relation from \protect\citealt{Essig_2019} is calculated by Eq.~\eqref{eqn:tc1} and presented in dashed lines as a comparison. $\alpha$ denotes the exponent index of each parameter in the power law of Eq. \protect\eqref{eqn:tc2}.}
	\label{fig:tc}
\end{figure*}

We sample isolated halo simulations in three groups, within each fixing two parameters of $[\sigma/m, c_{\rm 200c}, M_{\rm 200c}]$ and varying the other one. The scaling relations of $t_{\rm c}$ against these three parameters are presented in Fig. \ref{fig:tc}, together with the one from \citealt{Essig_2019} calculated using Eq.~\eqref{eqn:tc1} with $\beta=0.75$. Similar to Eq.~\eqref{eqn:tc2}, we use power laws for each parameter in the scaling relation:

\begin{equation}
     t_{\rm c} \propto (\sigma/m)^{\alpha_{\sigma/m}} M_{200c}^{\alpha_{M_{200c}}} c_{200c}^{\alpha_{c_{200c}}}.
\end{equation}

We can confirm that the scaling of $t_{\rm c}$ follows almost perfectly a power law as observed in \citealt{Essig_2019}, with the slopes of all our parameters slightly smaller in magnitude than their fitting results. The differences in our slopes are physical in origin. A correction term $K^{3/2}=[\ln{(1+c_{200c})} - c_{200c}/(1+c_{200c})]^{3/2}$ has been neglected in Eqn. \eqref{eqn:tc2} from \citealt{Essig_2019} for simplification, which is a relatively small factor of order unity when $c_{200c}$ is small but becomes larger for the $c_{200c}$ we consider in this set of simulations. Adding this factor of $K^{3/2}$ back to Eqn. \eqref{eqn:tc2} leads to a larger $\alpha_{c_{200c}}$ of \citealt{Essig_2019}, and thus better agreement with our simulation results (Hai-Bo Yu 2021, private communication). Besides,
\citealt{Essig_2019} mainly focuses on small, low concentration halos with relatively low cross sections where $\hat{\sigma} = (\sigma_{\rm T}/m)\rho_{\rm s} r_{\rm s}<0.1$, while our sample of halos roughly spans the range of $0.04\lesssim\hat{\sigma}\lesssim4$, because we need higher concentrations and more frequent scattering to drive core-collapse within the simulation time limit of 10 Gyr. As can be seen in Fig. 2 of \citealt{Essig_2019}, the slope of $t_{\rm c}$ indeed gradually decreases with $\hat{\sigma}$, until reaching the minimum of $t_c$ at about $\hat{\sigma}=5$. This is because the heat conduction is actually suppressed when the dark matter scattering enters the highly-frequent regime and thus becomes more localized \citep{Balberg_2002, Agrawal_2017, Essig_2019}.  Other than this we conclude that our scaling relation of $t_{\rm c}$ is in agreement with that of \citealt{Essig_2019}. 

We combine the $t_{\rm c}$ of all the halos in the three subplots of Fig. \ref{fig:tc} and numerically fit (with \texttt{scipy.optimize}) its scaling relation as 

\begin{equation}\label{eqn:tc-200c}
    t_{\rm c}({\rm 200c}) = 8.23\times 10^7 \left( \frac{\sigma/m}{\rm cm^2/g}\right)^{-0.74} \left( \frac{M_{\rm 200c}}{M_\odot}\right)^{-0.24} c_{\rm 200c}^{-2.61}\ {\rm Gyr}, 
\end{equation}
or for the '200m' definition
\begin{equation}\label{eqn:tc-200m}
    t_{\rm c}({\rm 200m}) = 3.58\times 10^8 \left( \frac{\sigma/m}{\rm cm^2/g}\right)^{-0.74} \left( \frac{M_{\rm 200m}}{M_\odot}\right) ^{-0.24} c_{\rm 200m}^{-2.67}\ {\rm Gyr}.
\end{equation}

Core-collapse occurs in a shorter time for more massive (thus larger velocity dispersion), more concentrated (thus larger inner density) halos with higher cross sections, because the heat conduction is boosted by the self-interaction before it is suppressed when the mean-free-path of self-interaction becomes too small. The reversal in the high-frequency scattering regime actually hints that the ultimate fate of a core-collapsing (sub)halo may not be collapse to a singularity, but instead the formation of a dense core in thermal equilibrium that has lost thermal contact with the outer parts of the (sub)halo, unless the core is so dense that the relativistic instability takes over \citep{Shapiro_1986, Pollack_2015, Feng_2021, Feng_2021b}. We also refer readers to more recent works on SIDM and the formation of black holes, such as \citealt{Feng_2021}, \citealt{Choquette_2019}, \citealt{Shapiro_2018}, \citealt{Cruz_2021} and \citealt{dc_2017}. 

As we argue later in the following sub-sections of Sec.~\ref{sec:results}, this trend in core-collapse time with concentration, cross section, and halo mass observed in isolated halos is preserved to some extent for subhalos as well.

\subsection{Tidal effects on subhalo core-collapse}\label{sec:result2}

\begin{figure}
    \includegraphics[width=\columnwidth]{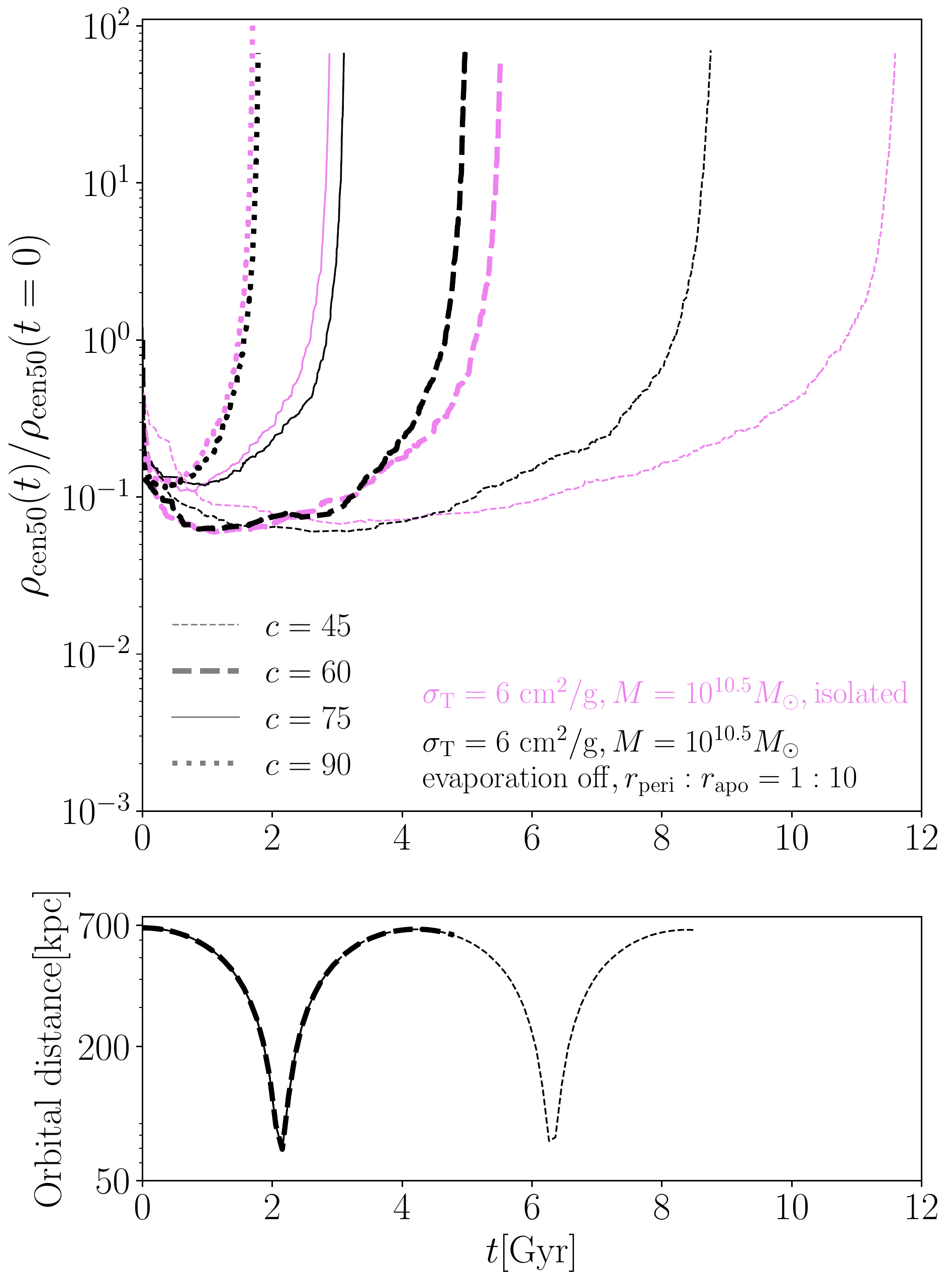}
    \caption{The central densities $\rho_{\rm cent50}$ as a function of time, for halos in isolation (pink lines) or in a host tidal field (black lines), with four groups of initial concentrations $c=45, 60, 75, 90$, together with the host-sub separations of the subhalos. Note that $\rho_{\rm cen50}$ in the top panel has been manually smoothed with exponential smoothing after the core-formation phase ($t\gtrsim 1\ \rm Gyr$). }
    \label{fig:uds-tidal}
\end{figure}

\begin{figure*}
    \centering
    \begin{subfigure}{0.32\textwidth}
        \centering
        \includegraphics[width=\textwidth, clip,trim=0.3cm 0cm 0.3cm 0cm]{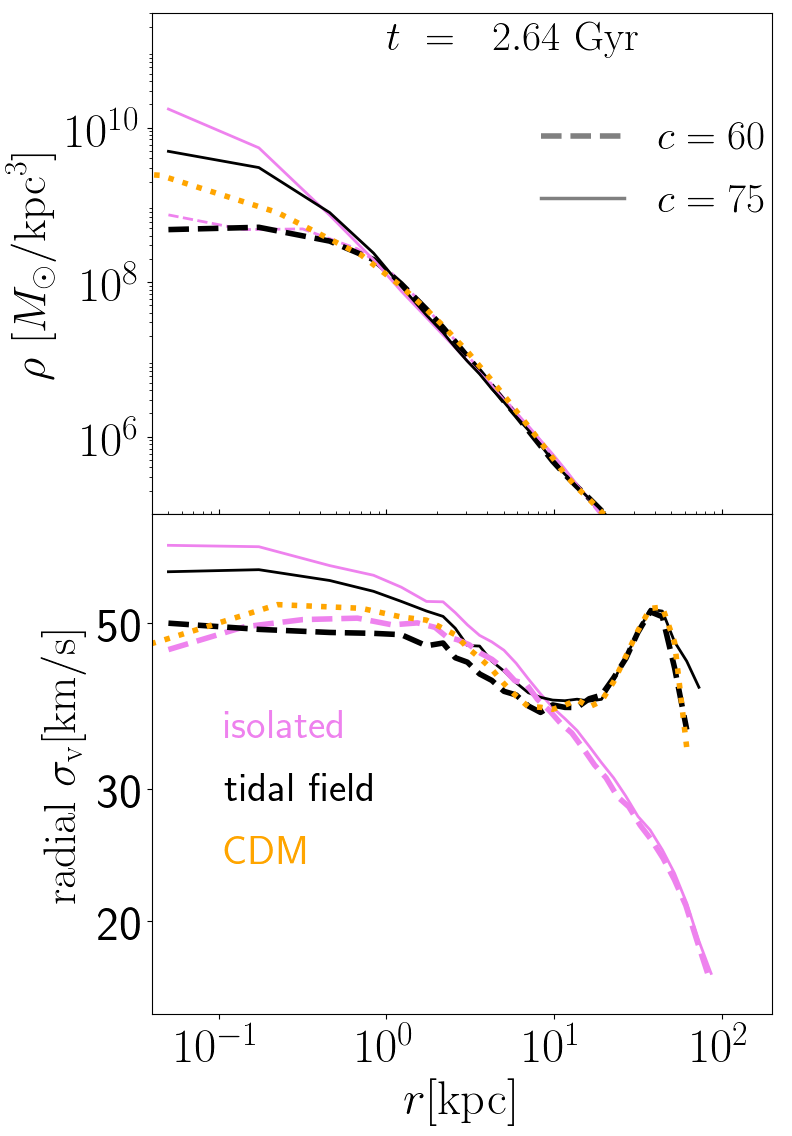} 
        \caption{}
    \end{subfigure}
    ~
    \begin{subfigure}{0.32\textwidth}
        \centering
        \includegraphics[width=\textwidth, clip,trim=0.3cm 0cm 0.3cm 0cm]{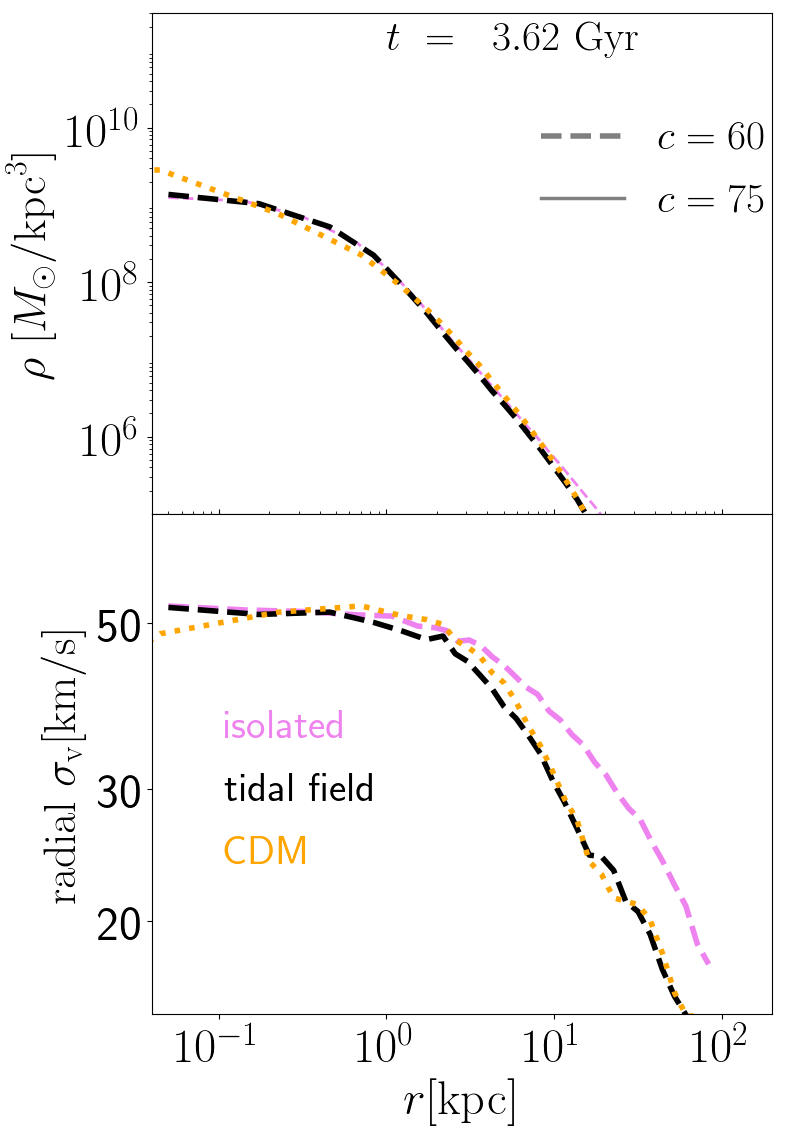} 
        \caption{}
    \end{subfigure}
    ~
    \begin{subfigure}{0.32\textwidth}
        \centering
        \includegraphics[width=\textwidth, clip,trim=0.3cm 0cm 0.3cm 0cm]{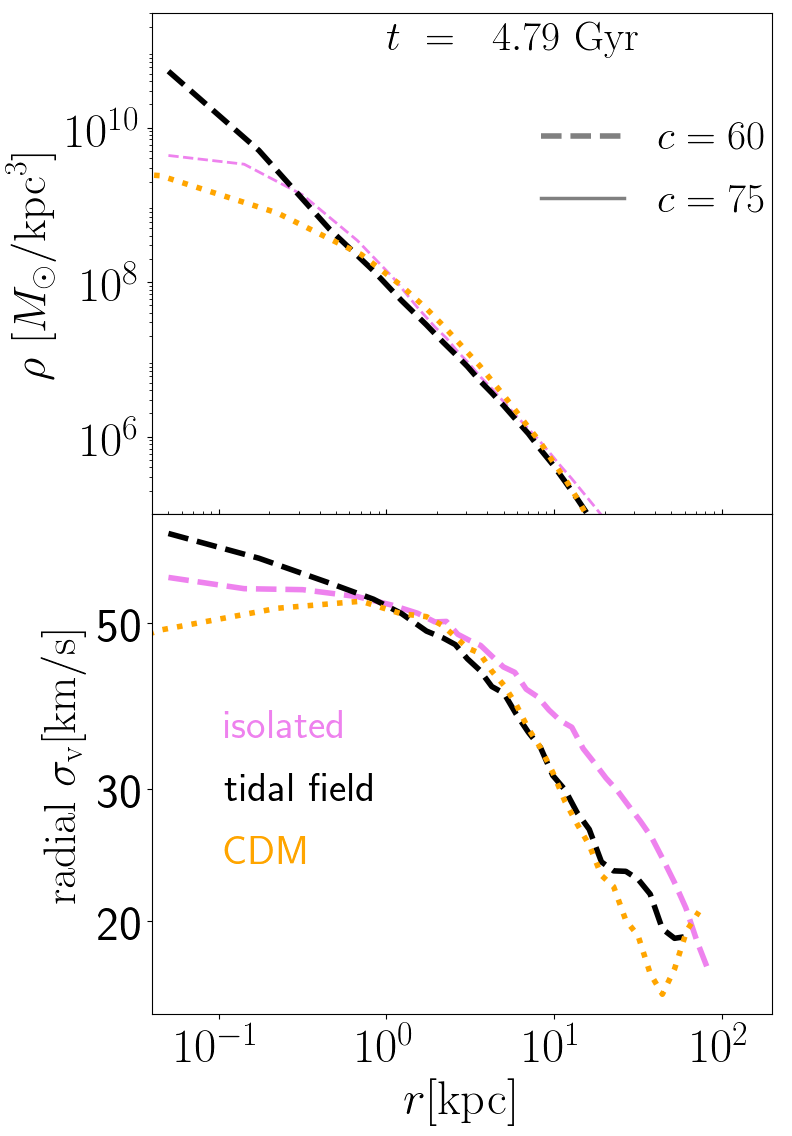} 
        \caption{}
    \end{subfigure} 
    \caption{The density profiles and radial velocity dispersion profiles of SIDM halos in isolation (pink lines) or in a host potential (black lines), with initial concentration $c=60$ (dashed lines) or 75 (solid lines), and fixed cross section $\sigma_{\rm T}/m = 6\ \rm cm^2/g$ and $M_{\rm sub} = 10^{10.5}M_\odot$ (with analytic host mass $10^{13.5}M_\odot$). A same CDM halo in the tidal field is shown in the yellow dotted line as a comparison. The three time stamps are: a) core-collapse time of the $c=75$ isolated halo, which is earlier than the core-collapse of the $c=75$ subhalo in a tidal field; b) after the first pericenter, both $c=60$ halos are relaxed again; c) core-collapse of the $c=60$ subhalo in a tidal field, which is earlier than its counterpart for the same halo in isolation. Readers are referred to Appendix C for a full evolution movie.}
	\label{fig:tidal-pro}
\end{figure*}

As has been observed in both semi-analytical treatments of truncating the subhalo \citep{Nishikawa_2020, Correa_2021} and N-body simulations of SIDM subhalos with an analytical host potential \citep{Kahlhoefer_2019, Sameie_2020}, tidal effects that detach dark matter from the outskirts of an SIDM subhalo enhance the formation of the negative temperature gradient, and thus accelerate the core-collapse process (see also Sec. \ref{sec:intro} for more context). In this section we run simulations as in \citealt{Sameie_2020}, with a similar setup of host and subhalo as in Sec. \ref{sec:val2} ($M_{\rm host}=10^{13.5}M_\odot, M_{\rm sub} = 10^{10.5}M_\odot, \sigma_{\rm T}/m=6\ \rm cm^2/g, r_{\rm peri}:r_{\rm apo}=1:10$). 
Our goal is to understand the role that the tidal field plays in subhalo core-collapse, which is more complex than shown in prior work, and which depends on the timescales of orbiting and core-collapse.

We use the (sub)halo central density $\rho_{\rm cen50}$ as the indicator of whether and when the (sub)halo core-collapses, and show the comparison of subhalos and the same halos in isolation with black and pink colors in Fig. \ref{fig:uds-tidal}.  $\rho_{\rm cen50}(t)$ is updated at each simulation timestep, and saved for output when a $\pm 20\%$ change is detected, i.e. the data points are equally spaced along the $\rho$-axis but not $t$-axis for the top panel in order to capture the feature where the central density exponentially grows in a short time.  In this figure, we consider four sets of simulations with initial concentrations $c=45, 60, 75, 90$. Our results confirm the tidal acceleration on subhalo core-collapse for relatively low $c$, in our cases of $c=45$ or $c=60$. However, the net tidal effect on subhalo core-collapse is \emph{reversed} at high concentration ($c = 75$), and the effect of tides nearly vanishes at our highest concentration ($c=90$).  This trend was not observed in previous studies. 
To understand the tidal deceleration on core-collapse that we first saw in the $c=75$ case, we select the two groups $c=60$ and $c=75$ and scrutinize the detailed evolution of their densities and radial velocity dispersion as shown in Fig. \ref{fig:tidal-pro}.  We also include CDM simulations for comparison.

The leftmost figure of Fig. \ref{fig:tidal-pro} presents the first pericenter encounter, where the tidal effects are the strongest. 
For isolated halos (pink), their velocity dispersion is high at the center and dropping outward, with a flat core at the center for the lower concentration case and a steep slope for the higher concentration, a sign that core-collapse has started for the latter.
However, for the subhalos (black lines for SIDM, orange for CDM), we observe a rise in the velocity dispersion from the center to a peak at large radii due to tidal heating, for both $c=60$ and $c=75$. The sharp drop of radial velocity dispersion beyond that peak, at large $r$, is due to the heated particles being stripped away by the tidal force. The small peak-like feature at around $r=10\ \rm kpc$ is moving outward with time (see the animation in Appendix \ref{appendix:anime} for details), showing the outflow of inner dark matter particles that have been heated up by tidal heating, which is also shared by the CDM subhalo (yellow line). This well-synchronised movement of the peak among CDM and SIDM models indicates that this outflow is indeed in the form of particles rather than heat transfer via dark matter self-interaction. 

Tidal heating affects the evolution of core-collapse in subhalos.  The core-collapse of SIDM halos relies on the formation of the negative temperature gradient, which makes core-collapse a self-accelerating process. But when tidal heating is strong, the negative temperature gradient of the subhalo is disrupted, because dark matter at larger radius of the subhalo are heated more (see Sec. \ref{sec:result3} for a more quantitative description). This blocks the heat from being transferred outward. Thus the core-collapse is delayed for the $c=75$ halo when it is in the tidal field as we see in its density profile (upper panel of Fig. \ref{fig:tidal-pro}a). However, tidal effects have a net acceleration on the core-collapse of the $c=60$ subhalo in the same potential field, because after the pericenter passage the whole subhalo regains virial equilibrium and forms a steeper velocity gradient due to the loss of dark matter heated up and stripped by tidal forces, as seen in Fig. \ref{fig:tidal-pro}b, and eventually core-collapses earlier than its isolated counterpart as seen in Fig. \ref{fig:tidal-pro}c. We can see that while tidal stripping accelerates the core-collapse of a subhalo, tidal heating plays a mixed role, that it delays the core-collapse when the subhalo approaches the pericenter, but becomes an accelerant after the pericenter. But, tidal heating and stripping, both rooted from the same tidal force, are not completely separable. Thus, tidal effects as a whole have a mixed impact on the core-collapse of SIDM subhalos, and simple truncation of the subhalo density profile does not capture all of it, as tidal heating is left out.

As shown in Fig. \ref{fig:uds-tidal}, for the plunging orbit of $r_{\rm peri}:r_{\rm apo} = 1/10$, the mixed impact of the tidal field on the central density is a result of the competition between time scales of the subhalo intrinsic (when in isolation) core-collapse $t_{\rm c}$ and its orbital motion period $t_{\rm orb}$.  The tidal field accelerates subhalo core-collapse when $t_{\rm c} > t_{\rm orb}/2$, decelerates when $t_{\rm c} \approx t_{\rm orb}/2$, and has little effect when $t_{\rm c} < t_{\rm orb}/2$. 

\subsection{Tidal-field-only vs. full evaporation effects on subhalo evolution: a case study with circular orbits}\label{sec:result3}

\begin{figure}
    \includegraphics[width=\columnwidth]{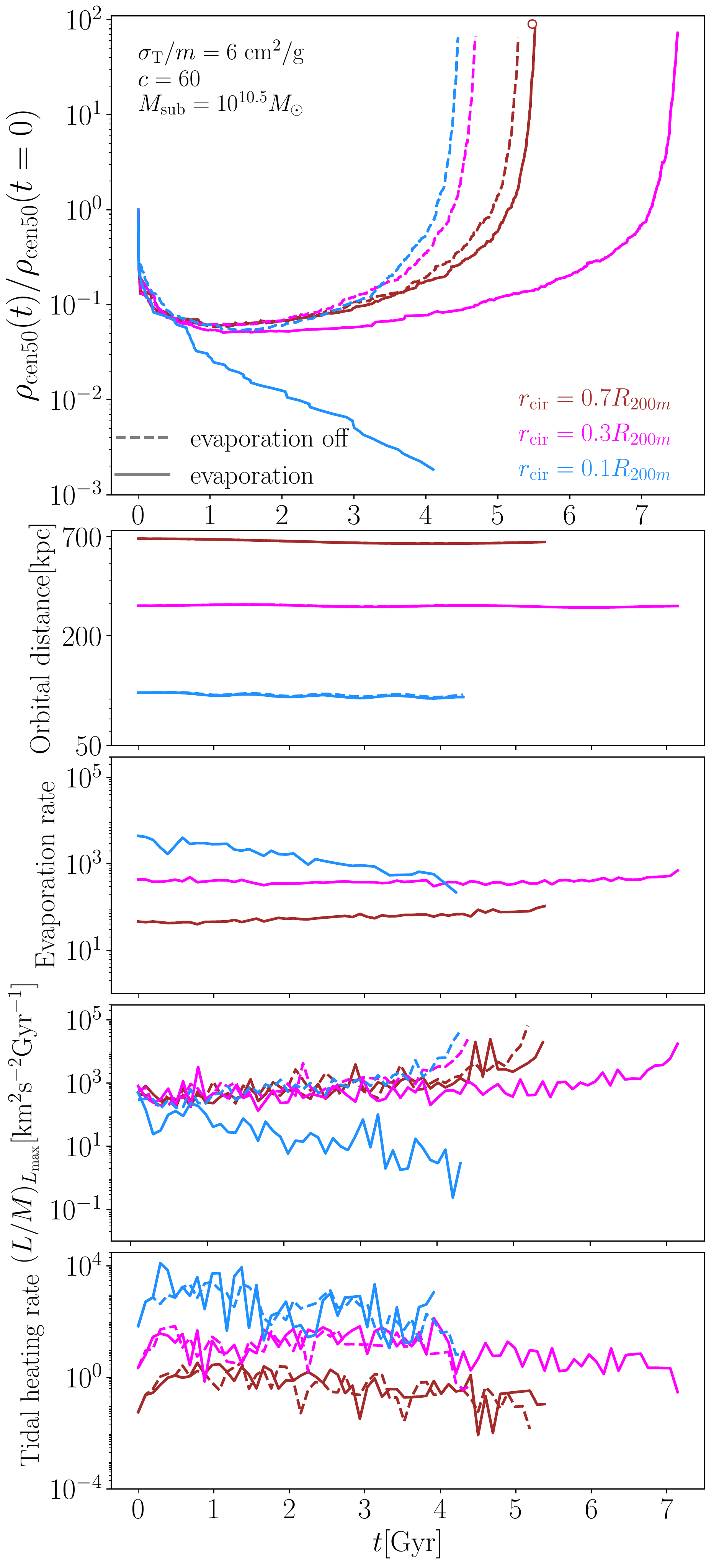}
    \caption{Evolution of subhalos on different circular orbits, from top to bottom: the central density of the subhalo $\rho_{\rm cen50}$, the host-subhalo separation, the evaporative heating rate in the subhalo, the heat outflow at the subhalo velocity core boundary as a cooling term, and the average tidal heating rate within the subhalo core. Circular orbits with three radii are chosen: brown lines for $r_{\rm cir}=0.7R_{\rm 200m}$, magenta lines for $r_{\rm cir}=0.3R_{\rm 200m}$, and blue lines for $r_{\rm cir}=0.1R_{\rm 200m}$; two semi-analytical models for SIDM subhalo evolution are shown, dashed lines for tidal-field only and solid lines for cases with full evaporation. The corresponding core-collapse time for a same halo, but in isolation without a host potential, is shown as the brown circle in the top panel. Note that the bottom three panels of heating/cooling terms have the same units. }
    \label{fig:uds-rr1}
\end{figure}

In this section, we simulate a set of subhalos ($c=60,~\sigma_{\rm T}/m = 6\ \rm cm^2/g,~M_{\rm sub}=10^{10.5}M_\odot$) on circular orbits.  By simulating circular orbits, we remove the periodic behavior of the host-subhalo evaporation and tidal effects, such that the comparison between different physical processes is clearer. A more complete and realistic study of subhalos with varying parameters will be shown in the next two sections. Here we also compare the cases with only tidal field (evaporation off) against those with full evaporation. 

In the top panel of Fig. \ref{fig:uds-rr1}, we show the central density $\rho_{\rm cen50}$ of different subhalos as a function of time, where we use dashed lines to denote simulations with evaporation turned off, and solid lines for cases with full evaporation.  The three colors mark subhalos on three different circular orbits. It can be seen that the host-subhalo evaporation always delays the subhalo core-collapse, with an increasing impact as the radius of the circular orbit $r_\mathrm{cir}$ decreases. For our orbit with the largest radius of $r_\mathrm{cir} = 0.7R_{\rm 200m}$, evaporation has only a minimal effect; while at the orbit with the smallest radius of $r_\mathrm{cir} = 0.1R_{\rm 200m}$, evaporation is so strong that the whole subhalo is gradually evaporated and never core-collapses. 

We further explore the interplay between the two orbit-related effects and SIDM-induced heat flow in the lower panels of Fig. \ref{fig:uds-rr1}. Here, we present supplementary information about how relevant physical processes affect subhalo evolution, from the top to bottom: the orbital distance between the host and subhalo, the host-subhalo evaporation heating rate (Eq. \ref{eq:apd2}), the average internal cooling rate within the surface of maximum luminosity of the subhalo (luminosity as in Eq. \ref{eq:L}), and the tidal heating rate (Eq. \ref{eq:th}). All of these heating/cooling rates are in units of energy per unit mass per unit time, as indicated in the luminosity panel.

The evaporation heating rate is the product of the event rate of a subhalo particle scattering off a (virtual) host halo particle and the average energy transferred to the subhalo during one such scattering event. Note that the energy gain of the subhalo per event may not equal the energy transferred to the evaporated subhalo particle. In fact, it is very possible that the after-evaporation velocity of a subhalo particle exceeds the escape velocity, leading to its unbinding and mass loss of the subhalo. Thus the heat transferred during the host-subhalo particle scattering is partially carried away by the escaping subhalo particle. A more careful analysis of the net effect of evaporation on the subhalo should involve at least the mass loss, the adiabatic expansion induced by the mass loss, heating that does not lead to unbinding, and re-scatterings between escaping particles and the rest of the subhalo, which is complicated and beyond the scope of this work. We propose two limits for the evaporation heating rate: the scenario where all evaporated subhalo particles remain bounded to the subhalo, as the maximal heating limit; and the scenario where all evaporated subhalo particles are instantaneously expelled from the subhalo, as the minimal heating limit (complete thermalization limit vs. instantaneous expulsion limit; see Appendix \ref{appendix:heating} for a detailed discussion). Because of the large difference between the typical velocity of the subhalo particles and that of the host particles, most of the host-subhalo evaporation scatterings should lead to expulsion of the subhalo particles, thus here we choose the minimal heating case to evaluate the evaporation heating rate. The event rate (probability per unit time) is given by $P_{h}/\delta_t$ of Eq.~\eqref{eqn:ph}, and the heat transferred to the subhalo per such event per unit mass equals the change in the binding energy within the subhalo core due to the mass loss, $<-\phi-v^2/2>$, where $\phi$ and $v$ are the gravitational potential and velocity of subhalo particles relative to the subhalo center, averaged within the subhalo core radius.  

We follow the formalism of \citealt{Essig_2019} to trace the internal heat flow of the subhalo, where the luminosity $L(r)$ as a function of the temperature gradient is the key factor. The luminosity is given by 
\begin{equation}\label{eq:L}
    L/4\pi r^2 = -(\kappa_{\rm lmfp}^{-1} + \kappa_{\rm smfp}^{-1})^{-1} \partial T / \partial r ,
\end{equation}
 where $\kappa_{\rm lmfp}\approx 0.27 \beta nv^3\sigma_{\rm T} k_B/G m$ and $\kappa_{\rm smfp}\approx 2.1 vk_B / \sigma_{\rm T}$ are the heat conductivity for the long-mean-free-path and short-mean-free-path regimes, $v$ is the one dimensional velocity dispersion, $n=\rho/m$ is the number density of dark matter, $k_{\rm B}$ is the Boltzmann constant, and $T = mv^2/k_{\rm B}$ is the temperature, assuming dark matter follows monatomic thermal dynamics. $L(r)$ is the total heat outflow at the surface with radius $r$ per unit time, thus we take $L(r)/M(<r)$ to be the average cooling rate on all dark matter within the surface $r$. We trace the time evolution of $L(r)/M(<r)$ evaluated at the radius of maximal $L$, which generally coincides with the size of the velocity core of the subhalo (not the density core, see Fig. 6 of \citealt{Nishikawa_2020} for an  insightful set of plots illustrating the difference), as the strength of the cooling of the subhalo core.

For the tidal heating rate, we follow the semi-analytical evaluation of  \citealt{Pullen_2014} (their Eq. 15), which accounts for the breakdown of the impulse approximation of the gravitational encounter when the orbit and encounter time scales are comparable, and also includes corrections for higher order heating effects:

\begin{equation}\label{eq:th}
    \frac{{\rm d}E}{{\rm d}t} (x, t) = \left[1+\left(\frac{T_{\rm shock}}{T_{\rm orb}}\right)^2 \right]^{-2.5} x^2 g_{a,b}(t) G_{a,b}(t),
\end{equation}
where $T_{\rm shock}=r_{\rm orb}/v_{\rm orb}$ and $T_{\rm orb}$ are the gravitational shock and orbit timescales, with $r_{\rm orb}$ and $v_{\rm orb}$ being the orbital distance and velocity of the subhalo, and $x$ is the radius of the subhalo at which the differential tidal heating rate is calculated.  We consider two tensors:  $g_{a,b}$ is the tidal tensor (Eq. 10 of \citealt{Pullen_2014}); and $G_{a,b}$ is the time integral of $g_{a,b}$:

\begin{equation}
    g_{a,b} = \frac{GM_{\rm host}(<r)}{r^3} \left( \frac{3r_a r_b}{r^2} - \delta_{a, b}\right) - 4\pi G\rho_{\rm host} \frac{r_a r_b}{r^2},
\end{equation}
with $r=r_{\rm orb}$ being the orbital distance between the subhalo and the host center. We can see that the strength of tidal heating grows as $x^2$, meaning that although it heats up the inner part of the subhalo and delays the central density growth, it heats up the outer layers even more. The stronger heating at large subhalo radius could either directly hinder the formation of the negative temperature gradient (larger cooling by $L$) or help with it by boosting stripping efficiency, at different phases of an orbital period, as we have observed in Sec. \ref{sec:result2}. But this complex, mixed role of the tidal field on subhalo core-collapse via effects on subhalo outliers has been absorbed in the luminosity term we show above. Thus, in Fig. \ref{fig:uds-rr1} and similar figures below, we evaluate the average tidal heating rate within the radius of $L_\mathrm{max}$, the same radius as the cooling term $L/M$, to highlight the direct heating term of the tidal field within the subhalo central core:

\begin{equation}
    \frac{{\rm d}E}{{\rm d}t}(t) = \frac{\int _0 ^{r_{L_{\rm max}}} \frac{{\rm d}E}{{\rm d}t}(x,t) 4\pi x^2dx}{\int _0 ^{r_{L_{\rm max}}} 4\pi x^2 dx} = \frac{3}{5} \frac{{\rm d}E}{{\rm d}t}(r_{L_{\rm max}}, t)
\end{equation}
 
With all the three heating/cooling terms explained and evaluated at the same radius of the subhalo, the velocity core radius, we are able to compare them and their effects on subhalo evolution, as shown in the three panels of Fig. \ref{fig:uds-rr1}. The evaporation rates are flat on circular orbits (except for the blue subhalo, whose average binding energy within the core decreases), because they are only affected by SIDM cross section, the orbital parameters, the host properties and the average binding energy within the subhalo core to the first order. The heat outflow rates at the core radius are flat most of the time as well, until the late time exponential growth, which corresponds to the steep temperature gradient when the subhalo is core-collapsing. The $L/M$ panel suffers from overall greater fluctuations than the evaporation panel, mainly because $L$ is the first-order derivative of dark matter 1-D velocity dispersion, which is already a noisy term itself, limited by the simulation resolution and also the resolution of radial binning when we measure the subhalo velocity profile. The tidal heating rates are also relatively flat with time because of the circular orbit, with a slight trend of decreasing near the core-collapse time, as $r_{L_{\rm max}}$ where we measure the tidal heating rate shrinks. The wiggles in the tidal heating panel are similarly due to the finite resolution in measuring $L$, since the tidal heating rate is measured at the same radius of maximal $L$. Tidal heating is also less significant than the other two terms for the parameters of subhalos that we are considering, except for the case with the smallest orbital radius $r_{\rm cir} = 0.1R_{200m}$ (blue color), where although the data we present yield to noise, we can see that the tidal heating is comparable to the other two heating/cooling terms at order of magnitude level. 

These three subhalos on circular orbits serve as representative examples to analyze the interactions of different physical processes and the subhalo central density. For the subhalo with the largest orbital radius $r_{\rm vir}=0.7R_{200m}$ (brown color), both the evaporation and tidal effects are relatively weak on it---outweighed by the subhalo internal cooling term $L/M$ by one and two orders of magnitude respectively, and thus the evolution of its central density is dominated by its internal heat flow. As a result, the core-collapse time of the subhalo in brown color, either with full evaporation or only with the tidal field, is close to the $t_c$ of the same halo in isolation (the small circle in the $\rho_{\rm cen50}$ panel of Fig. \ref{fig:uds-rr1}). For the subhalo with the smallest orbital radius $r_{\rm vir}=0.1R_{200m}$ (blue color), the total heating outweighs cooling by about one order of magnitude from the beginning. This leads to the subhalo in blue color having a monotonically decreasing central density, and it is completely disrupted at around 4 Gyr. The subhalo with the intermediate orbital radius $r_{\rm vir}=0.3R_{200m}$ (magenta color) is an interesting, marginal case. The cooling term $L/M$ and the evaporation heating are close in magnitude, with tidal heating being insignificant. Therefore, we see that the central density of the subhalo in magenta color grows slowly for a long time after its core-formation phase, and the subhalo core-collapses long after the $t_c$ of its counterpart halo in isolation. This is the result of the close competition between the total heating and cooling processes.

Furthermore, we find additional evidence of the mixed effects of tidal heating and tidal stripping.  In the 1D model of \citealt{Nishikawa_2020}, in the absence of evaporation, core-collapse is increasingly accelerated for decreasing tidal radius.  In the case of circular orbits, we expect the smallest orbits to have the largest amount of tidal stripping, and the smallest tidal radii.  In the dashed lines in Fig.~\ref{fig:uds-rr1}, which show the central density evolution without evaporation, core-collapse proceeds nearly identically for our two innermost orbits, despite the fact that tidal stripping is most severe for the innermost (blue) orbit.  The tidal heating rate, however, is also significantly stronger for this orbit than our middle (magenta) orbit.  This has potentially important implications for the close-to-host-center subhalos that are especially relevant to substructure lensing. 

In this section, our primary conclusion is that there is a strong link between the evolution of the subhalo central density and the heating/cooling terms, that the central density grows (and finally undergoes core-collapse) when the cooling term outweighs the total heating terms inside the velocity core, and grows faster when the net cooling is stronger. We discuss more general cases with eccentric orbits in the next sub-section, where periodical behavior of relevant heating/cooling terms are added back.

\subsection{Full simulation suite including evaporation}\label{sec:result4}

\begin{figure}
    \includegraphics[width=0.99\columnwidth]{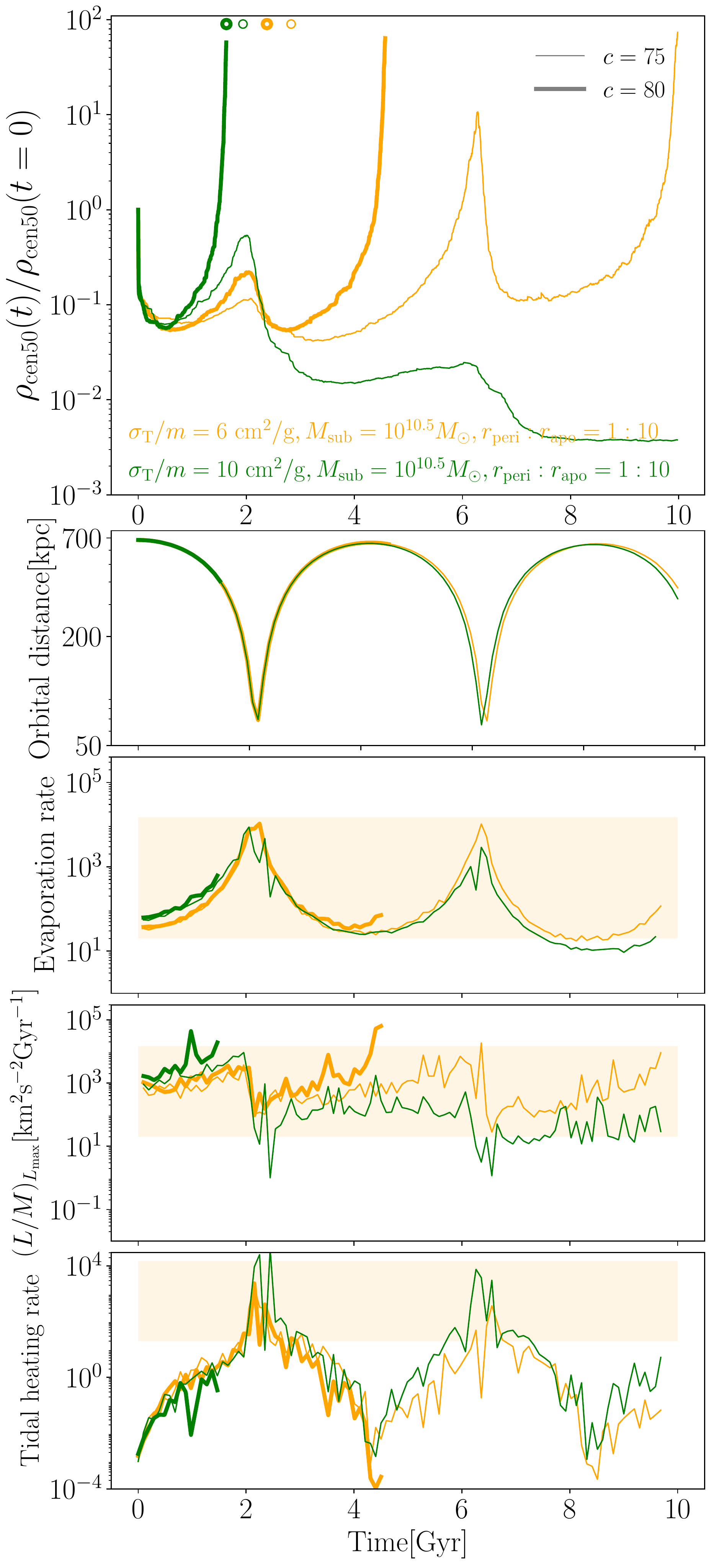}
    \caption{Similar to Fig. \ref{fig:uds-rr1}, but with eccentric orbits and fixed apocenter. The self-interaction cross section is $\sigma_{\rm T}/m=6\ \rm cm^2/g$ (orange) or $10\ \rm cm^2/g$ (green), with $c$ chosen to be 75 (thinner line) or 80 (thicker line), near the critical concentration for subhalo core-collapse. The corresponding intrinsic core-collapse time scales $t_{\rm c}$ calculated by Eq. (\protect\ref{eqn:tc-200m}), the expected core-collapse time when these halos are in the field rather than subhalos, are shown in corresponding small circles in the top panel. Note that we have highlighted the range of $(L/M)_{L_{\rm max}}$ of the orange thin line as the light orange shadow in the three lower panels, such that the comparison of the heating/cooling energy scales becomes visually easier. }
    \label{fig:uds-sig}
\end{figure}

\begin{figure}
    \includegraphics[width=0.99\columnwidth]{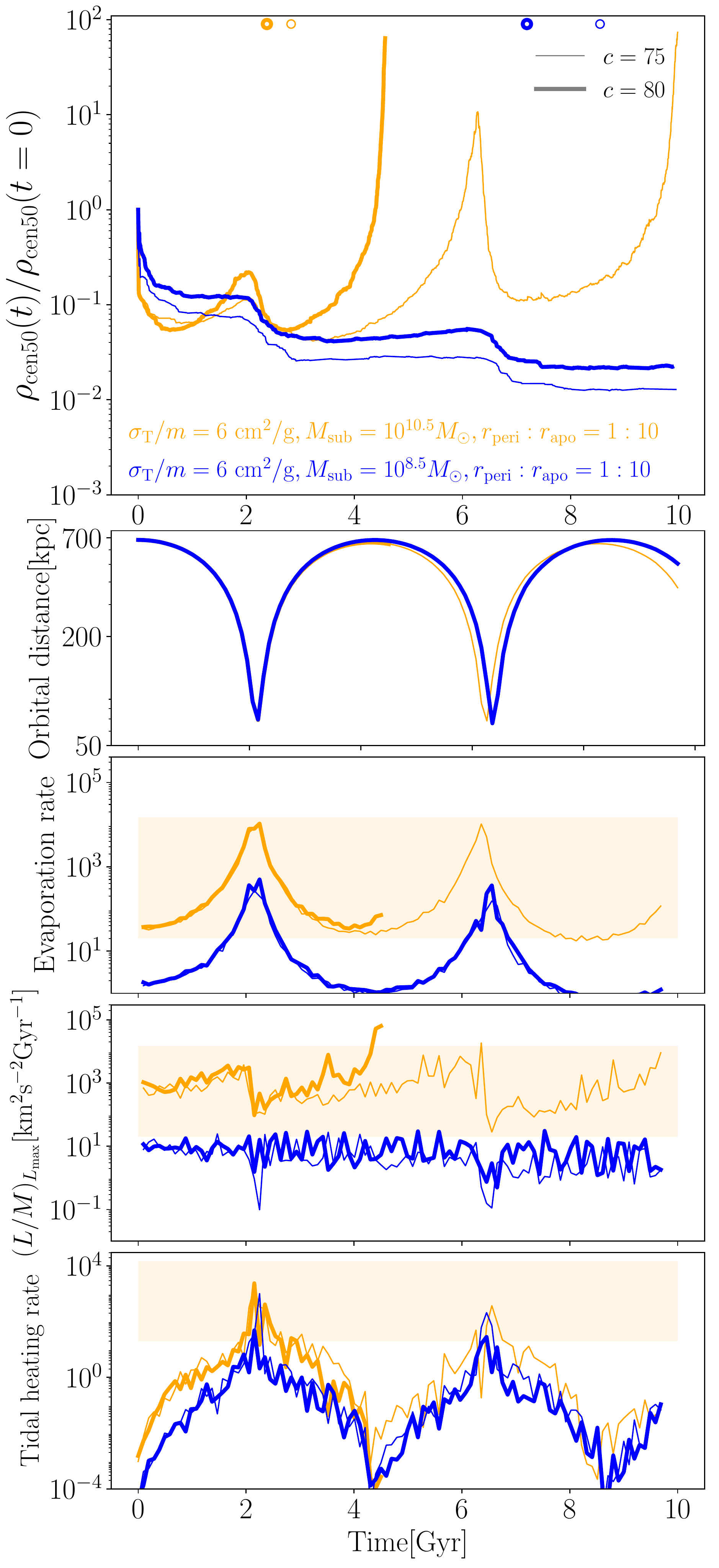}
    \caption{Same as Fig. \protect\ref{fig:uds-sig}, but varying the subhalo mass $M_{\rm sub}$ instead of cross section.}
    \label{fig:uds-m}
\end{figure}

\begin{figure}
    \includegraphics[width=0.99\columnwidth]{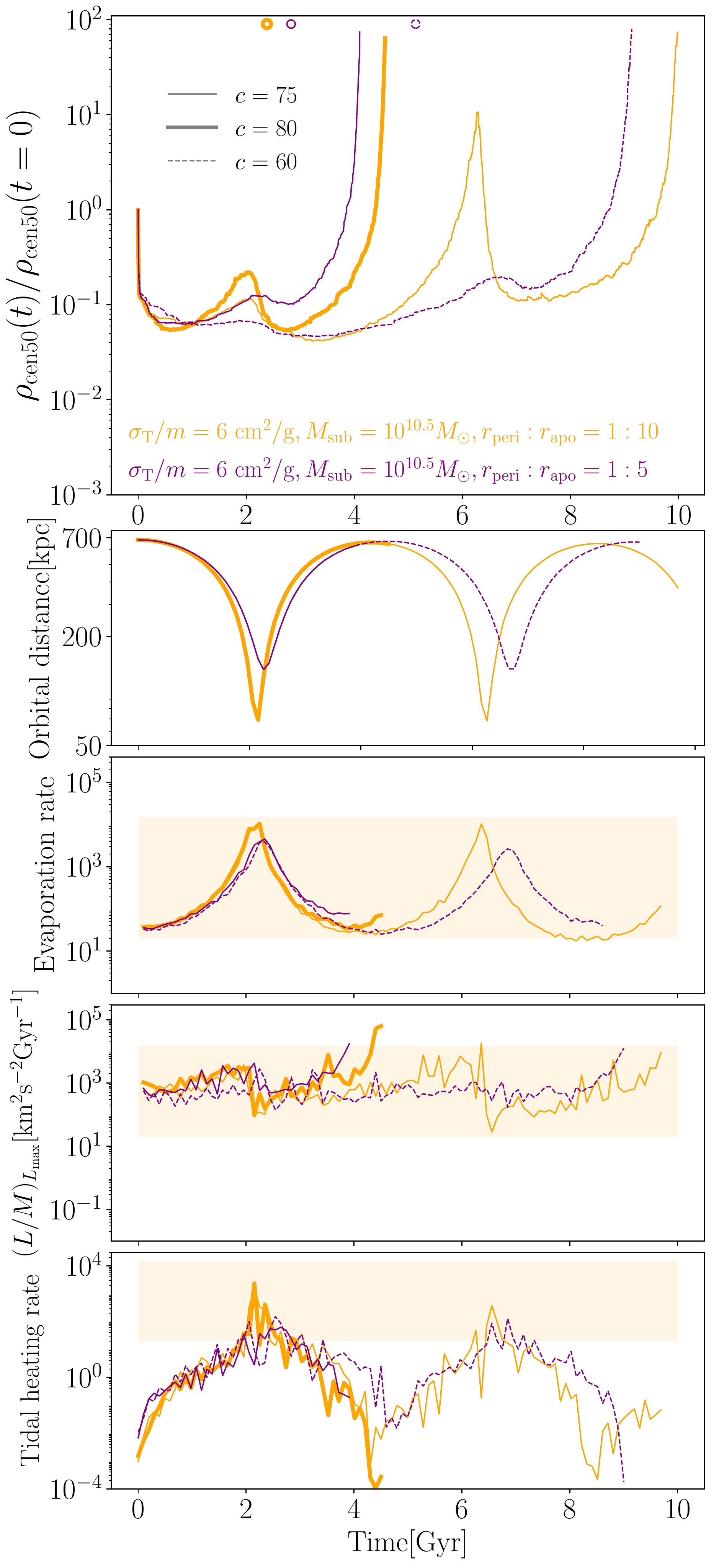}
    \caption{Same as Fig. \protect\ref{fig:uds-sig}, but varying orbit characterized by $r_{\rm peri}:r_{\rm apo}$ instead of cross section. }
    \label{fig:uds-orbit}
\end{figure}

In the code validation runs (Sec. \ref{sec:val2}), we deliberately selected a parameter set $[M_{\rm sub} = 10^{10.5} M_\odot, \sigma_{\rm T}/m = 6\ {\rm cm^2/g}, r_{\rm peri}:r_{\rm apo} =1/10,c=80]$ such that this particular subhalo is right on the edge of  core-collapse. In other words, the unphysically high $c=80$ is close to the minimal (critical) concentration needed for subhalo core-collapse with other parameters specifying the subhalo and its orbit held fixed, as we shall see later. In this section we vary these four parameters around the critical concentration to explore the transition cases.  We find that whether a subhalo core-collapses or not is highly sensitive to these parameters. Similar to the previous section, to fully understand the physics behind subhalo core-collapse, we track the evolution of the subhalo central density $\rho_{\rm cen50}$, together with a breakdown of the strength of relevant physical processes: host-subhalo evaporation and tidal heating as heating terms, and heat outflow at the velocity core as a cooling term. In Fig. \ref{fig:uds-sig} we show the cases where $c$ and $\sigma_{\rm T}/m$ are varied, in Fig. \ref{fig:uds-m} $c$ and $M_{\rm sub}$, and in Fig. \ref{fig:uds-orbit} $c$ and $r_{\rm peri}:r_{\rm apo}$.

In Fig. \ref{fig:uds-sig}, we present four subhalo simulations: SIDM cross sections $\sigma_{\rm T}/m=6~{\rm cm^2/g}$ (orange) or $10~{\rm cm^2/g}$ (green), and initial concentrations $c=75$ (thin lines) and $c=80$ (thick lines), with the same fixed subhalo mass $10^{10.5} M_\odot$ (i.e. subhalo-host mass ratio 1:1000) and $r_{\rm peri}:r_{\rm apo}=1:10$ (apocenter fixed at $0.7R_{200m}\approx 700\ \rm kpc$, unless otherwise specified). We choose $c=75$ or 80 because these subhalos are right on the edge of undergoing core-collapse within 10 Gyrs.  For example, in the $\sigma_{\rm T}=10~{\rm cm^2/g}$ with $c=80$ case, we clearly see the stages of the subhalo core-formation first, where $\rho_{\rm cen50}$ drops by one order of magnitude, then core-collapses, during which the central density increases from the lowest-density core by three orders of magnitude in a short time.   However, for the subhalo with the same SIDM cross section $\sigma_{\rm T} = 10\ \rm cm^2/g$ but slightly smaller initial concentration $c=75$, the core-collapse process happens more slowly and less violently, and is halted before the first pericenter. This hints that whether a subhalo will finally core-collapse or not on cosmological timescales may have a sharp transition in parameter space. A similar disruption of the core-collapse is observed for the cases of the smaller cross section $\sigma_{\rm T}/m=6~{\rm cm^2/g}$ (orange) near the first pericenter, but compared to the higher cross section cases, the subhalo survives the strongest evaporation and its core-collapse process is resumed at late times for $c=80$. For the subhalo with $\sigma_{\rm T}=6~{\rm cm^2/g}$ and $c=75$, the core-collapse process is further disrupted again at the second pericenter, where its central density has already grown by a factor of 10, and similarly resumed later, reaching our core-collapse criterion at the very end of 10 Gyr. Thus we record $c=75$ as the critical concentration for core-collapse for $\sigma_{\rm T}/m=6~\rm cm^2/g$ and $c=80$ for $\sigma_{\rm T}/m=10~\rm cm^2/g$, when the mass ratio is 1:1000 and the orbit is characterized by $r_{\rm peri}:r_{\rm apo}=1:10$. This seemingly counter-intuitive behavior that the subhalo with a lower cross section could have an easier core-collapse, contrary to what we have seen for isolated halos, highlights the power of host-subhalo evaporation in disrupting the core-collapse process of the subhalo. 

The expected core-collapse time for corresponding halos in isolation from the host are shown in small circles of Fig. \ref{fig:uds-sig}, calculated from Eq. \eqref{eqn:tc-200m}. The subhalo with $\sigma_{\rm T}=10~{\rm cm^2/g}$ and $c=80$ core-collapses at about the same time we expect for the same halo in isolation, because it happens well before the first pericenter encounter, and the host-induced effects have not become significant yet. For other cases though, their core-collapse time scales are much longer than the isolated halos, showing the impact of the host halo on SIDM subhalo evolution.

All three heating/cooling terms in the lower three panels of Fig. \ref{fig:uds-sig} are, as expected, well-synchronized with the central density and orbital distance plots. The evaporative heating rate, proportional to $\sigma_{\rm T}/m$, anti-correlates with the orbital distance, since both the host density and the bulk velocity of the subhalo increase as the orbital distance shrinks (see Eq.~\eqref{eqn:ph}). The internal cooling term on the subhalo core $(L/M)|_{r_{L_{\rm max}}}$ tracks the central density, which soars when the subhalo tries to core-collapse, and is disrupted together with the central density when the evaporation is strong. Tidal heating also follows the corresponding orbit period, but with a clear extra drop when the subhalo is core-collapsing, because of the shrinking of $r_{L_{\rm max}}$ at which the tidal heating rate is evaluated. The net cooling/heating of these three terms again determines the growth/decay of the central density $\rho_{\rm cen50}$, consistent with our primary conclusion from the previous section with circular orbits, with the evaporative heating and internal cooling being more dominant than tidal heating. 
 
Apart from the heating/cooling perspective, although not as accurate and complete, it is also straightforward to understand the core-collapse of an SIDM subhalo in terms of the competition of time scales. When the subhalo intrinsic core-collapse time $t_{\rm c}$, i.e. the isolated case fit by Eq. (\ref{eqn:tc-200m}), is smaller than half the orbit period, the subhalo may eventually core-collapse, but when its $t_{\rm c}$ is longer than $t_{\rm orb}/2$, its core-collapse process may either be permanently disrupted (the green thin line of $\sigma_{\rm T}/m = 10\ \rm cm^2/g$ in Fig. \ref{fig:uds-sig}), or temporarily halted near the pericenter and resumed at late times near the apocenter (the orange thick and thin lines of $\sigma_{\rm T}/m = 6\ \rm cm^2/g$), depending on the strength of evaporation. 
The latter scenario predicts a \emph{large diversity} of subhalo central density.
 
We show similar plots in Fig. \ref{fig:uds-m}, varying the subhalo mass $M_{\rm sub}$ instead of $\sigma_{\rm T}/m$. The group of $M=10^{8.5}M_\odot$ with the same $c=75$ or 80 as before is not close to core-collapse, compared to the group of $M=10^{10.5}M_\odot$ (blue lines vs orange), due to the long $t_{\rm c}$ in isolation (see blue circles vs orange) for low-mass halos. 
The long intrinsic core-collapse time scale $t_{\rm c}$ of a less massive halo is due to the shallower potential well.  This yields a smaller velocity dispersion and lower DM self-interaction rate within it (see Eq.~\eqref{eqn:tc-200m} and discussions in Sec. \ref{sec:result1}).  
Another way to understand it is through the $L/M$ panel of Fig. \ref{fig:uds-m}, in which we clearly see that the heat outflow of the $10^{8.5}M_\odot$ subhalo is weaker than that of the $10^{10.5}M_\odot$ subhalo by about two orders of magnitude. This is because in the long-mean-free-path regime for cored SIDM halos, the luminosity $L$ scales approximately with $\sigma_v^5$ (see Sec. \ref{sec:result3}). The evaporation heating rate is also lower for smaller subhalos of $10^{8.5}M_\odot$, because the average binding energy is lower, but only by about one order of magnitude. As a result, unlike the case with the $10^{10.5}M_\odot$ subhalo, the cooling by $L/M$ for the $10^{8.5}M_\odot$ subhalo is visibly smaller than the total heating for a big chunk of the orbit. Therefore, the critical $c$ for less massive subhalos to have core-collapse at fixed cross section will be larger, as we will show in the next section.

Similar results with varying orbits are presented in Fig. \ref{fig:uds-orbit}, where, compared to the fiducial group, we fix the apocenter of the subhalo orbit and alter $r_{\rm peri}:r_{\rm apo}$ between the default $1:10$ and the variant $1:5$. We can see that on this less plunging orbit, both the evaporation and tidal effects are weakened, with evaporation still being the dominant heating term. The net disruption on the subhalo central density near the pericenter is weaker for the less plunging orbit (see the sharp drop of $\rho_{\rm cen50}$ in the top panel, purple vs orange). As a result, the critical $c$ needed for core-collapse on the $1:5$ orbit becomes 60, while its counterpart is 80 for the $1:10$ orbit, for a fixed cross section of $\sigma_{\rm T}/m=6\ \rm cm^2/g$.

In this section we see that, as we concluded in the previous section, the core-collapse behavior of an SIDM subhalo on realistic orbits is also a result of the competition between the three relevant physical processes: internal heat transport, host-subhalo evaporation and tidal effects. Because of this, the evolution of a subhalo is very sensitive to its initial parameters, and shows a large diversity of subhalo central density, especially when near the critical parameter sets of ending up in core-collapse. In the next sub-section we map out a boundary in the parameter space for subhalo core-collapse.
 
\subsection{Mapping parameter space for subhalo core-collapse}\label{sec:result5}

\begin{figure}
    \begin{subfigure}{\columnwidth}
        \centering
    	\includegraphics[width=\textwidth]{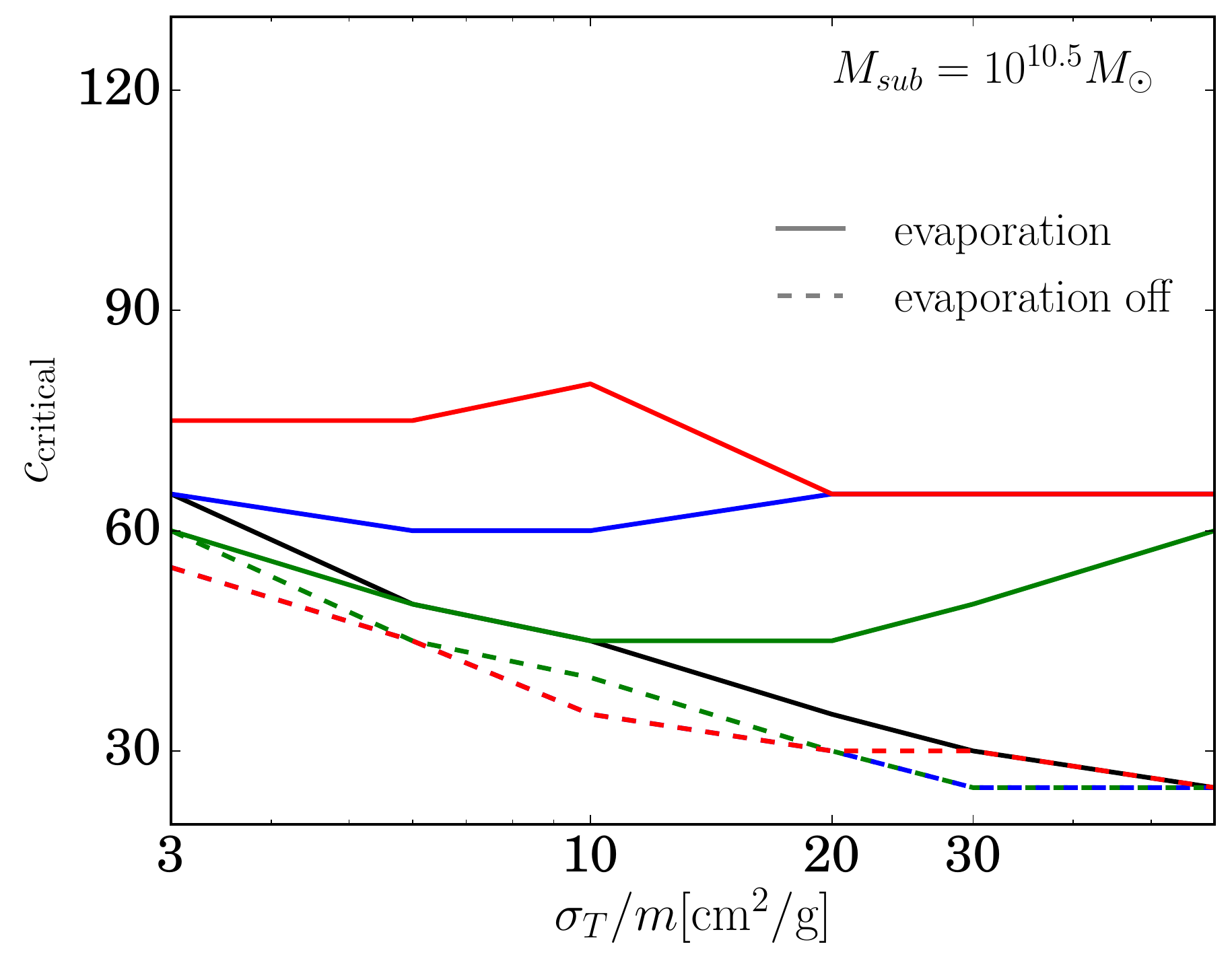}
	    \caption{}
	    \label{fig:param1}
	\end{subfigure}
	\hfill
	\begin{subfigure}{\columnwidth}
	    \centering
	    \includegraphics[width=\textwidth]{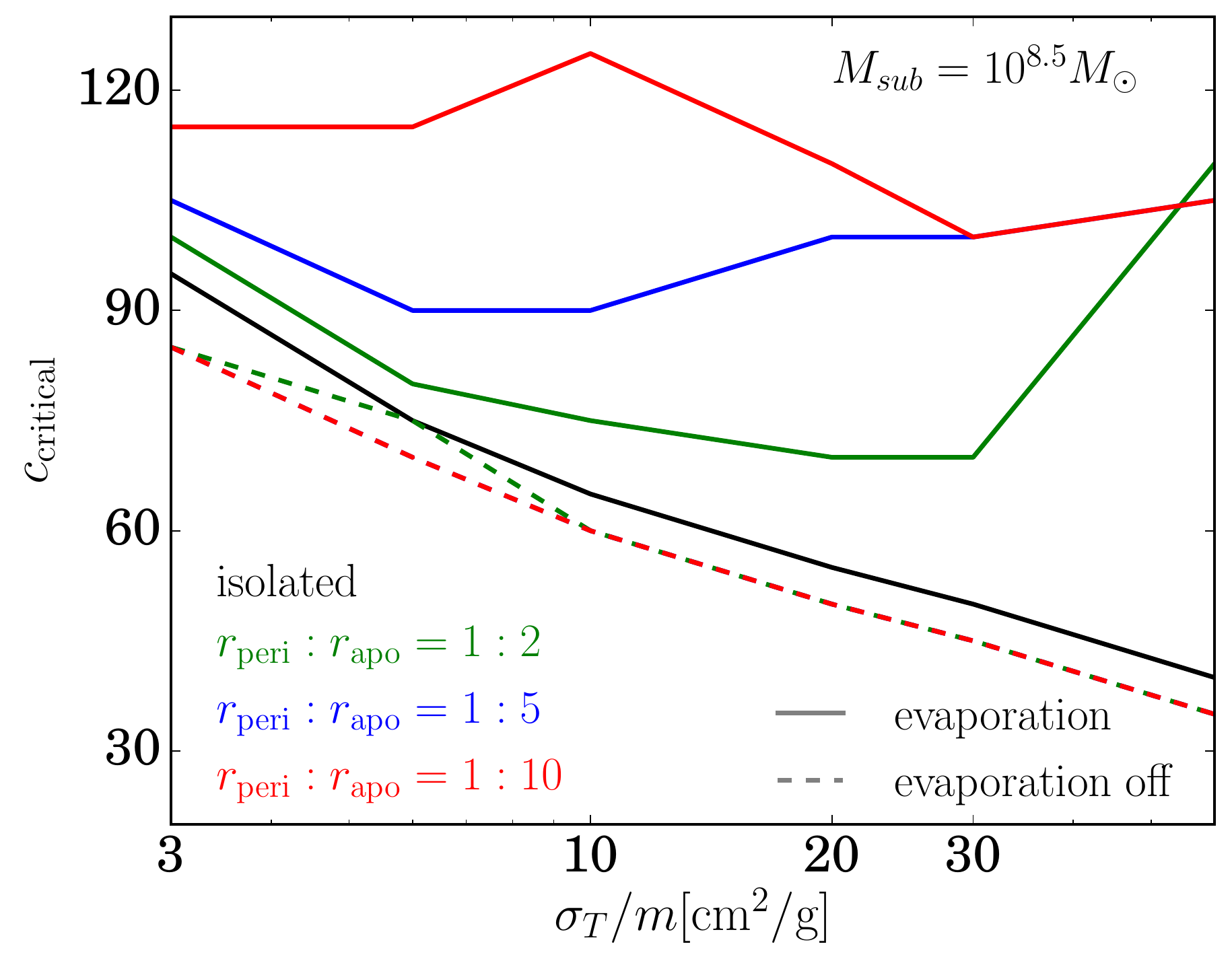}
    	\caption{}
    	\label{fig:param2}
	\end{subfigure}
	\caption{a) The critical (minimum) initial concentration $c$ needed for the subhalo of $M_{\rm sub} = 10^{10.5}M_\odot$ to  core-collapse (defined by central density growing by two orders of magnitude, see Sec. \ref{sec:method}) within 10 Gyrs. Different orbits $r_{\rm peri}:r_{\rm apo} = 1:2, 1:5, 1:10$ are shown in green, blue and red lines, while isolated halos are in black. The cases with evaporation switched on or off are shown in solid and dashed lines. b) Same as a), but for $M_{\rm sub}=10^{8.5}M_\odot$.}
    \label{fig:param}
\end{figure}

From the previous sections, we find that the critical initial concentration for driving subhalo core-collapse is sensitive to other parameters:  $M_{\rm sub}$, $\sigma_{\rm T}/m$ and $r_{\rm peri}:r_{\rm apo}$ (or, more generally, to the orbit). In this section, we map out the core-collapse criterion for isolated halos, subhalos in the absence of SIDM evaporation, and subhalos with the full evaporation model in the parameter space of $[c, M_{\rm sub}, \sigma_{\rm T}/m, r_{\rm peri}:r_{\rm apo}]$. 

Properties of the host, the subhalo's apocenter and the subhalo's evolution prior to infall matter as well, but we defer a more careful analysis of those effects to future work.  The host halo mass $M_{\rm host}$ is relevant because it determines the velocity dispersion of host particles, which affects the evaporation rate, and mediates tidal heating and stripping.  However, in this study, we fix $M_{\rm host}=10^{13.5}M_\odot$ as a representative system for substructure lensing \citep{Birrer_2017}, which is promising for constraining the likelihood of finding core-collapsed subhalos in observations. 
The apocenter distance $r_{\rm apo}$ controls the orbital time and thus is another relevant parameter, is fixed at $R_{\rm apo}=0.7R_{200m}\sim700\ \rm kpc$, following the subhalo population model of \citealt{Pennarrubia2010}. The evolution of the subhalo with dark matter self-interaction prior to its inall, which we denote as the 'pre-evolution', determines the actual initial state (e.g. NFW, cored, or early stage of core-collapse) of the subhalo at infall. In this work we effectively set the pre-evolution time $t_{\rm pre}$ to be 0 Gyr, since we include subhalos with extremely large cross sections or initial concentrations, of which the core-collapse time scales in isolation are small. We include a simple demonstration of the effect of different $t_{\rm pre}$ in Appendix \ref{appendix:preevolve}. $M_{\rm host}$, $r_{\rm apo}$ and $t_{\rm pre}$ should as well be realistically sampled in future works of population modelling of SIDM subhalos, but in this work we fix them for the purpose of reducing computational burden. 

The mass resolution in these simulations is chosen to be $M_{\rm p}=10^5M_\odot$ for $M_{\rm sub}=10^{10.5}M_\odot$ and $M_{\rm p}=10^3M_\odot$ for $M_{\rm sub}=10^{8.5}M_\odot$, ten times better resolution than the one used in the validation simulations, with robustness of the resolution tested in Sec. \ref{sec:val2}. This higher resolution is meant to reduce the uncertainty introduced by different realizations of the subhalo. With about $10^{5.5}$ simulation particles, the uncertainty in (sub)halo central density (< 1 kpc for a $10^{10.5}M_\odot$ halo, most relevant for core-collapse) of different realizations is smaller than the difference in central densities of (sub)halos with 10\% difference in initial concentration. 
Thus we claim a 10\% error bar on the critical $c$ plots in Fig. \ref{fig:param} (not shown explicitly for plot readability), and believe that this estimate is conservative. Of these simulations, the non-core-collapsed subhalos cost a few tens of CPU hours while the core-collapsed ones cost up to a thousand or more, due to the extreme refinement of the simulation timestep (four to five orders of magnitude relative to cored subhalos) on account of the high central density.

In Fig. \ref{fig:param} we show the critical $c$ needed for subhalos to have core-collapsed in 10 Gyr, against the self-interaction cross section $\sigma_{\rm T}/m$, for subhalo mass of $10^{10.5}M_\odot$ in Fig. \ref{fig:param1} and $10^{8.5}M_\odot$ in Fig. \ref{fig:param2}. Subhalos on different orbits $r_{\rm peri}:r_{\rm apo}=1:2, 1:5, 1:10$ are shown in green, blue and red colors, while same halos in isolation are in black lines. Cases with or without our evaporation model are plotted in solid or dashed lines.

As we can see in both figures of Fig. \ref{fig:param}, for isolated halos (black), $c_{\rm critical}$ monotonically decreases as $M_{\rm sub}$ and $\sigma_{\rm T}/m$ increase, as expected from Sec. \ref{sec:result1}. For subhalos with only tidal effects (dashed lines), we find that their critical concentrations $c$ are always smaller or equal to those of the isolated ones, with a decreasing trend of $c_{\rm critical}$ as orbit gets more plunging. This is due to the tidal acceleration of core-collapse. The `tidal deceleration' we find in Sec. \ref{sec:result2} is not reflected in this comparison between the isolated halos and tidal-only subhalos, because it happens in specific cases with $t_{\rm c}\lesssim t_{\rm orb}/2$, while for the boundary cases with $t_{\rm c}\sim 10$ Gyr, all subhalos have experienced at least a full orbital period. For the same reason as isolated halos, the critical $c$ of subhalos in a tidal field monotonically decreases with SIDM cross section and subhalo mass. 

The cases get more complicated with full evaporation. On top of the analysis of tidal-only subhalos above, the strength of evaporation increases with SIDM cross section and $r_{\rm peri}:r_{\rm apo}$. Thus larger $c_{\rm critical}$ is required for more plunging orbits, with an only exception at the high cross section end $\sigma_T/m=60\ \rm cm^2/g$ for the less massive subhalos $M_{\rm sub}=10^{8.5}M_\odot$, which is because the initial concentration is high enough to counter the evaporation before peri-centers, but tidal effects have come into play as an acceleration factor to core-collapse. But $c_{\rm critical}$ does not monotonically scale with $\sigma_{\rm T}/m$ as in the cases of tidal-only or isolated (sub)halos, because both the subhalo intrinsic core-collapse and evaporation are boosted with larger $\sigma_{\rm T}/m$. Thus we only numerically map out critical boundaries in the parameter space of subhalo core-collapse with evaporation, which, as we can see in Fig. \ref{fig:param}, are more complex than the other two models.

The most important take-away message from Fig. \ref{fig:param}, however, is that core collapse is not feasible for \textit{subhalos} in constant cross section SIDM cosmologies.  Core-collapse only occurs within the Hubble time for extremely high initial concentrations (compared to the cosmological CDM subhalo concentration of $c_{200m}\sim 24$ for $M_{200m}=10^{10.5}M_\odot$ and $c_{200m}\sim 40$ for $M_{200m}=10^{8.5}M_\odot$, see \citealt{Duffy_2008}, \citealt{Diemer_2019}, \citealt{Wang_2020}). These concentrations are unphysically high for standard inflationary power spectra, although on the small halos' end they may be resulted from non-standard cosmologies such as a distorted primordial power spectrum \citep{Gosenca_2017, Delos_2018, Delos_2018b} or an early matter-dominated era \citep{Erickcek_2015, Delos_2019, Barenboim_2021}.
However, without evaporation, core-collapse can occur with typical halo concentrations if the cross section at velocity scales typical of particles within the subhalo is high.  This suggests that a strongly velocity-dependent cross section --- such that the cross section relevant to evaporation is low even as the cross section relevant to heat transport within the subhalo is high \citep{Turner_2020,Correa_2021, Jiang_2021} --- may allow for the development of core-collapsed subhalos in a strong-lens-mass host.  Alternatively, turning on another degree of freedom in the cross section, such as allowing energy dissipation in scattering \citep{Essig_2019, Huo_2020}, may allow our constraints here to be evaded.  We will explore these extra degrees of freedom, especially in the context of popular velocity-dependent models, in future work.

\section{Summary and discussion}\label{sec:s&d}

Small dark matter substructures are promising targets for constraining microphysical models and cosmological phenomenology of dark matter, and can be directly studied with substructure lensing techniques. Self-interacting dark matter, a class of dark matter model that allows for scattering between dark-matter particles, is even more promising to be constrained by substructure lensing, because core-collapsed (sub)halos may form as a result of the thermalization and heat outflow within the (sub)halo \citep{Gilman_2021, Yang_2021}. These core-collapsed subhalos make excellent lenses due to their highly concentrated central densities and convergence, bringing larger distortions to the strongly lensed images of the host halos.  However, for SIDM, because of the computational expense, traditional cosmological simulations and staged simulations with live host halos are barely able to resolve small subhalos with mass $\lesssim 10^8 M_\odot$, which is within the sensitivity scope of coming substructure lensing observations \citep{Vegetti_2014, Hezaveh_2016, Nierenberg_2014, Nierenberg_2017, Gilman_2018, Gilman_2020a, Hsueh_2020}. This is even more costly for SIDM subhalos that are core-collapsing.

In this work, we present a hybrid semi-analytical + N-body method for staged simulations of SIDM subhalo evolution with the host halo modeled as an analytical static density distribution, where for the first time the host-subhalo evaporation is properly incorporated to ensure full consistency. In principle, we are able to trace the detailed evolution of arbitrarily small SIDM subhalos with this new method.  
Our method is implemented as a patch to the SIDM module of the hydro-gravity code \texttt{Arepo} \citep{Springel_2010, Vogelsberger_2012, Vogelsberger_2014}, and the evolution of subhalos in such analytical hosts is accelerated by orders of magnitude compared to a full live host simulation. 

We validate the method, according to both the subhalo mass loss history and density/mass profile evolution, with live host runs. The mass-loss history of the subhalos in our method is consistent with the live host runs at the level of $\pm 10\%$ for sub-host mass ratio 1:1000. We show that the lingering discrepancy can be mostly explained by a small amount of dynamical friction that is missing in our analytic host model. 
Thus our method works best for relatively small subhalos ($\lesssim 1:1000$ host halo mass; for larger subhalos, simulation with a live host is the most accurate, and inexpensive as well) and we expect that the accuracy of our method increases for even smaller subhalos ($\lesssim 10^{10} M_\odot$), which are the main targets of substructure lensing in the near future. The detailed evolution of the subhalo density/mass profiles also show that our method is able to track both cored and core-collapsing subhalos at a remarkable level. We tested  constant and  velocity-dependent cross section models for code validation.  Both are in good agreement with the fully live host simulations.  We will explore the physics of core-collapse of velocity-dependent SIDM subhalos in substructure lenses in future work.

To explore the conditions for SIDM subhalos to core-collapse and study the relevant physical processes that drive such central density evolution, we ran a set of simulations scanning the subhalo parameter space of SIDM constant cross section, subhalo mass, peri-center distance and subhalo initial concentration, $[\sigma/m, M_{\rm sub}, r_{\rm peri}:r_{\rm apo}, c]$. We fixed the host mass to a typical strong lens system \citep{Birrer_2017}, and subhalo orbit apocenter at a typical infall radius \citep{Pennarrubia2010}.  We use the subhalo central density $\rho_{\rm cen50}$ as the indicator of whether and when the subhalo core-collapses, and find that $\rho_{\rm cen50}$ grows if and only if there is a net cooling (energy loss) out of the three cooling/heating terms: the heat outflow from the subhalo core as a cooling term, and host-subhalo evaporation and tidal heating as heating terms. The SIDM subhalo central density decreases first during the core-formation era, then increases as a result of the heat outflow from the core \citep{Colin_2002}. However, this increase is disrupted when the subhalo comes near its pericenter, where both the evaporation and tidal heating are the strongest. Thus we have observed three possible fates of an SIDM subhalo in our simulations: 1. that it either core-collapses before the first peri-center, 2. is largely evaporated at the first peri-center and can never core-collapse, 3. or is disrupted but not fatally enough such that its central density may resume the increase again after the peri-center passage.  In the last case, we may expect multiple cycles of growth and loss in the central density evolution,  suggesting a large diversity of subhalo density profiles.

We find that whether a subhalo core-collapses in a Hubble time, defined by its central density growing by two orders of magnitude (comparable to the criterion of \citealt{Essig_2019}), is highly sensitive to the subhalo initial parameters, especially the concentration $c$. Even a small change in $c$ may lead to completely different evolution histories. Therefore, we sample a few hundred subhalos from the four-dimensional parameter space of $[\sigma/m, M_{\rm sub}, r_{\rm peri}:r_{\rm apo}, c]$ to map the boundaries of critical conditions for subhalos to have core-collapse. We find that when host-subhalo evaporation is included, the critical/minimum $c$ is always higher by a factor of a few compared to the cosmological $c$ ($\gtrsim60$ for most $\sigma_T/m$ and orbits we consider vs. $\sim 24$ for $M_{\rm sub}=10^{10.5}M_\odot$), no matter what values we choose for the other parameters. However, for field halos or subhalos without the evaporation, the gap between the critical $c$ and the cosmological ones is greatly reduced ($\sim30$ vs $\sim 24$, for sufficiently high cross sections $\sigma_T/m \gtrsim 30\ \rm cm^2/g$), in agreement with \citealt{Correa_2021}. This suggests that it is nearly impossible to have completely core-collapsed subhalos in the simplest case of SIDM with constant cross sections. More degrees of freedom, such as velocity dependence of $\sigma_{\rm T}/m$ \citep{Vogelsberger_2012, Zavala_2013, Nadler_2020, Banerjee_2020, Turner_2020, Correa_2021} or energy dissipation during the two-body scattering \citep{Schutz:2014nka, Essig_2019, Vogelsberger_2019, Huo_2020}, must be introduced to achieve subhalo core-collapse. Subhalos with constant cross sections and only tidal effects (but not evaporation) can be seen as being the limit of an extremely velocity-dependent cross section, in the limit where the relative speed between the subhalo and host is significantly different from the subhalo internal velocity dispersion. Because the relevant heating/cooling processes are sensitive to different velocity scales, the core-collapse rate of (sub)halos can be used to constrain velocity-dependence of SIDM models as well. We plan to include such advanced models of SIDM in future work, which are simple extensions to our current method (as shown in the validation for velocity-dependent SIDM in Sec. \ref{sec:valv}).

Our work suggests additional pathways for future progress to connect substructure lensing observations with predictions: 

\begin{itemize}

  \item \emph{Particle simulation near the hydrodynamic limit.} By sampling dark matter as macro simulation particles, it is straightforward to implement the two-body scattering processes in particle-based simulations such as \texttt{Arepo}, but they are limited by the finite mass/force resolution when the central density increases exponentially, when the cores transition to the short-mean-free-path limit in core-collapse. Our core-collapse criterion (Eq. \ref{eq:cc}) is comparable to the short mean-free-path criterion of Knudsen number $Kn\sim1$ (i.e. mean free path comparable to gravity scale height) in \citealt{Essig_2019}.  We terminate the simulation at this point because the timestep required to follow the scattering accurately becomes too small. Thus we suggest that transitioning to a hydrodynamical simulation model or analytical treatment should be implemented when an SIDM halo is approaching the core-collapse stage. 
  
  \item \emph{Softening length}
  As we have mentioned in Sec. \ref{sec:val2}, although the length scale of an ultra-dense core of a core-collapsing (sub)halo could become smaller than the gravitational softening length, only a small numerical error is introduced near the core-collapse time, since the core-collapse is a fast, runaway process. However, we also note that this small error could possibly have a significant effect on when and whether a subhalo core-collapses for certain rare marginal cases, such as the one with $[M_{\rm sub} = 10^{10.5} M_\odot, \sigma_{\rm T}/m = 6\ {\rm cm^2/g}, r_{\rm peri}:r_{\rm apo} =1/10,c=75]$ (thin orange line in Fig. \ref{fig:uds-sig}), where the cooling and heating processes have a close competition right at the second pericenter, and a small offset along the time-axis could lead to dramatic difference.  This could happen when the core-collapse time of the subhalo is well-synchronised with its peri-center arrival, thus should be only on rare occasions. The force resolution for CDM subhalos should be generally good enough for SIDM subhalos.

  \item \emph{Baryonic effects.}  
  The inclusion of baryons in simulations can dramatically change predictions of halo properties relative to those from dark-matter-only simulations.
  When baryonic physics is included in SIDM simulations, the halo central density may be either more cuspy (\citealt{Kamada_2017, Sameie_2018, Robertson_2019, Sameie_2021}), or more cored (\citealt{Robles_2017}), or display a variety of cusps and cores (\citealt{Creasey_2017, Robertson_2018, Robles_2019, Despali_2019, Burger_2021}), potentially depending on the halo mass, environment (\citealt{Robles_2019}), and SIDM cross section (\citealt{Sameie_2021}).  This will affect the effective concentration of both host halos and subhalos. Higher (lower) concentrations of host halos would strengthen (weaken) the tidal effects (further strengthened with a central galaxy potential) and evaporation, and thus disrupt more (less) subhalos before they can core-collapse. In contrast, higher (lower) effective concentrations of subhalos lead to more (less) core-collapse in them \citep{Feng_2021}. The net effect of baryons can thus be complicated and diverse, and systematic studies of hydrodynamic simulations with SIDM are needed to unpack it. For relatively small subhalos within the sensitivity scope of upcoming substructure lensing observations, the baryon fraction will be small enough to be safely ignored, but the baryonic effects on the host remain significant. Since substructure lensing is most sensitive to subhalo lenses concentrated towards the main lens, modelling baryonic effects correctly and consistently becomes even more important. 
  
  \item \emph{A diversity of (sub)halo central densities.} Core-collapse of SIDM subhalos has been proposed as a plausible solution to multiple small-scale problems of CDM, such as the cusp-core problem and the too-big-to-fail problem. In this paper we show that fully core-collapsed subhalos are nearly impossible to form for SIDM with constant cross sections, given the cosmological (sub)halo concentrations, but this does not rule out SIDM as a solution to the small-scale problems. 
  With periodical ups and downs of the central density as we have observed (e.g. the green and orange thin lines in Fig. \ref{fig:uds-sig}), the subhalos could be even more diversified and interesting, even though subhalos that once entered early core-collapse stage but are evaporation-disrupted later are not counted as completely core-collapsed in our critical parameter space. 
  This might be able to explain the mystery in \citealt{Turner_2020}, that two out of nine subhalos in a Milky Way analogue host halo have cored, decreasing central densities, while the other seven show clear sign of entering core-collapsing phase \citep[see also][]{Correa_2021}. To fit the diversity of observed subhalos into SIDM, either with constant cross sections or velocity-dependent ones, a proper population modelling of the subhalos with realistic initial parameters is needed. We plan to sample multiple subhalos from realistic merger trees as an extension to our single subhalo evolution model in this work \citep{Yang_2020}, where the possible interactions between the subhalos are also naturally included. 

\end{itemize}

This work presents our first step in studying the evolution of SIDM subhalos, including the core-collapsing scenario, in the context of strong lensing systems. In the near future, we can extend our work in multiple directions, such as: a more holistic study of SIDM subhalo population with realistic assembly histories -- potentially generating mocks of substructure lensing observations as well; implementing baryonic effects, either with hydrodynamic simulations or with analytical approximations; and further simplifying our method to a purely semi-analytical approach.  These are all necessary steps to generate large ensembles of mock observations to constrain the nature of dark matter with future substructure lensing observations.

\section*{Acknowledgements}
We thank Kimberly Boddy, Hai-Bo Yu, Akaxia Cruz, Sten Delos, Anthony Pullen, Shengqi Yang, Ethan Nadler, Ivan Esteban, Daniel Gilman, Leonidas Moustakas, Charles Mace, Ekapob Kulchoakrungsun, Birendra Dhanasingham, Lingyuan Ji and Yueying Ni for useful discussions. 

This work was supported in part by the NASA Astrophysics Theory Program, under grant 80NSSC18K1014. 

MV acknowledges support through NASA ATP grants 16-ATP16-0167, 19-ATP19-0019, 19-ATP19-0020, 19-ATP19-0167, and NSF grants AST-1814053, AST-1814259,  AST-1909831 and AST-2007355.

Simulations in this work were conducted using the Pitzer Cluster (including the CCAPP condo) at the Ohio Supercomputing Center.

\section*{Data Availability}

The data supporting the plots within this article are available on
reasonable request to the corresponding author.





\bibliographystyle{mnras}
\bibliography{sidm_subhalo} 




\appendix

\section{Dynamical friction}\label{appendix:dyn}

For a subhalo with mass $M_{\rm s}$ and orbital velocity $\textbf{v}_{\rm s}$, the dynamical friction it experiences is modelled by Chandrasekhar's formula \citep{BT_2008, Petts_2015}:

\begin{equation}
    \frac{d\textbf{v}_s}{dt} = -4\pi {\rm G}^2 M_{\rm s} \rho \log({\Lambda}) f(v_s) \frac{\textbf{v}_{\rm s}}{v^3_{\rm s}},     
\end{equation}
where $\rho$ is the host density at the position of the subhalo and $v_{\rm s} = |\textbf{v}_{\rm s}|$. The function $f(v_s)$ is the velocity distribution of the host/background matter, integrated from 0 to $v_s$. When assuming Maxwellian distribution of the host dark matter velocity, $f(v_s)$ is given by 

\begin{equation}
    f(v_s) = {\rm {Erf}} \left(\frac{v_{\rm s}}{\sqrt{2}\sigma_{\rm h}}\right) - \sqrt{\frac{2}{\pi}} \frac{v_{\rm s}}{\sigma_{\rm h}} \exp\left(-\frac{v^2_{\rm s}}{2\sigma_{\rm h}^2}\right),
\end{equation}
where Erf is the Gauss error function and $\sigma_{\rm h}$ is the 1D velocity dispersion of the host dark matter at the position of the subhalo.

\citealt{Petts_2015} adopted a physically motivated model for $\log({\Lambda})=\log \left( \frac{b_{\rm min}}{b_{\rm max}} \right)$, in which both the minimum and maximum impact parameters $b_{\rm min}$ and $b_{\rm max}$ are dynamically updated as the subhalo moves in its orbit:

\begin{equation}
    \log(\Lambda) = \log\left[\frac{\min \left(\frac{\rho(R)}{\rho'(R)}, R\right)}{\max\left(r_{\rm hm}, \frac{{\rm G}M_{\rm s}}{v^2_{\rm s}}\right)}\right],
\end{equation}
where $R$ is the subhalo's distance from the host center, $\rho'(R)$ is the first derivative of host density at $R$, and the subhalo mass $M_{\rm s}$ is defined at the tidal radius. 
In our simulation with dynamical friction, we are able to evaluate $R$, $\rho(R)$, $\rho'(R)$ and $v_{\rm s}$ at any timestep. But since the subhalo mass $M_{\rm s}$ is not known until the simulation is finished and halo finder is used, we have to pre-load $M_{\rm s}(t)$ measured from a companion simulation without such dynamical friction correction as a function of time ($M_{\rm s}(t)$ has been iterated once with the dynamical friction model to be more accurate). $r_{\rm hm}$ refers to the half-mass radius within which the enclosed mass is $M_{\rm s}/2$. 

\begin{figure}
    	\includegraphics[width=\columnwidth]{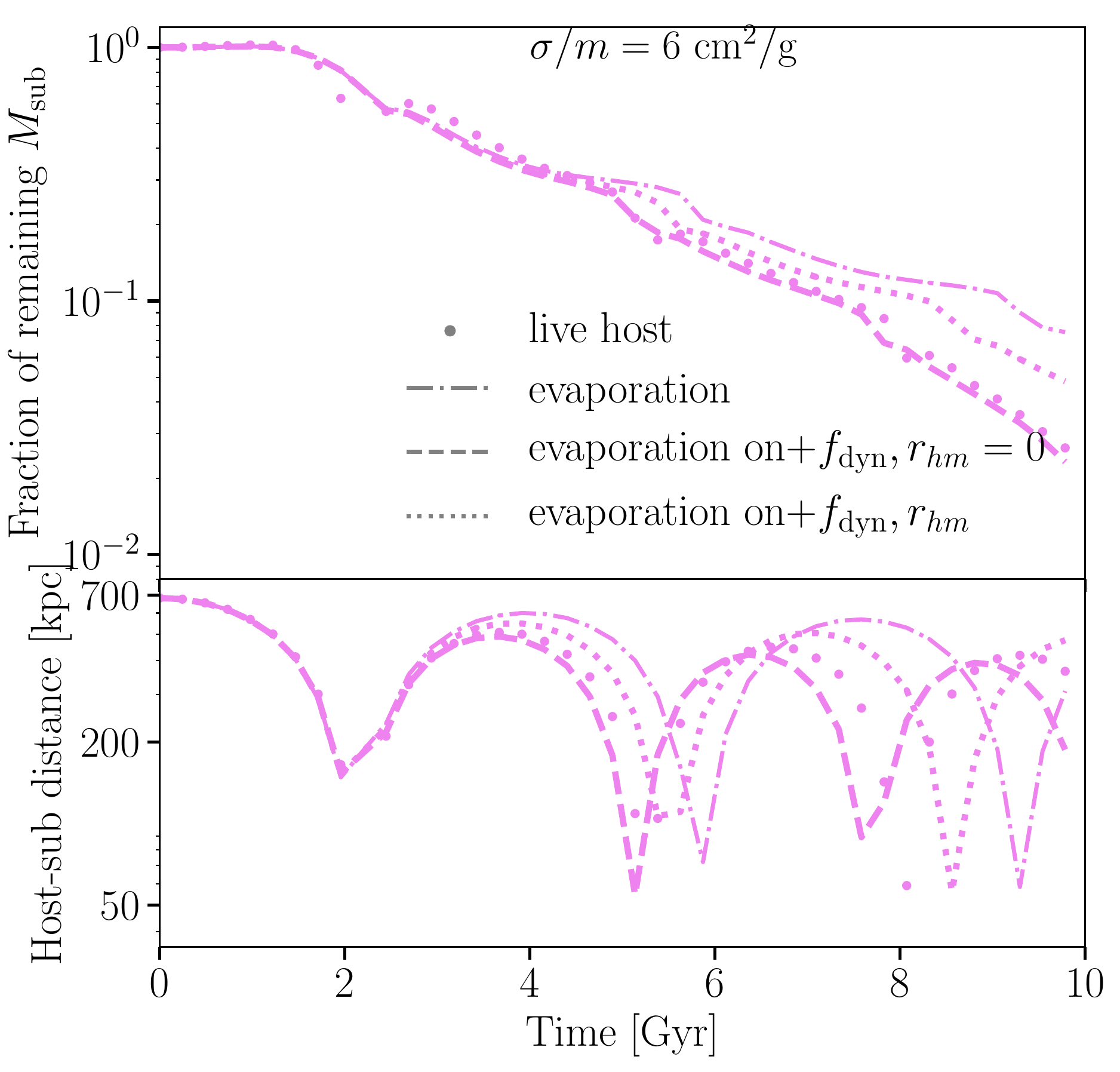}
    \caption{The mass loss history and orbital distance between subhalo and host, for subhalos with $c=40, M_{\rm sub}=10^{10.5}M_\odot, \sigma_{\rm T}/m=6\ \rm cm^2/g$, same as the violet data sets in Fig. \protect\ref{fig:val-m1}. }
    \label{fig:100-df}
\end{figure}

In practice we find that with this model of dynamical friction, the subhalo orbit in our semi-analytical framework reaches a better agreement with the live host simulation than without the inclusion of this dynamical friction term, but still find an under-decay of the orbit in the semi-analytic framework. Instead, if we set $r_{\rm hm}=0$, the discrepancy between orbits becomes smaller, with the modelling orbits having an over-decay (see Fig. \ref{fig:100-df} for a comparison). In neither case can we get dynamical friction perfectly correct and the live host orbit lies between these two cases. Thus the `true' mass loss history from our method, assuming we track the orbit decay perfectly, should lie between the dashed and dotted lines in Fig. \ref{fig:100-df} as well. Therefore, we only use the modelled dynamical friction as a demonstration, showing that the discrepancy in subhalo mass loss history is indeed reduced, if the orbits are in better agreement. 

We conclude that the missing dynamical friction is a primary systematic in our semi-analytical method of subhalo evolution. Thus, our model works best for relatively small subhalos, with a mass $\lesssim$ 1/1000 of the host, for which the dynamical friction is small, as we have shown in the main text.  And in fact, such small halos are the reason for our semi-analytic model development in the first place.

\section{Animation: density profile evolution in code validation}\label{appendix:anime-val}

\begin{frame}{}
  \animategraphics[loop,controls,width=0.9\linewidth]{3}{./val-pro/val-profile-}{0}{90}
\end{frame}

This animation shows the full evolution of density/mass profiles for the subhalo simulation from Sec. \ref{sec:val2}, corresponding to Fig. \ref{fig:val-pro}. Adobe Acrobat Reader is required to make the animation run.

\section{Animation: tidal heating}\label{appendix:anime}

\begin{frame}{}
  \animategraphics[loop,controls,width=0.9\linewidth]{3}{./tidal-pro/profile-}{0}{31}
\end{frame}

This animation shows the disturbed velocity dispersion profiles of subhalos in a tidal field, when these subhalos approach pericenter and tidal heating comes into effect. Readers are referred to Sec. \ref{sec:result2}, Fig. \ref{fig:uds-tidal}
and Fig. \ref{fig:tidal-pro} for more details. Adobe Acrobat Reader is required to make the animation run.

\section{Evaporation heating: complete thermalization limit vs instantaneous expulsion limit}\label{appendix:heating} 

The net effect of host-subhalo evaporation is complicated, since after scattering with a host particle, a subhalo particle could have a velocity that is either larger or smaller than the escape velocity of the subhalo (see also \citealt{Kummer_2018} for a quantitative discussion). When a subhalo particle's after-evaporation velocity is smaller than the escape velocity, the heat gained from this scattering event is completely absorbed by the subhalo. When its velocity is greater than the escape velocity, the major effect on the subhalo would be the adiabatic expansion caused by this instantaneous mass loss, with possible re-scatterings with other subhalo particles on its path-of-escape as a secondary effect, leading to partial thermalization. 

Thus we propose two natural limits when considering the heating of evaporation (within the subhalo core): the complete thermalization where all evaporated subhalo particles remain bound to the subhalo, thus all heat transferred to the evaporated particles are absorbed by the subhalo; and the complete instantaneous explusion limit, where all evaporated particles immediately leave the subhalo, with no re-scatterings. The complete thermalization scenario is the upper limit of evaporation heating, while the instantaneous expulsion is the lower limit.

In the complete thermalization limit, the evaporation heating per unit mass per unit time is given by the product of the event rate (probability per unit time) and the average change in subhalo particles' kinetic energy during one evaporation event:

\begin{equation}\label{eq:apd1}
    \frac{\delta E_{\rm evp(therm)}}{\delta m \delta t} = \frac{P_h}{\delta_t} \frac{<v_f^2-v_i^2>}{2}, 
\end{equation}
where $P_h/\delta t$ is given by Eq. \eqref{eqn:ph}, and $v_i$ and $v_f$ are the initial and final velocities before and after the host-subhalo particle scattering, relative to the subhalo center. We evaluate the average $<v_f^2-v_i^2>$ over the 50 particles that have the maximum local densities, to represent the center particles of the subhalo.

In the instantaneous expulsion limit, the evaporation heating per unit mass per unit time is given by the product of the event rate and the increase in subhalo center's binding energy per unit mass:

\begin{equation}\label{eq:apd2}
    \frac{\delta E_{\rm evp(expel)}}{\delta m \delta t} = \frac{P_h}{\delta_t} <-\phi-\frac{v^2}{2}>, 
\end{equation}
where $\phi$ is the gravitational potential of a subhalo particle and $v$ is its velocity relative to the subhalo center. The explusion leads to a subhalo particle's binding energy increases from ($\phi + v^2/2$) to 0. 

Note that for the expulsion limit we evaluate the averaged binding energy within the subhalo core, unlike in the thermalization limit where we fix the number of particles to take average over. This is because in the thermalization limit, the kinetic energy change is nearly universal subhalo-wide, because the velocity difference of the host-subhalo particle pair is much larger than the velocity dispersion of the subhalo. But in the expulsion limit, a subhalo particle's binding energy is sensitive to its radius from the subhalo center. Thus we must choose the core radius, $r_{\rm Lmax}$, as the boundary within which we evaluate the averaged binding energy.

\begin{figure}
    	\includegraphics[width=\columnwidth]{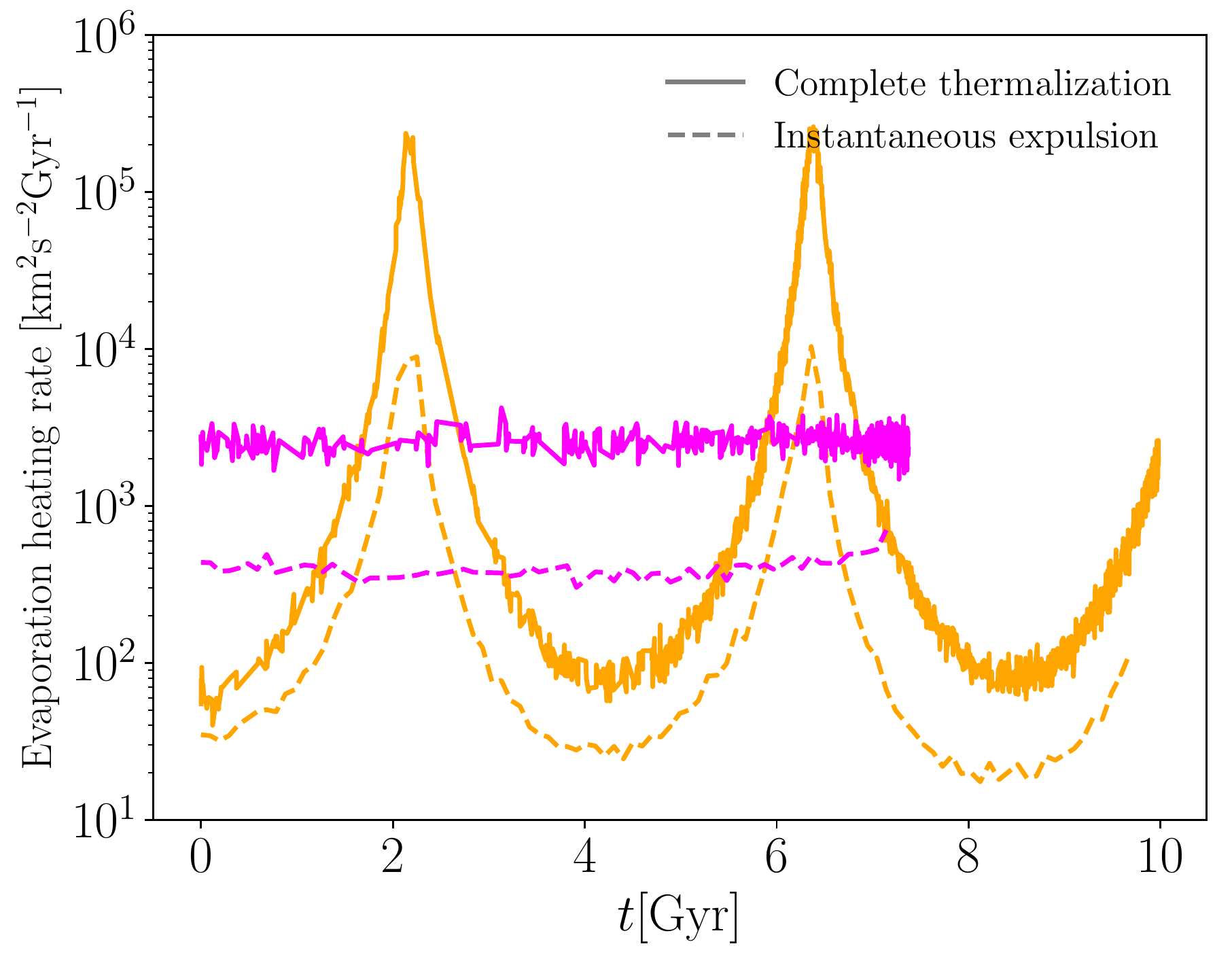}
    \caption{The evaporation heating rate, in the complete thermalization limit (solid lines) and instantaneous expulsion limit (dashed lines). The subhalo in orange color has a parameter set of [$\sigma_T/m = 6\ \rm cm^2/g$, $M_{\rm sub}=10^{10.5}M_\odot$, $c=75$, $r_{\rm peri}:r_{\rm apo} = 1:10$, $r_{\rm apo}=0.7R_{200m}$], same as the thin orange line in Fig. \protect\ref{fig:uds-sig}; and the subhalo in magenta color has [$\sigma_T/m = 6\ \rm cm^2/g$, $M_{\rm sub}=10^{10.5}M_\odot$, $c=60$, $r_{\rm peri}:r_{\rm apo} = 1:1$, $r_{\rm apo}=0.3R_{200m}$], same as the magenta solid line in Fig. \protect\ref{fig:uds-rr1}. }
    \label{fig:evp-df}
\end{figure}

As we can see from the comparison of these two limits in Fig. \ref{fig:evp-df}, the maximal heating of complete thermalization limit is indeed always larger than the minimal heating in the instantaneous limit. The difference ranges between a few times to slightly larger than one dex. For eccentric orbits, the difference between these two limits peaks at the pericenter. This is because for the maximal heating case, the change in kinetic energy per evaporation event, $<v_f^2-v_i^2>/2$ in Eq. \eqref{eq:apd1}, peaks at the pericenter where the subhalo's orbiting velocity is at maximum; but for the minimal heating case, the change in binding energy, $<-\phi-\frac{v^2}{2}>$ in Eq. \eqref{eq:apd2}, is not much affected by the orbit, or even slightly decreases at the pericenter, since the subhalo's gravitational potential gets shallower with more particles evaporated.

We expect the exact evaporation heating rate to lie between the complete thermalization limit and instantaneous expulsion limit as shown above, but it should be closer to the minimal heating of instantaneous expulsion scenario. Because of the large gap between the typical velocity of the subhalo particles and that of the host halo particles, most of the host-subhalo evaporation scatterings should lead to expulsion of the subhalo particles. Re-scatterings with other fellow subhalo particles on the escape-path of these evaporated ones should be relatively rare, as the outskirts of the subhalo are always in the long-mean-free-path regime of SIDM.

\section{Pre-evolution of subhalos}\label{appendix:preevolve}

In this work, all subhalos are initialized with NFW profiles according to the input concentrations and subhalo masses at the infall time. However, in reality, subhalos would have evolved with dark matter self-interaction for some time prior to infall, which we denote as the 'pre-evolution'. This pre-evolution time is another relevant free parameter that sets the 'realistic' initial condition of the subhalo at infall, but due to the computational cost we are not able to do a thorough analysis over another degree of freedom in this work. Below we show a simple demonstration of the effect that different pre-evolution time have on the subhalo.

\begin{figure}
    	\includegraphics[width=\columnwidth]{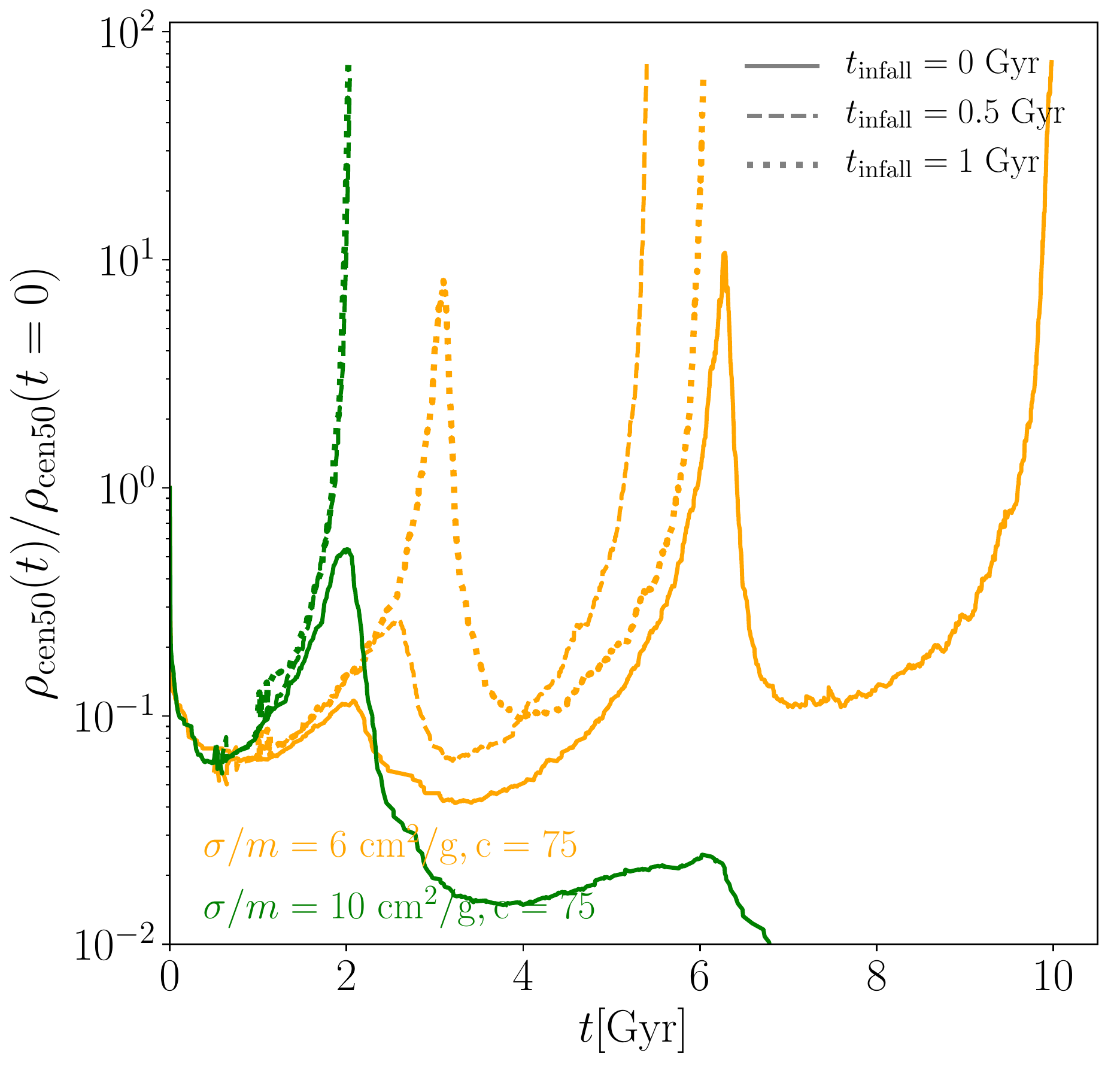}
    \caption{The evolution of central density of subhalos with different infall times. }
    \label{fig:pre-evolve}
\end{figure}

In Fig. \ref{fig:pre-evolve} we show the central density as a function of time for the two subhalos with $[M_{\rm sub} = 10^{10.5}, \sigma_T/m = 6\ \rm cm^2/g, c=75, r_{\rm peri}:r_{\rm apo} = 1:10]$ and $[M_{\rm sub} = 10^{10.5}, \sigma_T/m = 10\ \rm cm^2/g, c=75, r_{\rm peri}:r_{\rm apo} = 1:10]$ (the orange thin line and green thin line in Fig. \ref{fig:uds-sig}), with pre-evolution time of 0 Gyr, 0.5 Gyr and 1 Gyr separately. We can see that for the $\sigma_T/m = 10\ \rm cm^2/g$ subhalo, pre-evolving the subhalo for 0.5 or 1 Gyr is very different from not pre-evolving it --- the latter is completely disrupted while core-collapse of the cases with pre-evolution is not quite delayed. For the $\sigma_T/m = 6\ \rm cm^2/g$ subhalo, pre-evolving it for 0.5 or 1 Gyr still sees delay in the core-collapse time, but not as much as the case without pre-evolution ($\sim$1 orbital period vs. $\sim$2 orbital periods).

With realistic pre-evolution time, the critical concentration for subhalo core-collapse could be reduced, especially for SIDM with large cross sections, since the intrinsic core-collapse time for halos in isolation is short. our next paper (ongoing research) will be focused on simulating an ensemble/population of subhalos, with realistic assembly histories into the host. Thus the pre-evolution problem will be naturally dealt with in our next work.  

\bsp	
\label{lastpage}
\end{document}